\documentclass{aa}  

\usepackage{graphicx}
\usepackage{gensymb}
\usepackage{txfonts}
\usepackage[flushleft]{threeparttable}

\usepackage[hidelinks,colorlinks=true,linkcolor=blue,citecolor=blue]{hyperref}
\usepackage{color}
\usepackage{xcolor}
\usepackage{longtable,tabu}
\usepackage{lipsum}
\usepackage{adjustbox}
\usepackage{stfloats}
\usepackage{scrextend}
\usepackage{subcaption}
\usepackage{amssymb}
\usepackage{caption}
\usepackage{amsmath}
\usepackage{amssymb}
\usepackage{mathrsfs}
\usepackage{multicol}
\usepackage{supertabular}
\usepackage{orcidlink}
\usepackage{booktabs}

\newcommand{\statsmodels}{{\sc \tt statsmodels}\xspace}
\newcommand{\spright}{{\sc \tt spright}\xspace}
\defcitealias{Luque2022}{LP22}
\newcommand{\normrho}{$\rho_{\oplus\text{,s}}$}

\begin{document}

   \title{From super-Earths to sub-Neptunes: Observational constraints and connections to theoretical models}

\titlerunning{}
\authorrunning{L. Parc, et al.}
   \author{L{\'e}na Parc\inst{\ref{inst1}}\orcidlink{0000-0002-7382-1913}
          \and
          Fran\c{c}ois Bouchy
          \inst{\ref{inst1}}\orcidlink{0000-0002-7613-393X}
          \and
          Julia Venturini
          \inst{\ref{inst1}}\orcidlink{0000-0001-9527-2903}
          \and
          Caroline Dorn
          \inst{\ref{inst2}}\orcidlink{0000-0001-6110-4610}
          \and
          Ravit Helled
          \inst{\ref{inst3}}\orcidlink{0000-0001-5555-2652}
          }

   \institute{
        Observatoire de Genève, Université de Genève, Chemin Pegasi 51, 1290 Versoix, Switzerland.\label{inst1}
        \and
        Institute for Particle Physics and Astrophysics, ETH Zürich,  Otto-Stern-Weg 5, 8093 Zürich, Switzerland.\label{inst2}
        \and
        Center for Theoretical Astrophysics and Cosmology, Institute for Computational Science, University of Zürich, Winterthurerstrasse 190, 8057 Zürich, Switzerland.\label{inst3} \\
        \email{lena.parc@unige.ch}    
             }

   \date{Received March 9, 2024; revised May 28, 2024; accepted June 12, 2024}

 
  \abstract{The growing number of well-characterized exoplanets smaller than Neptune enables us to conduct more detailed population studies. We have updated the PlanetS catalog of transiting planets with precise and robust mass and radius measurements and use this comprehensive catalog to explore mass-radius (M-R) diagrams. On the one hand, we propose new M-R relationships to separate exoplanets into three populations: rocky planets, volatile-rich planets, and giant planets. On the other hand, we explore the transition in radius and density between super-Earths and sub-Neptunes around M-dwarfs and compare them with those orbiting K- and FG-dwarfs. Using Kernel density estimation method with a re-sampling technique, we estimated the normalized density and radius distributions, revealing connections between observations and theories on composition, internal structure, formation, and evolution of these exoplanets orbiting different spectral types. First, the substantial 30\% increase in the number of well-characterized exoplanets orbiting M-dwarfs compared with previous studies shows us that there is no clear gap in either composition or radius between super-Earths and sub-Neptunes. The "water-worlds" around M-dwarfs cannot correspond to a distinct population, their bulk density and equilibrium temperature can be interpreted by several different internal structures and compositions. The continuity in the fraction of volatiles in these planets suggests a formation scenario involving planetesimal or hybrid pebble-planetesimal accretion. Moreover, we find that the transition between super-Earths and sub-Neptunes appears to happen at different masses (and radii) depending on the spectral type of the star. The maximum mass of super-Earths seems to be close to 10~M$_\oplus$ for all spectral types, but the minimum mass of sub-Neptunes increases with the star's mass, and is around 1.9~M$_\oplus$, 3.4~M$_\oplus$, and 4.3~M$_\oplus$, for M-dwarfs, K-dwarfs, and FG-dwarfs, respectively. The precise value of this minimum mass may be affected by observational bias, but the trend appears to be reliable. This effect, attributed to planet migration, also contributes to the fading of the radius valley for M-planets compared to FGK-planets. While sub-Neptunes are less common around M-dwarfs, smaller ones (1.8 R$_\oplus$ < R$_p$ < 2.8 R$_\oplus$) exhibit lower density than their equivalents around FGK-dwarfs. Nonetheless, the sample of well-characterized small exoplanets remains limited, and each new discovery has the potential to reshape our understanding and interpretations of this population in the context of internal structure, composition, formation, and evolution models. Broader consensus is also needed for internal structure models and atmospheric compositions to enhance density interpretation and observable predictions for the atmospheres of these exoplanets.}


   \maketitle
%

\section{Introduction}

Over the last three decades, planetary sciences have undergone a revolutionary transformation, with the identification of over 5\,500 exoplanets orbiting stars within our galaxy. The majority, over 4\,100, were discovered using the transit method, thanks to the efforts of space telescopes conducting large-scale surveys like CoRoT \citep{Baglin2006}, \textit{Kepler}/K2 \citep{Kepler,K2}, and TESS \citep{TESS}. Notably, planets with sizes smaller than Neptune ($<$ 4 R$_\oplus$), are prevalent. Approximately every star is known to host such planets, detected at an orbital period of less than 100 days \citep{Kepler,Batalha2011}.

M-type stars, being the most abundant stellar type in our galaxy \citep{Winters2015}, appear to have a higher occurrence rate of exoplanets \citep{Dressing2013,Bonfils2013,Mulders2015,Gaidos2016}, especially small ones \citep{Bonfils2013,Dressing2015}. Their planets are easier to detect and study due to the smaller size and lower mass of their host star. The full sky coverage of TESS compared to \textit{Kepler} has significantly increased the detection of planets around these stars. This was further enhanced by ground-based instruments and telescopes dedicated to infrared photometry such as TRAPPIST \citep{TRAPPIST}, SPECULOOS \citep{Delrez2018}, or ExTrA \citep{ExTrA} that not only validated the TESS/ \textit{Kepler} candidates but also discovered new M-planets. Additionally, the development of high-resolution near-infrared spectrographs like CARMENES \citep{CARMENES}, MAROON-X \citep{MAROONX}, SPIRou \citep{SPIRou}, or NIRPS \citep{Bouchy2017} enabled efficient follow-up of these transiting exoplanets to obtain mass measurements. These surveys have led to a substantial increase in the number of planets around low-mass stars with precise measured mass and radius. This, in turn, facilitates statistical studies on larger populations, allowing more reliable conclusions.

The mass-radius (M-R) diagram stands out as a powerful tool for investigating the demographics of exoplanets. It serves various purposes, such as deriving mass-radius relationships for predicting the estimated masses of transiting objects in radial velocity follow-up campaigns \citep{Weiss2014, Wolfgang2016,Chen2017,Bashi2017,Kanodia2019,Otegi2020,Edmonson2023,Parviainen2023}. Despite the degenerate nature of the problem, the M-R diagram plays a crucial role in exploring the composition and internal structure of exoplanets. They enable us to study the radius and mass distributions of exoplanets, offering insights into their bulk density, and facilitating comparisons with theoretical models. 

One of the best-known features of the radius distribution of small planets is the "radius valley" \citep{Fulton2017}. It separates two sub-populations of super-Earths and sub-Neptunes by a gap at $\sim$ 1.5-2 R$_\oplus$, with the exact location depending on the mass of the host star \citep{Fulton2018,Berger2020,Ho2023}. Different theories contribute to the understanding of the origin of the radius valley. Some of them focus exclusively on processes happening after the disk dissipation, such as photoevaporation \citep{Owen2017,Jin2018} and core-powered mass-loss \citep{Ginzburg2018,Gupta2019}. Other theories combine formation and evolution processes into a single framework to explain the radius valley \citep{Venturini2020, Venturini2024, Burn24}.

In the view of the pure evolution models (e.g., photoevaporation or core-powered mass-loss), some of the planets lose their atmospheres, evolving into super-Earths, while those retaining their atmospheres are classified as today's sub-Neptunes. Based on the assumption of a unique core composition and H/He atmospheres, such models find that the correct location of the valley can only be reproduced if the underlying cores are rocky, implying a formation inside the water ice line for both super-Earths and sub-Neptunes \citep{Owen2017, Gupta2019}. 
On the contrary, combined formation and evolution models predict that sub-Neptunes are typically water-worlds assembled beyond the water ice line \citep{Venturini2020, Venturini2024, Burn24}.
In particular, \citet{Venturini2020} showed that larger icy versus smaller rocky cores form naturally by pebble accretion, due to the difference of sticking properties between dry and icy pebbles. The separation between the two populations occurs exactly at the location of the observed gap. However, this bi-modality in mass and size distribution is hidden at the end of the formation phase due to the presence of gaseous atmospheres. Photoevaporation is needed to remove the gas and reveal the radius valley that was set by core formation in the first place. \citet{Venturini2024} expand these models for a range of stellar masses, pointing out the role of migration in the fading of the radius valley toward M-dwarfs, as well as the effect of the phases of water in the enhancement of the radius valley. \citet{Burn24} perform calculations similar to \citet{Venturini2020}, but for planetesimal accretion, highlighting the role of steam atmospheres in the emergence of the radius valley. However, these previous studies neglect any compositional coupling of water between the deep interior and the surface \citep{LuoDorn2024}.

The characterization of sub-Neptune-sized planets is an active area of study. These planets, unlikely to have a pure rocky composition \citep{Rogers2015,Fulton2017}, inhabit a degenerate region in the M-R parameter space. Their bulk densities can be equally well explained by solid rocky and iron cores with primordial gaseous hydrogen-rich atmospheres ("gas dwarfs", \citealt{Lopez2014,Rogers2023}) or by a water-rich interior and/or atmosphere ("water-worlds", \citealt{Leger2004,Dorn2021,Aguichine2021,Luque2022}). 

Around M-dwarfs, the existence of this radius valley remains unclear \citep{Cloutier2020,Luque2022,Ho2024} and the existence of a water-world population is still debated \citep{Luque2022,Parviainen2023}. \citet{Luque2022} propose the existence of a compositional gap separating rocky from water-rich planets. They claim that the small transiting planets orbiting M-dwarfs can be categorized into three distinct groups ("rocky planets", "water-worlds", and "puffy sub-Neptunes"), rather than constituting a seamless continuum. Among these groups, two are positioned along specific compositional lines in the M-R diagram: one representing Earth-like composition ("rocky planets", noted \normrho), and the other representing a composition of 50\% condensed water and 50\% silicates ("water-worlds"). An examination of the density and radius distributions within these three sub-categories indicates an absence of a radius valley. However, they find a compositional gap at a normalized density value of 0.65~\normrho, where there is an absence of planets and so no overlap observed between the distribution of "rocky planets" and that of "water-worlds". On the contrary, \citet{Parviainen2023} shows strong support against the existence of a water-world population around M-dwarfs using the Python module \spright \protect\footnote{\url{https://github.com/hpparvi/spright}} and an updated sample from \citet{Luque2022}. Indeed, they present \spright, a novel probabilistic mass-density-radius relationship for small planets. It employs a Bayesian model comparison to represent the joint planetary radius and bulk density probability distribution as a mean posterior predictive distribution of an analytical three-component mixture model. The three components represent rocky planets, water-rich planets, and sub-Neptunes, and the final numerical probability model is obtained by marginalizing over all analytical model solutions allowed by observations.

From a formation point of view, to explain this population of "water-worlds" around M-stars, theoretical studies by \citet{Alibert2017b}, \citet{Miguel2020}, and \citet{Burn2021} suggest that low-mass water-rich planets should be prevalent in close-in orbits around M-dwarfs. This is attributed to the proximity of water ice lines to the stars and the increased efficiency of inward migration at lower planetary masses around lower-mass stars. Alternatively, formation models that do not consider planetary migration show that these low-density planets orbiting M-dwarfs could consist of rocky cores surrounded by H/He envelopes \citep{Lee2022}.

In this work, we update the PlanetS catalog initially published by \citet{Otegi2020} of transiting planets with precise mass and radius measurements and the associated mass-radius relationships in Section~\ref{sect:planets_MRrelations}. Subsequently, we conduct a comprehensive examination of the M-R diagram for small planets around M-dwarfs in Section~\ref{sect:MR_Mdwarfs}. This includes deriving distributions of radius and density (Sect.~\ref{sect:M_distribs}) for analysis and discussing observed properties with theoretical models of internal structure (Sect.~\ref{sect:M_composition}) and formation and evolution (Sect.~\ref{sect:M_formation}). We further broaden our investigation to include planets orbiting FGK-type stars, allowing a comparative analysis of the distinctions between the population of planets orbiting M-dwarfs and those around earlier spectral types in Section~\ref{sect:MR_FGK}. The conclusions drawn from our study are presented in Section~\ref{sect:conclusions}.


\section{The PlanetS catalog: Precise and reliable mass and radius for transiting planets}\label{sect:planets_MRrelations}

\begin{figure*}[t]
\centering
\sidecaption
   \includegraphics[width=12cm]{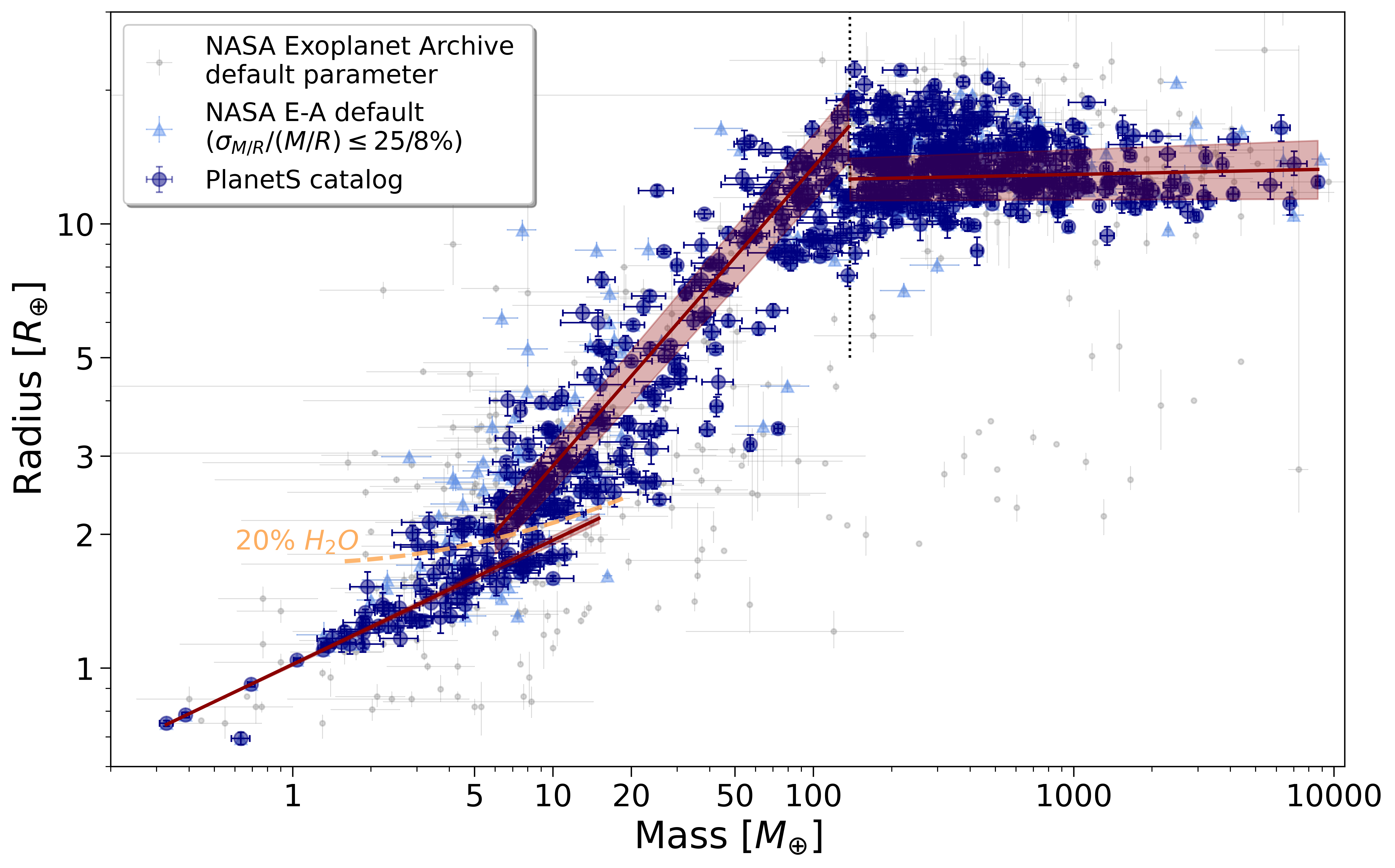}
     \caption{Mass-radius diagram comparing the data from the PlanetS catalog and the data from NASA Exoplanet Archive to highlight the effects of our selection criteria and our detailed study of references. The gray dots represented the default parameters from NASA, the light blue triangles the same but filtered with our criteria on uncertainties and the blue dots correspond to the PlanetS catalog. The resulting M-R relationships from Sect.\ref{sect:mrrelations} are shown as solid red lines with the confidence intervals represented by the shaded red areas. The transitions between the three populations are represented by the composition line of 20\% water at 650~K from \citet{LuoDorn2024} in dashed orange and the dotted black line at 138~M$_\oplus$.}
     \label{fig:PlanetSvsNASA}
\end{figure*}

\subsection{Update and redesign of the catalog}\label{sect:planets_catalog}

We present here the PlanetS catalog\protect\footnote{\url{https://dace.unige.ch/exoplanets/}\label{catalogfootnote}} of transiting planets with reliable parameters. Initially published in \citet{Otegi2020}, where it was limited to 120 M$_\oplus$, the catalog has since been extended to all planetary masses, then updated with new discoveries and revisited parameters of old systems, and modified to include more planetary and stellar parameters.

The overall aim of this catalog is to bring together the orbital, stellar, and planetary parameters of transiting exoplanets with precise and reliable mass and radius measurements, and to keep it up to date. To achieve this, as in \citet{Otegi2020}, the catalog update is based on the NASA Exoplanet Archive,\protect\footnote{\url{https://exoplanetarchive.ipac.caltech.edu}\label{NASAArchivefootnote}} as this is the most up-to-date and complete catalog in terms of parameters and references for each exoplanet detected. Selection criteria are also identical to \citet{Otegi2020}:

\begin{itemize}
    \item We consider planets with relative measurement uncertainties smaller than 25\% in mass ($\sigma_M{_p}/M_p \leq$ 25\%) and 8\% in radius ($\sigma_R{_p}/R_p \leq$ 8\%).
    \item Reference is checked before a planet is added to the catalog to ensure the reliability and robustness of the analysis of the photometric and spectroscopic data.
    \item Among several possible references, the most precise (without underestimated uncertainties) and the most recent is chosen. A single reference is chosen for a given system, even if the precision for one of the planets is better in another study, in order to maintain consistency.
\end{itemize}
A special case here are planets whose mass is determined by Timing Transit Variations (TTVs). The robustness of these determinations is more difficult to assess, as TTVs models often give several solutions. Like \citet{Otegi2020}, we only consider planets whose mass estimations have been shown to be robust against mass and eccentricity degeneracy \citep{Lithwick2012,HaddenLithwick2017,Leleu2023}. Indeed, inferred planet masses are considered "robust" if the 1$\sigma$ credible region of the default posterior (logarithmic mass and uniform eccentricity priors) excludes zero and includes the peak of the high-mass posterior (uniform mass and logarithmic eccentricity priors). This criterion is important, because if the mass posterior is strongly dependent on the prior, then the data are considered insufficient to lift this degeneracy. We therefore ignore references giving only one solution of TTVs or two solutions that disagree.

Thresholds for uncertainties have been maintained since \citet{Otegi2020} and correspond to median uncertainties at that time, with the same impact for mass and radius on density uncertainty. Nevertheless, advancements in detection and analysis techniques suggest that these medians would now hover around 15\% for relative mass error and 5\% for relative radius error (retaining this factor of 3 for the impact on density error). However, adopting this new criterion would result in a 40\% reduction in the total sample size, and sub-samples would be affected to an even greater extent, potentially compromising the statistical analysis's robustness. We have therefore chosen to keep the original thresholds, leaving open the possibility of reducing them for a given study. 

Finally, on the 11th of December 2023, among the 5\,557 known exoplanets, the PlanetS catalog contains 715 transiting exoplanets. Figure \ref{fig:PlanetSvsNASA} illustrates the M-R diagram derived from our catalog, comparing it to the NASA Exoplanet Archive database. The comparison is made using both the default parameter of the NASA Exoplanet Archive and this default parameter but filtered with the same uncertainty criteria as ours. This provides insight into the effects of our selection process: first, our uncertainty thresholds, and then our detailed study of references, enabling the removal of less reliable data points. Exoplanets identified using the default parameter from the NASA Exoplanet Archive often lack mass uncertainty information or provide only upper limit masses, leading to a wide dispersion in the data. While the uncertainty criterion of the PlanetS catalog can help address this problem, challenges persist with masses determined through TTVs, where verifying robustness is difficult. Additionally, planets analyzed spectroscopically may exhibit underestimated uncertainties in their masses, especially when radial velocity curves include only few data points. This underscores the importance of our comprehensive study of references in addressing these issues. 

Another important goal of the catalog modification was to offer the community a complete catalog in terms of parameters. To accomplish this, we sourced the primary orbital, planetary, and stellar parameters from the NASA Exoplanet Archive, with the reference paper selected. Any missing parameters were supplemented by cross-checking with the study due to potential import issues between the paper and the database. In addition, a cross-match between catalogs was performed to obtain stellar parameters from Gaia DR3\protect\footnote{\url{https://gea.esac.esa.int/archive/}\label{Gaiafootnote}} (\citealt{gaia2016}; \citealt{gaia2023}), as well as orbital obliquities from TEPCat\protect\footnote{\url{https://www.astro.keele.ac.uk/jkt/tepcat/}\label{TEPCATfootnote}} (\citealt{Southworth2011}). We assigned each star's spectral type using the table in \citet{Pecaut2013}, taking the star's effective temperature from the reference study.

We have also included classical calculations in the database - planet bulk density, insolation flux, equilibrium temperature, and expected radial velocity semi-amplitude - in order to have these values for all planets in a more homogeneous way (see appendix \ref{appendix:calculations}). 

In December 2023, the PlanetS catalog includes 715 transiting exoplanets along with over 40 parameters and their respective uncertainties. This comprehensive catalog is accessible through the Data \& Analysis Center for Exoplanet (DACE).\protect\footnote{\url{https://dace.unige.ch/}\label{dacefootnote}}

\subsection{Updated mass-radius relationships}\label{sect:mrrelations}

The PlanetS catalog created by \citet{Otegi2020} aimed to investigate mass-radius (M-R) relationships and the transitional boundaries between rocky and volatile-rich planets. The present goal is to update these M-R relationships with the updated PlanetS catalog that includes a broader range of masses and new discoveries. 

Numerous studies propose M-R relationships for the observed planetary population, each with unique characteristics in terms of sample selection, laws employed, or methods utilized to identify transitions between different planetary populations. 
For instance, \citet{Chen2017} suggest a mass-based separation for super-Earths and sub-Neptunes at 2 M$_\oplus$, while \citet{Weiss2014} propose a radius-based transition at 1.5 R$_\oplus$. \citet{Wolfgang2016} provide a best fit “average” relation for the sample of RV-measured transiting sub-Neptunes (R $<$ 4 R$_\oplus$), and quantify the astrophysical scatter of this relation. \citet{Otegi2020} modified the perspective on these relationships by introducing a discontinuity and segregating rocky planets from more volatile-rich ones using the equation of state for pure water. This introduced a physically grounded criterion for distinguishing between the two populations. However, with the updated catalog, the discovery of planets spanning the gap between more massive rocky planets and less massive volatile-rich ones has made the transition between these populations less clear. It is plausible that the transition between rocky and volatile-rich populations depends on other parameters, such as insolation or stellar characteristics. Still, recently, \citet{Edmonson2023} provided evidence supporting the consistency and validity of using the pure water composition line for this transition. Nevertheless, this composition line lacks physical validity, as a planet with such a composition would be unable to form and sustain its existence. In a different way, \citet{Kanodia2019} and \citet{Parviainen2023} present Python modules that can be used to fit a M-R relationship without assuming an underlying power law. \citet{Kanodia2019}, however, do not consider any separation between the different populations, whereas \citet{Parviainen2023} use composition lines to represent 3 populations, "rocky planets", "water-worlds", and "sub-Neptunes", constituting the three-component mixture model.

Regarding the transition to giant planets, \citet{Weiss2013} identified a transition at 150~M$_\oplus$ based on visual inspection of mass-radius and mass-density diagrams. \citet{Hatzes2015} conducted a similar analysis, examining the change in slope in mass-density relationships, and identified the transition at 95~M$_\oplus$. \citet{Chen2017} utilized a detailed forecasting model employing a probabilistic M-R relation, placing the transition at 130$\pm$22~M$_\oplus$. In contrast, \citet{Bashi2017} identified two empirical regimes on the M-R relation with a transition at 124$\pm$7~M$_\oplus$ and 12.1$\pm$0.5~R$_\oplus$. \citet{Edmonson2023} employed piecewise regression and located the transition at 115~M$_\oplus$. However, when considering the equilibrium temperature of the gas giants, it was found that, for giant planets, the radius only depends on the temperature.
Finally, \citet{Muller2023} have leveraged the updated PlanetS catalog to investigate M-R relationships in a purely statistical manner, finding the transitions to be at 7.80~M$_\oplus$ and 125~M$_\oplus$ to separate the three populations.

\begin{figure*}[t]
\centering
   \includegraphics[width=19cm]{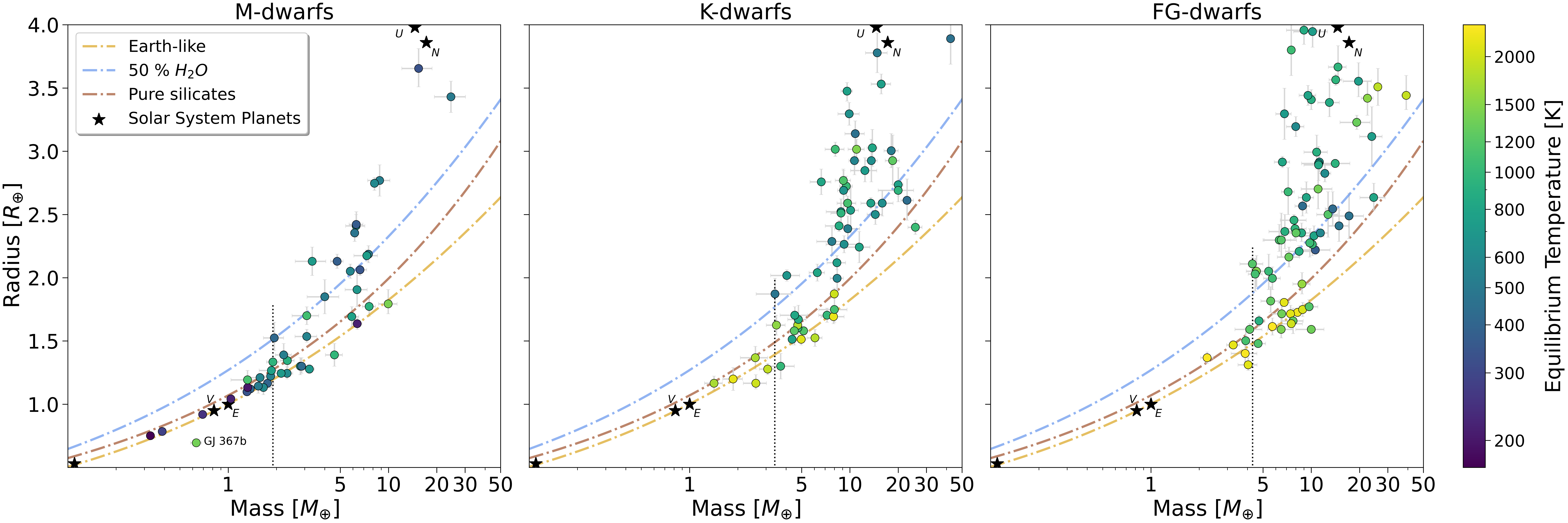}
     \caption{M-R diagram of small planets around M-dwarfs, K-dwarfs, and FG-dwarfs from the PlanetS catalog. The planets are color-coded by their equilibrium temperature calculate within the catalog with a bond albedo of 0 and a total heat redistribution. The composition lines of pure-silicates (brown) from \citet{Zeng2016}, Earth-like planets (yellow), and 50\% water (blue) from \citet{Zeng2019} are displayed. The vertical dotted lines correspond to the minimum mass of the sub-Neptunes across spectral types (at 1.9, 3.4, and 4.3~M$_\oplus$ for M-, K-, and FG-dwarfs) discussed in Sect. \ref{sect:sample_selection_FGKdwarfs} and \ref{sect:formation_MKFG}.}
     \label{fig:MR_MKFGdwarfs}
\end{figure*}

In this study, we decide to use the 20\% water line at 650~K from \citet{LuoDorn2024} to distinguish between rocky planets and those with higher volatility. This model is described in more details in Section \ref{sect:M_composition} and allows us to separate the two groups in a way that is consistent with what we observe visually. Finally, this 20\% is consistent from a formation point of view, with the water fraction corresponding to a body that has to have formed beyond the ice line of the planetary disk. We also limit ourselves to planets below 10 M$_\oplus$ for the rocky population as can be seen in the observations (see next Sections and Fig.~\ref{fig:MR_MKFGdwarfs}). For the transition between intermediate-mass and giant planets, we employ the piecewise regression Python package from DataDog,\protect\footnote{\url{https://github.com/DataDog/piecewise}\label{datadogfootnote}} which applies piecewise regression with automated breakpoints detection, coupled with a MCMC to take into account the uncertainties of the mass and radius measurements. We find the transition at 138$\pm^{+21}_{-42}$~M$_\oplus$. This is in line with the values found in the literature, and is consistent with planet formation theory \citep{Helled2023}. Indeed, \citet{Helled2023} suggested that the transition to gas giant planets occurs at about Saturn's mass, with an uncertainty of $\sim 20$ M$_\oplus$, this means that above this mass, the planetary composition is expected to be H-He dominated. Below this mass, however, the composition is expected to be very diverse, and while some H-He is expected, the ratio between H-He and heavy elements can vary significantly, depending on the exact formation conditions. As a result, for intermediate-mass planets there should not be a distinct density or radius as in the case of small and giant planets. Instead, for this planetary type the planetary mass is the physical property that is more natural to use when separating the different populations.

We fit the three populations using a Orthogonal Distance Regression (ODR) method, in which observational errors on both dependent and independent variables are considered, and the results are the following:
\begin{multline*}
R= \begin{cases}(1.02 \pm 0.01) M^{(0.28 \pm 0.01)} & , \text { if } \rho>\rho_{20\%\mathrm{H}_2 \mathrm{O}} \text { and } M<10M_{\oplus}  \\ (0.61 \pm 0.04) M^{(0.67 \pm 0.02)} & , \text { if } \rho<\rho_{20\%\mathrm{H}_2 \mathrm{O}} \text { and } M<138M_{\oplus} \\ (11.9 \pm 0.7) M^{(0.01 \pm 0.01)} & , \text { if } M>138M_{\oplus}\end{cases}
\end{multline*}
or
\begin{multline*}
M= \begin{cases}(0.98 \pm 0.01) R^{(3.57 \pm 0.13)} & , \text { if } \rho>\rho_{20\%\mathrm{H}_2 \mathrm{O}} \text { and } M<10M_{\oplus}  \\ (1.64 \pm 0.11) R^{(1.49 \pm 0.04)} & , \text { if } \rho<\rho_{20\%\mathrm{H}_2 \mathrm{O}} \text { and } M<138M_{\oplus} \\ (0.08 \pm 0.01) R^{(100 \pm 100)} & , \text { if } M>138M_{\oplus}\end{cases}
\end{multline*}
Despite the change in separation between the two groups, our two relations for small and intermediate-mass planets align with \citet{Otegi2020}. The increased number of planets in the "rocky" category suggests a slight elevation in the exponent of the power law, but this remains within the error bars of the equation from \citet{Otegi2020}. 
They also exhibit slight deviations from \citet{Edmonson2023}'s relations. This discrepancy can be attributed to differences in object selection between our two samples and the different separation criteria. In the gas giant regime, our relation is relatively close to the one presented in \citet{Bashi2017}. The observed population in this regime displays significant dispersion, with a notable dependence on incoming irradiation. Planets subjected to higher irradiation tend to exhibit hotter and more expanded atmospheres, contributing to the observed variability \citep{Batygin2011,Sestovic2018,Thorngren2018}.

However, it's worth remembering that no physical interpretations about composition, internal structure or formation should be drawn from these M-R relationships and purely arbitrary category separations. For example, the composition line of 20\% water depends on the planet's temperature (see Sect. \ref{sect:M_composition}) and so a more precise examination of all conditions must be made in order to assert the rocky or more volatile nature of an individual planet. A good use of these relationships is to estimate the mass and/or semi-amplitude of the radial velocity signal during radial velocity follow-up campaigns of planets detected by TESS or other photometric spacecrafts and instruments. For this purpose, we recommend employing the relationship for rocky planets up to 1.75~R$_\oplus$. From 1.75 to 2.1~R$_\oplus$, it's advisable to use both the "rocky" and "intermediate" relationships. Beyond that, the intermediate relationship serves as a good estimate. 
These values are derived from the minimum and maximum radius values of the two currently observed populations, with an added safety margin. It's important to note that the dispersion of the observed population remains extensive around these relations, indicating that the estimated mass may vary significantly for a given radius. Using the two relations "rocky" and "intermediate" with their uncertainties, for R = 1.75~R$_\oplus$, the mass varies from 3.45 to 7.85~M$_\oplus$ and for R = 2.1~R$_\oplus$, it varies from 4.49 to 14.31~M$_\oplus$.


\section{M-R diagram of small exoplanets around M-dwarfs}\label{sect:MR_Mdwarfs}

\subsection{Sample selection}\label{sect:sample_selection_Mdwarfs}

We use our robust and comprehensive PlanetS catalog to select the sample of small planets around M-dwarfs. We apply selection criteria similar to those of \citet{Luque2022} (hereafter \citetalias{Luque2022}) in order to compare our results. We choose planets with a radius of less than 4 R$_\oplus$, orbiting stars with an effective temperature of less than 4000~K. We obtain 46 planets with precise mass and radius measurements orbiting stars with an effective temperature ranging from 2566 to 3997~K corresponding to stellar masses from 0.09 (TRAPPIST-1) to 0.63~M$_\odot$ (TOI-1235). This sample contains planets all robustly confirmed with photometric and spectroscopic instruments, including new discoveries up to December 2023. By way of comparison, the sample of \citetalias{Luque2022} study contains 34 planets. A complete comparison between the two samples are detailed in the appendix \ref{appendix:comparison_samples_M}. 

The resulting M-R diagram for our sample is shown in the first panel of Fig.~\ref{fig:MR_MKFGdwarfs}, with the planets colored by their equilibrium temperature computed in the PlanetS catalog (appendix \ref{appendix:calculations_Teq}, using an albedo of 0). For comparison the two theoretical composition lines used by \citetalias{Luque2022} (Earth-like composition and 50\% water/50\% silicates from \citet{Zeng2019}) are drawn in Fig.~\ref{fig:MR_MKFGdwarfs}, but are discussed in more details in Section \ref{sect:M_composition}. We also draw the pure-silicates composition line \citep{Zeng2016} to represent the lowest density limit of a rocky planet.

A first examination of this M-R diagram shows us that there are exoplanets lying between the two composition lines of Earth-like and 50\% water shown in the figure. These are both new planets characterized after \citetalias{Luque2022}'s paper catalog (July 2021), but also some of the exoplanets whose properties they have re-analyzed (see Appendix \ref{appendix:comparison_samples_M}). The substantial 30\% increase in the number of planets within the sample has considerable implications for the conclusions drawn regarding planetary populations. It clearly demonstrates the importance of acquiring more data in a parameter space that is inherently incomplete and biased, such as the M-R diagram. 

A subset of exoplanets closely follows the Earth's composition line, featuring masses ranging from TRAPPIST-1 h (0.33 M$_\oplus$) to 10 M$_\oplus$ and associated radii of approximately 0.8-1.8 R$_\oplus$. Within this cluster, a prominent group of planets, known as super-Earths, exhibits masses around 2 M$_\oplus$ and radii around 1.2 R$_\oplus$. Following this, there are planets with larger radii (1.5-2.8 R$_\oplus$) and masses between 2 and 10 M$_\oplus$, identified as sub-Neptunes positioned above the pure silicate line. These sub-Neptunes are presumed to contain more volatile elements to account for their observed density. Subsequently, there are only two planets with radii around 3.5 R$_\oplus$ (K2-25 b : R$_p$ = 3.43 $\pm$ 0.12 R$_\oplus$ \citep{Stefansson2020} ; and TOI-1231 b : R$_p$ = 3.43$^\text{+0.16}_\text{-0.15}$ R$_\oplus$ \citep{Burt2021}) and very few objects surpassing 2.5 R$_\oplus$. 
Another noteworthy feature in this diagram is the ultra-dense sub-Earth GJ 367 b with a mass of 0.633~M$_\oplus$ and a radius of 0.699~R$_\oplus$ \citep{Goffo2023}. This planet boasts an exceptionally short orbital period (P = 0.32 days) and experiences intense irradiation (T$_{\text{eq}}$ = 1365 K). The phase curve from JWST revealed that it appears to be a hot, dark barren rock \citep{Zhang2024}. Positioned just below a theoretical 100\% Iron composition line, GJ 367 b stands out as the densest planet in our sample.

\subsection{Density and radius distributions}\label{sect:M_distribs}

\subsubsection{Histograms}\label{sect:hist_method}
To evaluate the probability distribution of a given variable, the first idea that naturally comes to mind is to use a histogram - it's well-known, simple to understand and works reasonably well. But histograms have clear limitations: the positioning and quantity of bins are arbitrarily determined and the estimated distribution lacks smoothness, whereas the true distribution has a continuous, smooth nature (\citealt{Boels2019}). Moreover, as they stand, they do not take into account the uncertainties of the chosen variable. 
The limits are even greater when a histogram is applied to a very small sample (see Appendix \ref{appendix:hist_limits}). So the first objective is to consider the overall sample to plot the histogram. 

\begin{figure}[t]
  \centering
    \resizebox{\hsize}{!}{\includegraphics{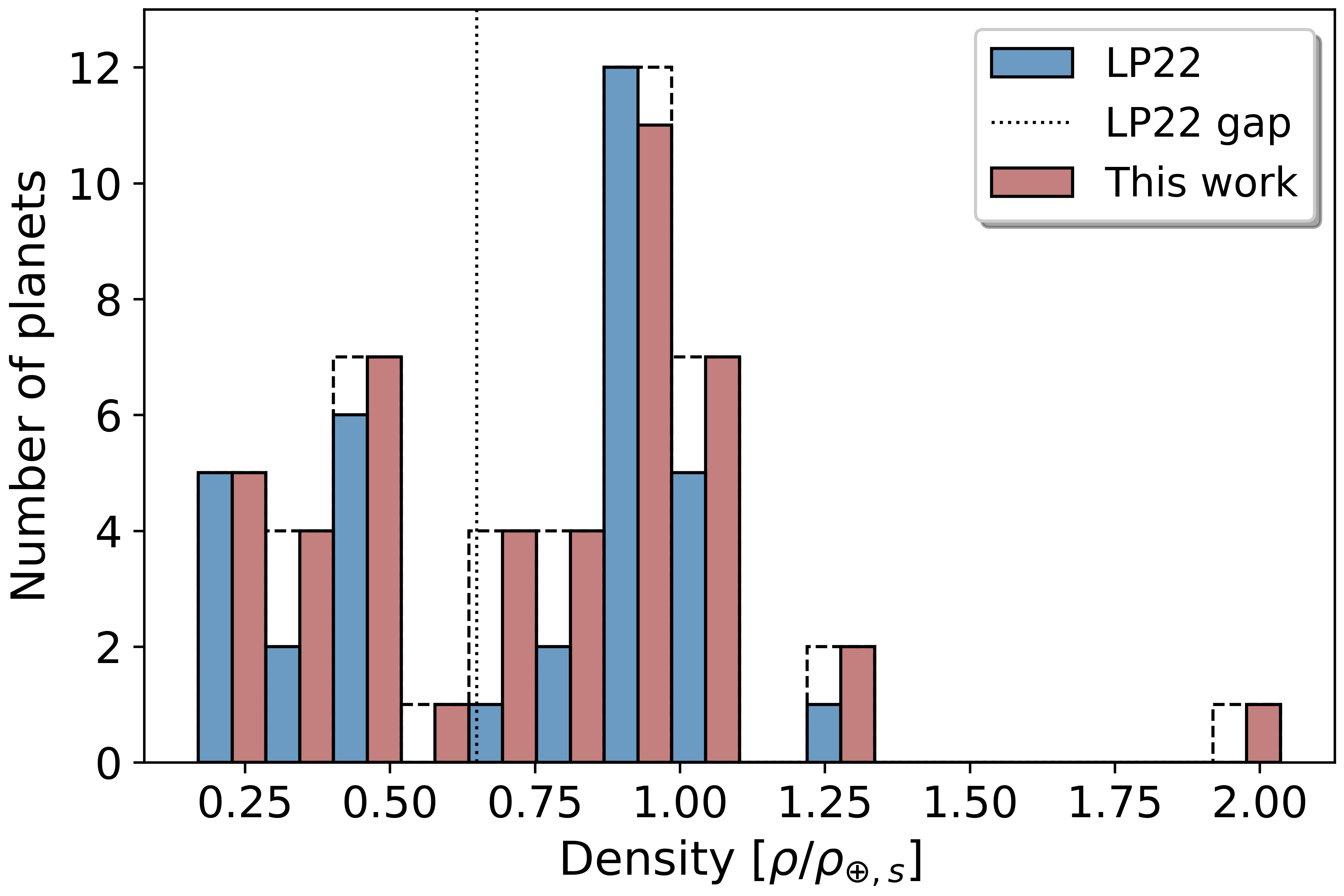}}
      \caption{Histogram of the  normalized density distribution of small planets orbiting M-dwarfs. In blue, for the \citetalias{Luque2022} sample. In pink, for this study. For reasons of clarity, the bins are separated in two with the two colors representative of the two samples. A dotted line is drawn at 0.65~\normrho~to represent the compositional gap suggested by \citetalias{Luque2022}.}
         \label{fig:Histo_density}
\end{figure}

In Figure \ref{fig:Histo_density}, we plot a comparison of the histograms of the normalized density of the \citetalias{Luque2022} sample and our filtered PlanetS catalog (Section \ref{sect:sample_selection_Mdwarfs}).  We normalize the planetary bulk density by the Earth-like composition: we divide by the density that a planet of the same mass would have with terrestrial composition (32.5\% iron and 67.5\% silicates, represented as \normrho~in this study). To make the figure easier to read, we've separated the colors of the two samples inside the bins. We draw the line at 0.65~\normrho~found to be the gap in composition by \citetalias{Luque2022}. 

As already seen in the M-R diagram (Fig.~\ref{fig:MR_MKFGdwarfs}), some planets populate intermediate compositions in a continuous way between rocky and volatile-rich, making the existence of the 0.65~\normrho~gap between the two populations less obvious. This conclusion can also be drawn from the histograms of the two samples compared in figure \ref{fig:Histo_density}. Indeed, by taking the entire population of \citetalias{Luque2022} planets instead of dividing them into the three categories, the gap becomes very faint, whereas for our sample it is filled by several planets. This implies that there is a continuous transition between the two population. While this representation is biased by the choice of the number of bins, the histogram shows two main components: one centered around 0.9~\normrho~and the other around 0.4~\normrho, with a similar width of about 0.4~\normrho. Once again, as noted in the previous section, we find the very dense outlier GJ 367 b.

\begin{figure*}[t]
\centering
   \includegraphics[width=17cm]{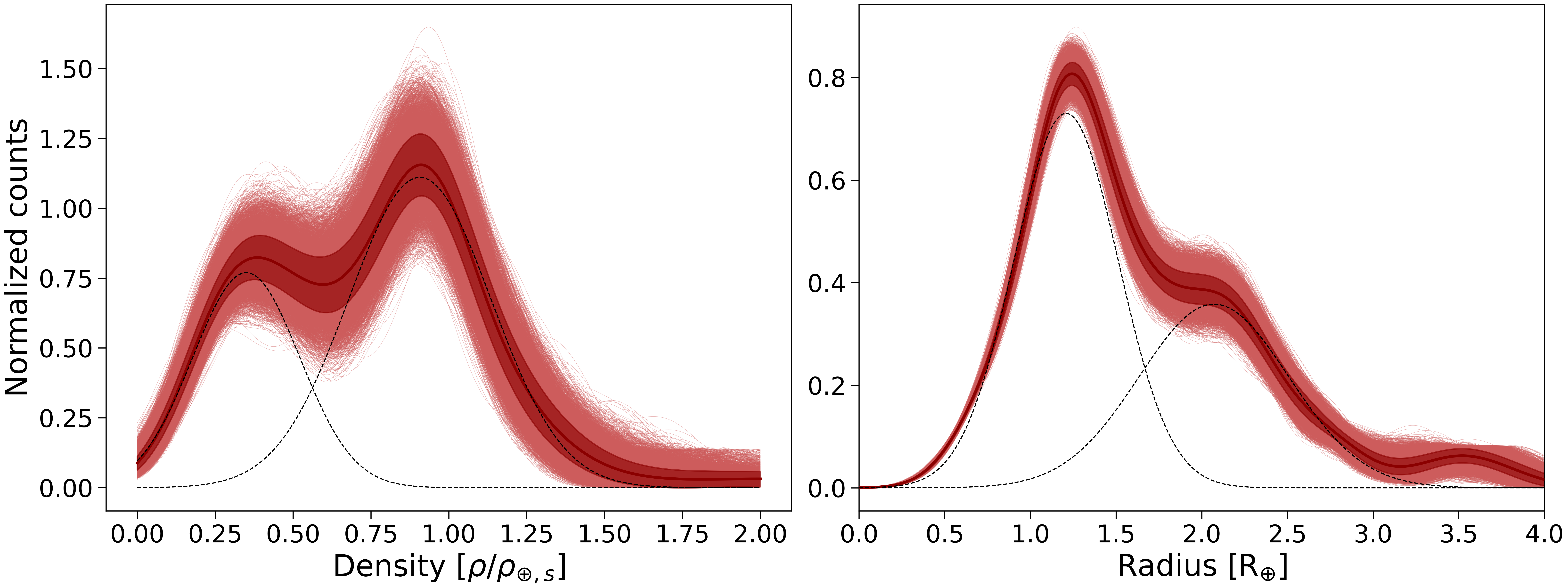}
     \caption{Kernel density estimates of the density (left) and radius (right) distributions of the small planets orbiting M-dwarfs in the PlanetS catalog. The 10,000 KDEs realizations are plotted in the background (light red), along with the resulting mean KDE (dark red solid line) and its standard deviation (dark red filled area). The black dashed lines are the two Gaussians that provide the best representation of each distribution (fitted to the mean KDE).}
     \label{fig:KDE_Mdwarfs}
\end{figure*}

\subsubsection{Kernel density estimation}\label{sect:kde_method}

In this study, we go beyond the simple histogram and look at the Kernel density estimation (KDE) of our sample. A KDE can be seen as an extension of the histogram. Like the latter, its aim is to estimate an unknown probability density function from a sample of data. 
However, so far, the uncertainties are not taken into account in the construction of the histogram. Despite our selections on mass and radius, our sample has a median relative error in density of 16\% and can go up to 31\%, which can shift a planet from one composition to another but also change the appearance of the total distribution. To take these uncertainties into account, we perform a re-sampling technique: a large number of new data sets (N = 10\,000 samples) are created, which lie within the normal distribution of the density measurements with the maximum uncertainty taken between the upper and lower uncertainty as standard deviation. For each of these new data sets, the KDE is calculated. We then calculate the mean and standard deviation of all the KDEs obtained. 

The choice of kernel and bandwidth are the two adjustable KDE parameters, and it is widely accepted in the literature that the choice of bandwidth is more important than the choice of kernel \citep{Silverman1981,Wand1995,Silverman2018}. A larger bandwidth makes the estimate smoother, while a smaller bandwidth captures more detail. The KDE is calculated using the \statsmodels Python package\protect\footnote{\url{https://github.com/statsmodels/statsmodels/}\label{statsmodelfootnote}} (\citealt{seabold2010statsmodels}) that allows for continuous data and for an automatic search of the optimal bandwidth. This optimal bandwidth, crucial for capturing data distribution accurately, is automatically determined through a combination of cross-validation and maximum likelihood operator (\citealt{heidenreich_bandwidth_2013}). Cross-validation assesses KDE's performance by partitioning data into validation sets, while the maximum likelihood operator identifies the bandwidth that maximizes the likelihood of observing the data. This integrated approach ensures an effective selection of bandwidth, enhancing the fidelity of density estimation. However, as this approach is computationally expensive, we search for the optimum bandwidth for 1,000 samples randomly selected from the 10,000 and use the resulting median bandwidth for all our resampled data. A test with 1\,000 vs. 10\,000 samples shows that the median bandwidth agrees in both cases.

We apply this method to the distribution of bulk density of the planets in our sample but also to the distribution in radius. Figure \ref{fig:KDE_Mdwarfs} shows the result: we have plotted the 10,000 realizations of the KDE calculation for the density and radius distribution of our small planets orbiting M-dwarfs, and we have indicated the mean KDE as well as the standard deviation of all KDEs.

First, concerning the density distribution, again normalized here by the Earth's composition as in the histogram, we can see the effect of measurement uncertainty allowing a wide dispersion in the possible shapes of the distribution. A bimodality in composition seems to appear when looking at this distribution: we fit to the mean KDE two Gaussians to model this bimodality and find a population centered around 0.35~\normrho with a sigma of 0.17 and another centered around 0.91~\normrho with a sigma of 0.23 (dashed black lines Fig.~\ref{fig:KDE_Mdwarfs}). However, the two distributions overlap considerably between 0.25 and 0.75~\normrho. 

Looking at the right tail of the distribution, we see that there are very few planets denser than Earth's composition. Consequently, there is no clear evidence of the existence of a population of Super-Mercuries in this sample. For planets with a radius and mass greater than  Earth, there is no bias in detecting such objects below the Earth's composition line. On the distribution, they should lie between 1.3 and 1.45 in normalized density. Concerning the left tail of the distribution, we're biased by our cut-off at 4 R$_\oplus$, but we also suffer from the lack of planets with a radius $\geq$ 2.5 R$_\oplus$, which we should be detecting and which should have a normalized density between 0.15 and 0.5~\normrho. This lack of sub-Neptunes relative to the maximum of our distribution, which is around the Earth's compositional density, is consistent with occurrence rate studies of transiting planets around M-dwarfs showing that sub-Neptunes are less abundant than super-Earths around especially mid-to-late type M-dwarfs \citep{MentCharbonneau2023}. However, the maximum of the distribution is highly biased by the incompleteness of the sample. The TRAPPIST-1 system and the few planets with mass less than 1 M$_\oplus$ bias this maximum around the terrestrial composition. Indeed, due to our detection limits in radial velocity and in transit (preventing us from going to longer periods), this part of the parameter space is highly incomplete. The resulting density and radius distributions of the planets around M-dwarfs without TRAPPIST-1 and the few planets with mass less than 1 M$_\oplus$ can be found as a dashed red line in Fig.~\ref{fig:KDE_MFGKdwarfs}.

Second, by examining the radius distribution produced, we can see that there is a maximum around 1.3 R$_\oplus$. This is again in agreement with \citet{MentCharbonneau2023} that super-Earths appear to be more abundant around M-dwarfs than sub-Neptunes. Examining the left-hand tail of the KDE, we are constrained by the limits of mass detection before those in radius. Indeed, we have 30 sub-Earth size planets around M-dwarfs validated by TESS and \textit{Kepler}/K2 (from NASA Exoplanet Archive). We are therefore not biased toward the right-hand side of the distribution. The abrupt decrease in the distribution around 2.5 R$_\oplus$, or "radius cliff", is a feature not biased by detection limits. The latter is still largely unexplored, and studies such as \citet{Kite2019} try to explain it by equilibrium processes between the atmosphere and a magmatic surface. It is also worth noting the bias regarding Earth-size planets around M-dwarfs detected by TESS. In fact, it appears that given current sensitivity limits, TESS may miss dozens of planets in transit around nearby late M-dwarfs, and that the planet occurrence rate is unlikely to increase and may even decrease for the latest M-dwarfs \citep{Brady2022,Dietrich2023}.

An important characteristic of this distribution is that it does not distinctly show the radius gap of \textit{Kepler} planets between super-Earths and sub-Neptunes: the radius valley, or Fulton gap (\citealt{Fulton2017}). Here, the radius valley fades, but there is still a break in the slope of the distribution around its value of 1.5-2 R$_\oplus$, a small plateau before another decrease in the distribution. This feature could then represent the transition between two populations of planets. Considering the median of the stellar masses of our sample ($\sim$0.33~M$_\odot$), one would anticipate the radius valley to be more around 1.3-1.5~R$_\oplus$ \citep{Fulton2018,Berger2020,Ho2023}. Interestingly, in our analysis, this range aligns more with the maximum of our distribution. We can once again fit two Gaussians to the mean KDE to model this bimodality and we find a population centered around 1.21 R$_{\oplus}$ with a sigma of 0.30 and another centered around 2.07 R$_{\oplus}$ with a sigma of 0.44 (dashed black lines Fig.~\ref{fig:KDE_Mdwarfs}). Here again, the two distributions overlap ($\sim$1.2-2 R$_{\oplus}$). Another small bump in the distribution around 3.5 R$_{\oplus}$ is due to the 2 planets around this value and reflects the lack of planets in this region.

Finally, the KDE allowed us to explore the properties of density and radius distributions, while not preventing us from analyzing the biases and incompleteness of our sample. While the KDE can enhance the portrayal of a distribution by considering its continuous characteristics, as the histogram, its effectiveness is constrained by the sample size of planets actually observed (see appendix \ref{appendix:hist_limits}). This limitation becomes even more pronounced when we consider the subtleties of planet formation and the evolutionary mechanisms that give rise to the real and complex distribution we seek to derive, compounded by observational biases that result in an incomplete sample. Only by taking these aspects into account can we now discuss the implications for the potential internal structure and conditions of formation and evolution of these planets.

\subsection{Impact on internal structure and composition}\label{sect:M_composition}

\begin{figure}[t]
  \centering
    \includegraphics[width=8cm]{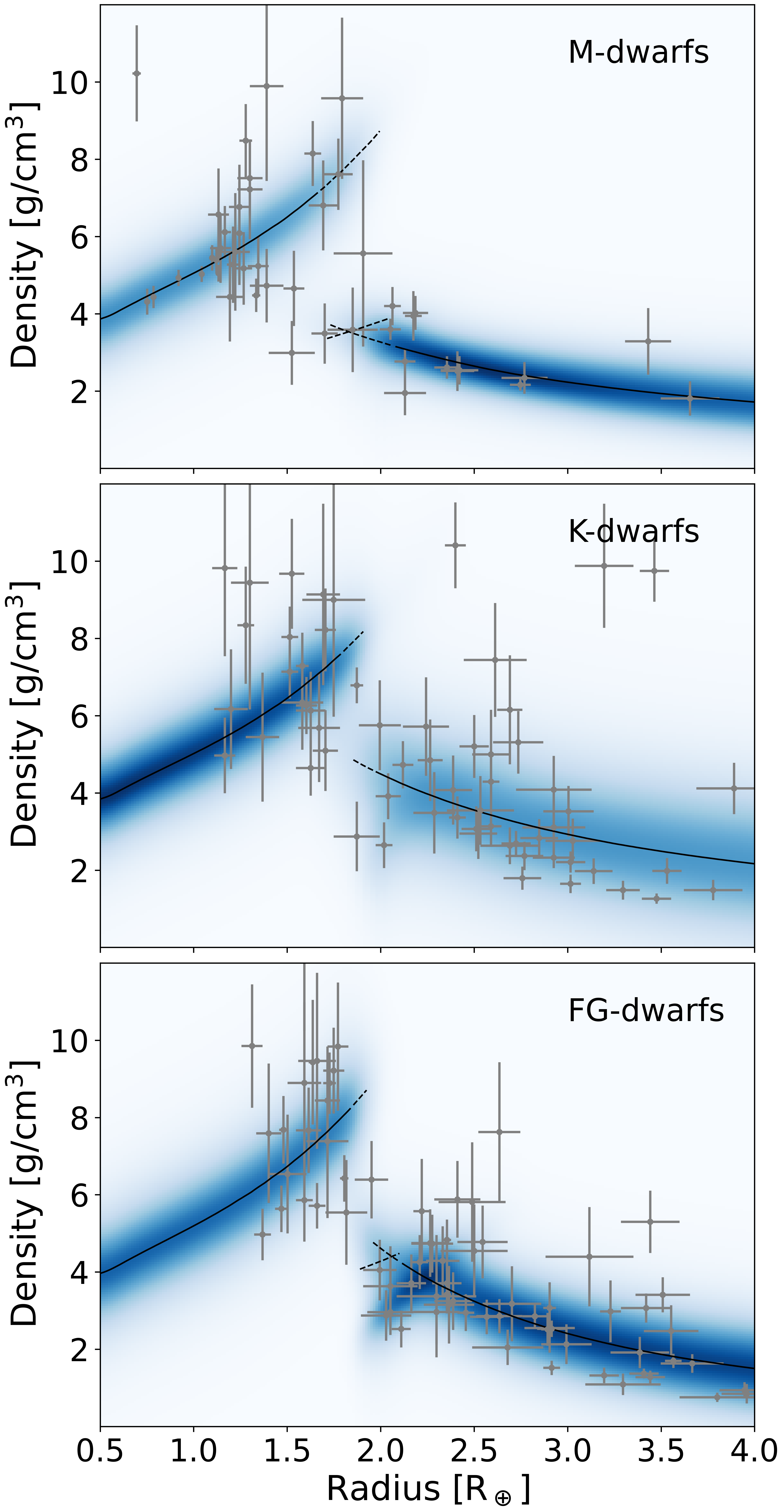}
      \caption{Numerical radius-density probability model applied to small planets orbiting M-, K-, and FG-dwarfs from the PlanetS catalog (Top, middle, and bottom panel, respectively) using \spright and the water-rich planet density model proposed by \citet{Aguichine2021}. Gray data points represent measured values with associated uncertainties for planets in our sample. The blue map depicts the logarithm of the posterior probability, and the black lines indicate posterior means for each of the three planet populations ("rocky", "water-worlds", and "sub-Neptunes"). The solid lines denote radius regimes where the component has a weight of unity, meaning all planets in this range belong to this specific component, while the dashed lines mark transition regimes between populations. This clearly shows that the "water-worlds" population is only considered as a transition between the other two populations, and not as a distinct population.}
         \label{fig:spright_Mdwarfs}
\end{figure}

\begin{figure*}[t]
\centering
\sidecaption
   \includegraphics[width=12cm]{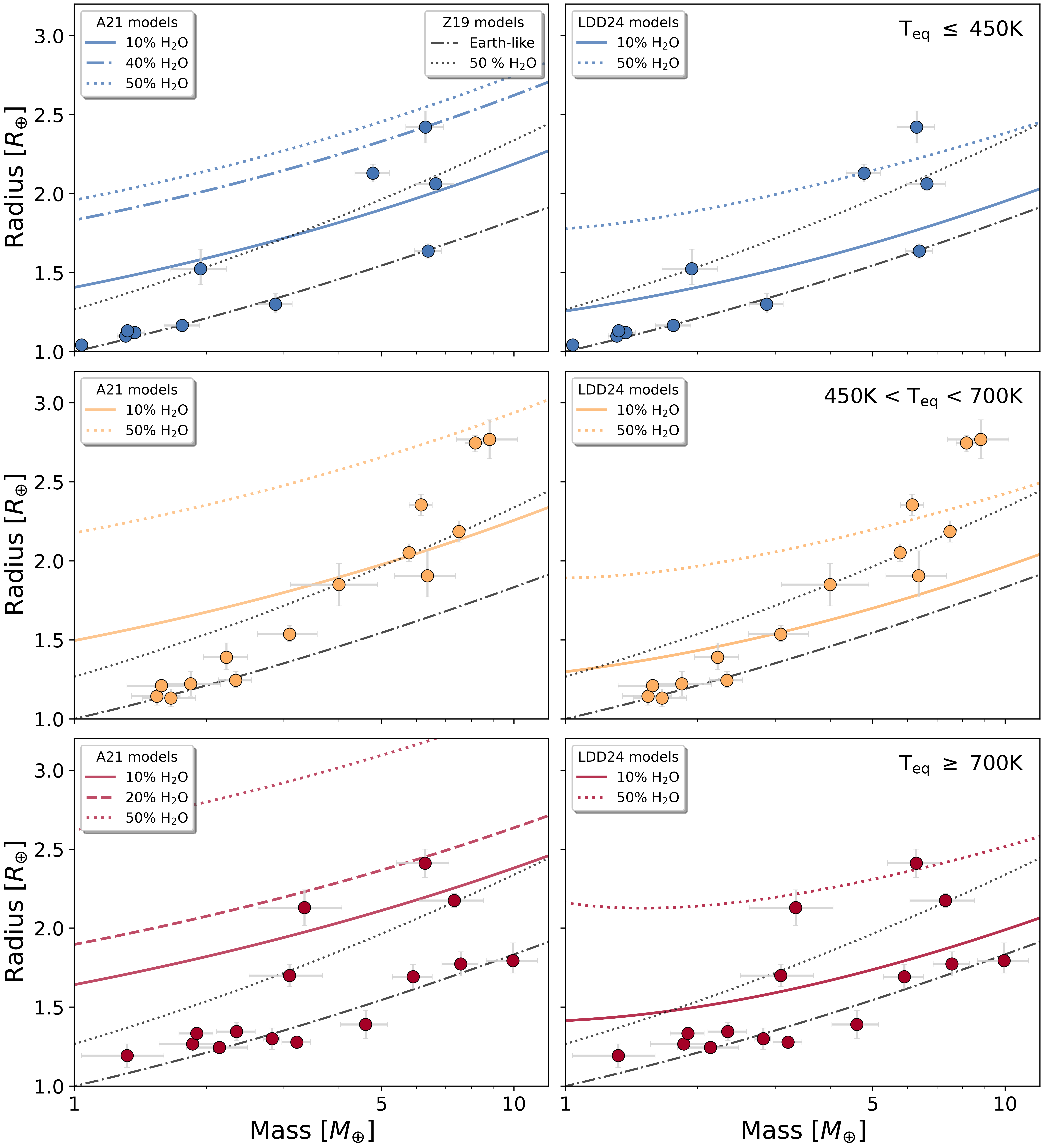}
     \caption{Mass-radius diagrams of small planets orbiting M-dwarfs, focusing on those larger and more massive than Earth, up to 3~R$_\oplus$. The three rows correspond to three equilibrium temperature intervals, with planets with T$_{\text{eq}} \leq$ 450~K at the top, 450~K < T$_{\text{eq}}$ < 700~K in the middle, and at the bottom T$_{\text{eq}} \geq$  700~K. The two columns correspond to the two internal structure models, from \citet{Aguichine2021} (A21) on the left and from \citet{LuoDorn2024} (LDD24) on the right. The composition lines are drawn from these models at the average temperatures of planets in the intervals, 400~K, 550~K, and 800~K for cold, warm, and hot respectively. Each panels feature \citet{Zeng2019}'s terrestrial and 50\% water composition lines represented by black dash-dotted lines and dotted respectively. }
     \label{fig:MR_compostion}
\end{figure*}

The exact positioning of a planet within the M-R diagram, and thus its bulk density, prompts a deeper exploration of its internal composition and structure in terms of the interplay between rocky and volatile elements. At any specific point within the diagram, numerous potential compositions for the planet come into consideration. Additionally, a crucial factor to consider is the irradiation received by the planet from its host star, which we quantify by the equilibrium temperature. This temperature estimation provides insights into the planet's thermal state and allows us to tailor our understanding of the pressure--temperature profiles of the planet's constituent elements. Hence, it transcends the traditional two-dimensional M-R problem and introduces equilibrium temperature as a pivotal parameter. Consequently, there emerges a unique combination of mass, radius, and equilibrium temperature to juxtapose against the composition lines that can be delineated on the M-R diagram.

It is at this juncture that \citetalias{Luque2022}'s study exhibits a shortcoming: when categorizing "rocky-worlds" from "water-worlds", while the rocky planets seem to have a terrestrial composition (63\% silicates, 27\% iron), they rely on a single composition line for the water-rich planets (comprising 50\% water and 50\% silicates). However, in this model, the water is in condensed form, which is not justified for equilibrium temperature above 400~K. Out of the 6 planets on \citetalias{Luque2022}'s water line, only one has an equilibrium temperature below this threshold: TOI-270 d. This exoplanet is no longer even situated on this composition line in our sample (see appendix \ref{appendix:comparison_samples_M}). Another limitation that could be seen in this composition is the absence of an iron core in the planet's structure.

Moreover, as highlighted earlier, the determination of a planet's internal composition and structure based solely on the M$_p$, R$_p$, and T$_{\text{eq}}$ parameters is profoundly challenging due to the high degree of degeneracy. Consequently, it is not feasible to definitively pinpoint a single composition and structure for the planet using this trio of parameters alone. \citetalias{Luque2022} also explores alternative compositions, such as terrestrial or aqueous compositions with and without an atmospheric layer of H/He using compositions from \citet{Zeng2019}. They conclude from the observations that a slight change in mass fraction of H/He envelopes or in their temperatures would lead to substantial discrepancies in the planets' radii and thus that these aligned planets are more likely to be composed primarily of water rather than being predominantly gas-rich in their composition. Furthermore, it's noteworthy that the temperature of these lines actually corresponds to the corresponding entropy temperature at 100 bar, not the equilibrium temperature. \citet{Zeng2019} assume a uniform specific entropy across exoplanets of different masses and this assumption is not justified particularly in case of exoplanets that have undergone thermodynamic processes such as cooling and mass loss as well as giant impact. These processes naturally lead to a decrease in a planet's specific entropy, a phenomenon influenced by multiple factors, including planetary mass and equilibrium temperature.

They also conducted a comparative analysis with the model proposed by \citet{Turbet2020}, which accounts for runaway greenhouse radius inflation by the potential presence of a hydrosphere expanded due to the intense irradiation experienced by these planets. Nevertheless, it is essential to note that the data presented in this study primarily encompass terrestrial-like planets with a mass ranging from 0 to 2 Earth masses, leading to substantial extrapolation in \citetalias{Luque2022}.

In this study, we want to highlight other models that could explain the composition and internal structure of these planets. \citet{Rogers2023} shows that these "water-worlds" could well correspond to rocky planets with an H/He atmosphere. The comparison between their models and the sample of planets from \citetalias{Luque2022} can be found in Figure 1 of their study. The differences in our sample are consistent with this comparison. However, they use a pure evolutionary model (mass-loss, cooling) and theoretical initial quantities of H/He not taken from global formation models (that include core formation, planet migration, collisions...).

\citet{Mousis2020}, \citet{Aguichine2021}, \citet{Vivien2022}, and \citet{Pierrehumbert2023} extend the work of \citet{Turbet2020} to more massive planets by considering the runaway greenhouse effect of water-rich planets, considering in particular a layer of water in supercritical form in the planetary structure. These models allow for a much larger planetary radius with a smaller amount of water than \citet{Zeng2019} model. This is because the latter considers an atmosphere with an isothermal profile (and a condensed form for water), whereas the others consider a radiative-convective profile. The recent study by \citet{Selsis2023} shows that the truth lies somewhere in between, isothermal near the surface and convective in the upper atmosphere.

More recently, \citet{Parviainen2023}, employing Bayesian model comparison, uses an updated sample from \citetalias{Luque2022} and shows evidence against the presence of a "water-world" population around M-dwarfs when considering the mass-density-radius relationships for small planets derived from the "water-worlds" models of \citet{Zeng2019} and \citet{Aguichine2021}. Consequently, our study aligns with \citet{Parviainen2023} findings, supporting the notion that "water-worlds" do not constitute a distinct and discernible population. Indeed, we use the Python module \spright presented in \citet{Parviainen2023} to fit our sample and obtain the water world population strength $\omega$ consistent with 0. The result is shown Fig.~\ref{fig:spright_Mdwarfs}.

However, all these models are based on the assumption that water is solely present on the surface. Recently, following \citet{Dorn2021}, \citet{LuoDorn2024} revisited this assumption by allowing water to be part of the planet's inner layers: mantle and core. This is following recent studies for the Earth, that indicate that the largest portion of Earth's water may be stored deep within its core (although not in molecular form). This discovery has substantially expanded the estimated total water inventory to 1.7\% by weight, signifying a considerable increase compared to previous assessments \citep{Li2020,Tagawa2021}. New ab initio calculations from \citet{LuoDorn2024} now show that the majority of a planet’s water budget (even more than 95\%) can be stored deep in the iron cores and the rocky mantles, not at their surfaces. As a consequence, the limited amount of steam present at the surface results in significantly reduced planetary radii for a given mass. This revelation underscores the critical role of internal water distribution in shaping planetary characteristics and prompts a reevaluation of our assumptions regarding the interplay between a planet's composition and its observable features.

In Figure \ref{fig:MR_compostion}, we have depicted the composition lines from the models of \citet{Aguichine2021} (left-column) and \citet{LuoDorn2024} (right-column) in conjunction with our small planets around M-dwarfs and the Earth-like and 50\% water composition lines from \citet{Zeng2019}. In comparison with Fig.~\ref{fig:MR_MKFGdwarfs}, we've zoomed in on the area of interest in the "water-worlds" discussion, so we are ignoring planets larger than 3~R$_\oplus$ and smaller than Earth. This representation allows us to estimate the composition bounds of the M-dwarf planets based on these theoretical models. To enhance the clarity of this comparison, we have categorized and color-coded the equilibrium temperature of the planets into distinct intervals: $\leq$ 450~K for the top panels, [450~K, 700~K] for the middle panels and $\geq$ 700~K for the bottom panels. The contrast in shapes between Aguichine and Luo's models, which incorporate steam water and solubility, and Zeng's model, which assumes condensed water, is immediately noticeable.

On one hand, Aguichine's models offer insights into the potential composition of planets with larger radii, expanded by the presence of water in vapor form in the atmosphere and the greenhouse effect induced by the planets' temperatures. With these models, it becomes apparent that planets corresponding to the 50\% water line of Zeng's model (and the water-world population of \citetalias{Luque2022}) can be interpreted as having a composition of approximately 10\% water in the atmosphere, with a terrestrial composition for the core. These planets are associated with equilibrium temperature ranges of 300 to 700~K. As we move from these 10\% composition lines toward terrestrial composition, the planets are slightly more irradiated, resulting in substantially lower quantities of water. On the contrary, larger planets may contain more water - 20, 40 or even 50\% - in their atmosphere.

Conversely, \citet{LuoDorn2024} model excels in depicting the diversity of mass, radius, and temperature combinations between Zeng's water and terrestrial compositions. The radius in this model is less influenced by the irradiation due to the presence of most of the water in the planet's mantle and core. For the sub-Neptunes, even the most water-rich scenario with 50\% cannot represent those planets with radii above 2.2~R$_\oplus$, as these planets  require some H/He in their atmosphere to explain their observed radius.
Also, \citetalias{Luque2022}'s water-world population would have variable water mass fractions between 10\% and 50\% with water distributed between the surface, mantle, and core for equilibrium temperatures between 350 and 850~K.


Finally, Aguichine's model, while incorporating irradiation effects, has limitations stemming from the lack of interaction between the atmospheric layers and the planet's interior. Notably, it does not consider solubility and solely focuses on surface water in form of the steam atmosphere. 
The model by \citet{LuoDorn2024} seems more coherent in this respect. But both models are static in time, and don't take into account the cooling and possible compression of the planet as it evolves. In this context, all the sub-Neptunes around M-dwarfs cannot be solely explained by the presence of water in their atmosphere and core; for some of them, an atmosphere containing H/He is needed to account for their observed bulk density.

In conclusion, despite being expected to represent the same population of planets with identical composition, these two models yield markedly different outcomes in terms of mass or final radius. Despite the inherent degeneracy in this parameter space, achieving greater consensus among theories of internal structure, formation, and evolution is imperative to accurately interpret observables and predict the likely atmospheric compositions to be observed.

        

In the end, discriminating between these diverse planetary models proves challenging when relying solely on observable parameters such as mass, radius, and equilibrium temperature. In addition, this latter is contingent upon numerous assumptions about the atmospheric and surface characteristics of the planets in question. A comprehensive understanding of the atmospheric compositions becomes imperative for achieving a meaningful distinction among these varied compositions.


\subsection{M-R trends in the context of formation and evolution models}\label{sect:M_formation}

Examining the M-R diagram alongside density and radius distributions offers valuable demographic insights into the formation and evolution of planets. Conversely, observational data serves as a crucial constraint, refining and enhancing the precision of these theoretical frameworks. 

As in \citetalias{Luque2022}, we find that the radius valley fades for M-dwarfs (Fig.\ref{fig:MR_MKFGdwarfs}); in other words, no distinct value in planet radius separates planets crossing the Earth-like composition from the 50\% water line. 
\citet{Venturini2024} expand the pebble accretion model for Sun-like stars of \citet{Venturini2020} to a range of stellar masses spanning 0.1 to 1.5 $M_{\oplus}$. They predict a filling of the radius valley for M-dwarfs, in line with our findings. The reason behind the fading radius valley is type-I planetary migration, which happens for lower planetary masses when the stellar mass diminishes \citep{Paardekooper2010}. Consequently, smaller icy planets (formed beyond the iceline) reach the inner planetary system compared to the Sun-like case, producing a larger overlap in mass and radius with the rocky planets formed within the iceline. This effect of planet migration acting on smaller planets for M-dwarfs was reported as well in \citet{Burn2021}. \citet{Ho2024} also identifies a shallower radius valley around lower-mass stars when revisiting planetary parameters from \textit{Kepler} 1-minute short cadence light curves. They conclude that photoevaporation models alone are insufficient to explain this phenomenon. Instead, they propose three alternative scenarios: variations in the stars' XUV output, dispersion in planetary core composition, and collision events.

\citetalias{Luque2022} explicates the dual compositional nature of their findings as a consequence of planetary formation through pebble accretion \citep{Venturini2020}. 
Indeed, the formation and evolution simulations of \citet{Venturini2020} for Sun-like stars revealed that the formed planets exhibit a stark bimodal distribution in Water Mass Fraction (WMF), manifesting typically as 0 or 0.5, with very few planets with intermediate values. 
However, for M-planets, \citet{Venturini2024} show that the number of planets with intermediate compositions (in between pure rock and 50\%water), increases compared to planets formed around Sun-like stars. The bimodality in water mass fractions still persist, but the gap occurs at 0 < WMF < 20\%. When considering that most of the real exoplanets are hot enough to host steam atmospheres, the radius of planets with $\sim20\%$ WMF is similar to the ones having condensed water at 50\% by mass as shown by \citetalias{Luque2022}.
Overall, in the mentioned pebble-based model, two groups of planets in M-R can be easily identified due to the mentioned dichotomy in composition. This is not really observed in our PlanetS Catalog for M-planets. On the other side of the theoretical possibilities, planetesimal-based models yield more mixed, intermediate compositions compared to pebble-based models \citep{Burn2021, Brugger2020}. Thus, our results could give support to a planetesimal or hybrid pebble-planetesimal \citep{Alibert18, Guilera20} accretion scenario. Alternatively, when N-body interactions are taken into account among hundreds of planetary embryos growing in the same disk, giant impacts can yield as well more intermediate compositions that blurs the radius and density valley \citep[e.g.,][]{Emsenhuber2021}

\section{M-R diagram of small exoplanets around FGK-dwarfs}\label{sect:MR_FGK}

\subsection{Sample selection}\label{sect:sample_selection_FGKdwarfs}

Similar to our approach with M-dwarfs, we are leveraging the PlanetS catalog to broaden our investigation to FGK-dwarfs for comparative analysis. To categorize planets based on their spectral type, we focus on stars with T$_{\text{eff}} \ge$ 4000~K, effectively excluding the previously examined M-dwarfs. Subsequently, we utilize the "Spectral Type" column in the catalog, determined through the table from \citet{Pecaut2013}, to specifically target K-, G-, and F-dwarfs for our study. Similarly, we select planets with a radius of less than 4 R$_\oplus$.

In the final dataset, we have a total of 61 planets orbiting K-dwarfs, 63 planets orbiting G-dwarfs, and 9 planets orbiting late F-dwarfs. To conduct a robust analysis, we combine the G- and F-dwarfs into a single sub-sample, considering that the F-dwarfs alone are not numerous enough to support a statistically reliable study. The sample sizes of planets around K- and FG-type stars are comparable to those around M-dwarfs, facilitating a statistically fair comparison.

A test was carried out to reduce the relative uncertainties on mass and radius in the PlanetS catalog from 25\% to 15\% on mass and from 8\% to 5\% on radius. This reduced the samples to 26, 32, and 31 planets for M-, K-, and FG-dwarfs respectively (i.e., 40\%, 48\%, and 57\%, respectively). In view of the results in Appendix~\ref{appendix:hist_limits}, we chose to keep the "full" samples, so as to be able to interpret the resulting distributions more confidently.

The effective temperature range of our stars goes from 4065 to 5320~K for the K-dwarfs and from 5325 to 6321~K for the FG-dwarfs. Our separation into spectral types obviously induces a classification by host star mass, useful for the discussions in the following sections. The median stellar masses are 0.33~M$_\odot$, 0.8~M$_\odot$, and 0.96~M$_\odot$, for M-dwarfs, K-dwarfs, and FG-dwarfs respectively. We'd expect a higher stellar mass value for FG-dwarfs, but since we're focusing on the PlanetS catalog's sample of planets of precise mass (Sect \ref{sect:planets_catalog}), we're biased toward the lower stellar masses where it's easier to detect and characterize a small planet. 

The resulting M-R diagram for the two sub-samples is shown on the center and left panels of the Fig.~\ref{fig:MR_MKFGdwarfs}, with the planets colored by their equilibrium temperature. 
Upon comparing our two new samples with the panel on the left, which depicts small planets around M-dwarfs, a noticeable detection bias is evident around FGKs. This bias is characterized by an increase in the minimum mass and minimum radius of the well-characterized planets in our catalog. Additionally, there is a notable prevalence of larger, more massive planets around FGKs compared to M-dwarfs. Remarkably, for all three subsamples, planets with an Earth-like composition seem to exhibit a consistent maximum mass of 10 M$_\oplus$. Another noteworthy mass limit to consider in the comparison of our three samples is the minimum mass of sub-Neptunes. It appears that as stellar mass increases, so does the minimum mass of these planets. We observe values of around 1.9~M$_\oplus$, 3.4~M$_\oplus$, and 4.3~M$_\oplus$, for M-dwarfs, K-dwarfs, and FG-dwarfs, respectively. These observations are major elements in the comparison of our observed population with synthetic populations derived from theoretical models of formation and evolution (see \ref{sect:formation_MKFG}).

\begin{figure*}[t]
\centering
   \includegraphics[width=17cm]{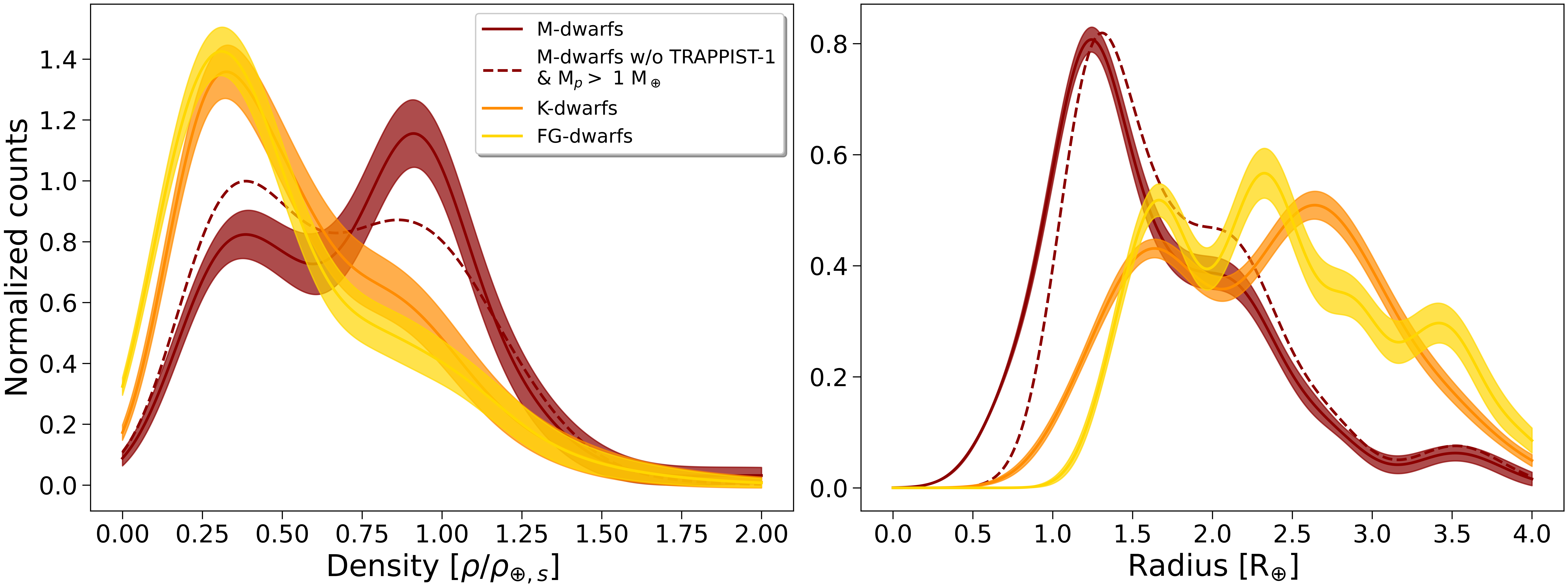}
     \caption{Kernel density estimates of the density (left) and radius (right) distributions of the small planets orbiting M- (red), K- (orange), and FG-dwarfs (yellow) from the PlanetS catalog. For visibility, only the resulting mean KDE (solid line) of the 10,000 realizations and its standard deviation (filled area) are plotted. The red dashed lines are the density and radius distributions of the sample orbiting M-dwarfs without the TRAPPIST-1 system and planets below 1 M$_\oplus$.}
     \label{fig:KDE_MFGKdwarfs}
\end{figure*}

\subsection{Comparing the density and radius distributions}\label{sect:FGK_distribs}

We employ the same methodology as detailed in Section \ref{sect:kde_method} to derive the KDEs of our two new planet samples around FG- and K-dwarfs. In Figure \ref{fig:KDE_MFGKdwarfs}, the outcomes of the KDE calculations are depicted, illustrating distinct density and radius distributions for M, K, and FG-stars. For greater visibility, the 10,000 samples of each test are not shown, and only the mean and standard deviation of the result are displayed.

Starting with the analysis of density distributions, a striking contrast emerges between the distribution for planets orbiting M-dwarfs and those around FG- and K-dwarfs. The distributions for FG- and K-dwarfs exhibit a notable similarity, while the distribution for M-dwarfs is nearly inverted concerning the occurrence of densities between rocky planets and those composed of more volatile materials. Examining the three distributions around 0.75-1 in normalized density, a noticeable discrepancy is observed; although they are considerably lower for FGKs, there exists a detection bias disfavoring super-Earths around FGKs compared to M-dwarfs.
In addition, we recall that the significant peak around Earth density is driven by the TRAPPIST-1 planets whose uncertainties are very low compared to the rest of our sample. Upon their removal, the density distribution assumes a shape more akin to that of FGK-stars (see dashed line Fig. \ref{fig:KDE_MFGKdwarfs}), revealing a relative decrease in the number of rocky planets with an increase in stellar mass, attributed to detection limits. In the same way as for M-dwarfs, there appears to be no discernible population of Super-Mercuries around FGK stars, although this is subject to the limits of transit detection as it involves small planet sizes around bigger stars. 
Nevertheless, at lower densities (around 0.25-0.4~\normrho) corresponding to larger and more massive planets, the impact of the detection limit diminishes. Consequently, a conspicuous shortage of volatile planets around M-dwarfs becomes evident when compared to their counterparts around FGK-stars.

Regarding the radius distributions, distinctions between M-dwarfs and FGK-dwarfs are once more apparent. This contrast is primarily influenced by the left tail of the distributions, reaffirming the detection bias favoring super-Earths toward the latest spectral types. Additionally, the less clear gap between the two populations of super-Earths and sub-Neptunes in the radius distribution around M-dwarfs contributes to this divergence. Indeed, our analysis reveals a bimodal distribution between super-Earths and sub-Neptunes for planets around FGK-type stars, consistently demonstrating a radius valley around 1.8 Earth radii \citep{Fulton2017,VanEylen2018}. However, it's important to note that the KDEs do not provide the granularity needed to discern variations in this value based on changes in stellar mass \citep{Fulton2018,Berger2020,Ho2023}. The plateau observed in the M-dwarfs distribution corresponds to the Fulton gap apparent in the FGK-stars distributions. 
Once more, the absence of sub-Neptunes (with radii between 2 and 4 R$_\oplus$) around M-dwarfs is clearly evident in the data. The oscillations in the radius distribution of the FG-dwarfs on the right-hand side are due to the bandwidth of the KDE, revealing both the radius valley and the subtleties of the larger-radius distribution. Additionally, it prompts inquiries regarding the evolution of the "radius cliff" concerning spectral type.

The augmented sample sizes of small planets around FGK-type stars provide a more robust foundation for precise estimation of density and radius distributions through KDEs. We were then able to carry out a comparative study of the evolutionary trajectories of these distributions across spectral types. This analysis is closely intertwined with considerations of detection limits and sample incompleteness, offering valuable insights into the potential implications for the internal structure, composition, formation, and evolution of planets orbiting diverse stellar types as discussed in the following sub-sections.

\subsection{Internal structure and composition across spectral type}\label{sect:FGK_composition}

After an extensive discussion of various models representing different compositions of small planets around M-dwarfs in Section \ref{sect:M_composition}, we extend our analysis to planets orbiting FGK-dwarfs. In the supplementary material of \citetalias{Luque2022}, an attempt is made to identify the same composition gap observed for planets around M-dwarfs in the case of planets around FG- and K-dwarfs. However, their sample size is considerably smaller compared to ours (82 vs. 133 planets). Nevertheless, akin to our findings, they also report scant evidence of a clear gap between a population of "rocky" planets and a population of "water-rich" planets. Notably, Figure \ref{fig:MR_MKFGdwarfs} in this work vividly illustrates the presence of planets falling between the two composition lines. As in Section \ref{sect:M_composition}, we use the \spright module from \citet{Parviainen2023} to fit our samples and find that it refutes the hypothesis of the existence of a distinct population of "water-worlds" around K- and FG-dwarfs (middle and bottom panels of Fig.\ref{fig:spright_Mdwarfs}).

Once more, this prompts a reconsideration of the validity of drawing a single equilibrium temperature line, especially given the broader range of temperatures exhibited by planets around FGK-stars, which are notably more irradiated than those around M-dwarfs. The degeneracy between the compositions, as discussed in Section \ref{sect:M_composition}, persists unchanged for these two new samples. 

A noteworthy observation is the greater prevalence of sub-Neptunes with larger radii (R$_\text{p}$>2.5~R$_\oplus$) around FGK-stars compared to M-dwarfs, a trend previously highlighted in the M-R diagrams and in the density and radius distributions in the preceding sections. 
This trend could be attributed to the intense XUV flux during the early stages of M-dwarf development, potentially leading to the atmospheric escape of sub-Neptunes. Indeed, the XUV luminosity of M-dwarfs is enhanced by factor 10-50 relative to solar-type stars during their prolonged premain-sequence evolution \citep{Shkolnik2014,Luger2015,Ribas2016,Peacock2020} and an increase in this XUV flux implies an increase in atmospheric escape \citep{Lammer2003,Vidal2004}. Furthermore, \citet{Kubyshkina2021} modeled the evolution of a wide range of sub-Neptune-like planets orbiting stars of different masses and evolutionary histories and found that atmospheric escape of planets with the same equilibrium temperature ranges occurs more efficiently around lower mass stars. This might also suggest richer H/He envelope compositions or atmospheres containing water, potentially inflated by the heightened irradiation received by planets around FGK-type stars. 

\begin{figure}[t]
  \centering
    \resizebox{\hsize}{!}{\includegraphics{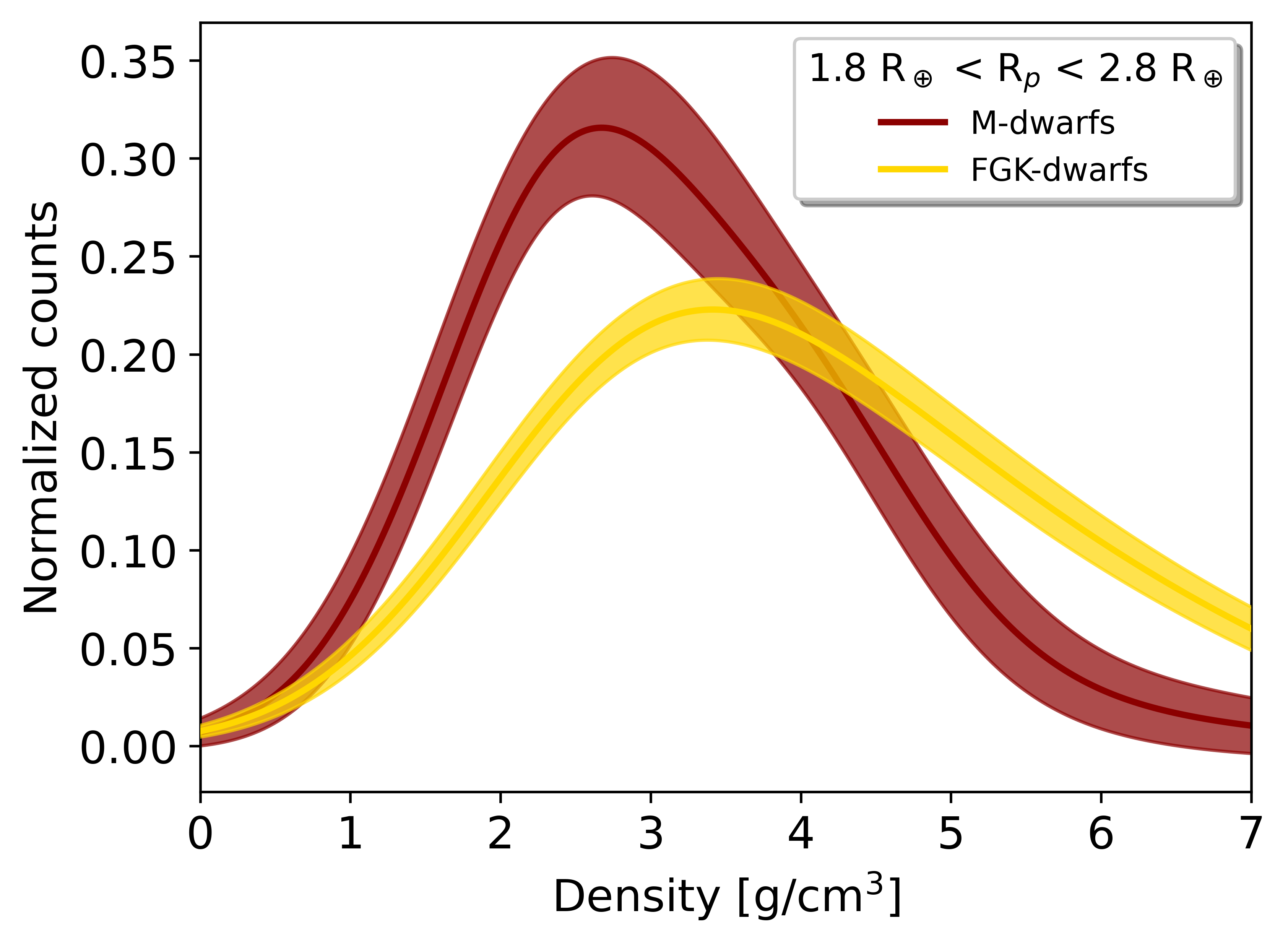}}
      \caption{Kernel density estimates of the density distributions of the small sub-Neptunes (1.8 R$_\oplus$ < R$_p$ < 2.8 R$_\oplus$) orbiting M- (red) and FGK-dwarfs (yellow) from the PlanetS catalog.}
         \label{fig:density_SN_comparison_M_vs_FGK}
\end{figure}

Despite evidence suggesting that M-dwarfs tend to form more rocky planets, another noteworthy observation pertains to the density distinction of the sub-Neptunes around M-dwarfs and FGK-dwarfs. Specifically, when focusing on the smaller sub-Neptunes (1.8 R$_\oplus$ < R$_\text{p}$ < 2.8 R$_\oplus$), indications emerge suggesting that those orbiting FGK-stars demonstrate a marginally elevated bulk density compared to those in orbit around M-dwarfs. 
The selection of this interval was made to exclude planets on the terrestrial composition line with a mass of 10~M$_\oplus$ and a corresponding radius of 1.79~$_\oplus$. Additionally, it aimed to include all planets around M-dwarfs while avoiding the lack of objects beyond 2.8 R$_\oplus$. This difference can be seen on the M-R diagrams (Fig.~\ref{fig:MR_MKFGdwarfs}) but also by plotting the density distributions of these samples using the same method as in Sect.~\ref{sect:kde_method}. The mean KDEs and their corresponding uncertainties are plotted in Fig.~\ref{fig:density_SN_comparison_M_vs_FGK}. These distributions do indeed show a shift toward the lowest densities for planets around M-dwarfs compared with those around FGKs.
To test whether this difference in densities is statistically significant, we utilize a Mann–Whitney U test \citep{Wilcoxon1945,Mann1947}, which studies whether the two data samples come from the same distribution. With a p-value of 0.013, we find that the densities of small sub-Neptunes orbiting M- and FGK-dwarfs belong to different distributions with over 95\% confidence. This can be because these planets are ice-rich, and hence likely migrated objects that accreted most of their solids beyond the ice line  \citep{Alibert2017,Venturini2020,Burn2021}.

Furthermore, by looking at the M-R diagrams, a notable increase in the dispersion of rocky planets around the terrestrial composition line is evident around FGK-dwarfs compared to M-dwarfs. This variability could be indicative of greater structural diversity in terms of core-to-mantle ratios or may reflect heightened uncertainties in the masses and radii of planets orbiting stars of earlier spectral types. 

\subsection{Formation and evolution: Trends from M-dwarfs to FGK-types}\label{sect:formation_MKFG}

Examining the formation and evolution of systems gains added complexity and interest when we incorporate a new dimension into the comparison between observations and theory: the mass of the host star. The variation in stellar mass introduces distinctive effects and adjustments within the models, like changes in the mass of the disk or in migration processes.
\citet{Venturini2024} extends the results of \citet{Venturini2020} by studying the influence of stellar mass. Similarly, \citet{Burn2021} perform this exercise for planetesimal accretion formation. 

The following demographic trends are worth to understand in the context of formation and evolution models:

\begin{enumerate}
    \item The minimum mass of planets falling near the 50\% condensed water composition seems to increase with stellar mass 
    (see black dotted lines Fig.\ref{fig:MR_MKFGdwarfs}).
    \item The maximum mass of rocky planets seems to be approx. 10 M$_{\oplus}$ for all spectral types.
    \item The radius valley gets filled when the stellar mass decreases.
    \item There is a depletion of sub-Neptunes with sizes $2.5\leq \text{R}_{\rm{p}} \leq 4 \, \text{R}_{\oplus}$ and $\text{M}_{\rm{p}} \geq 10 \, \text{M}_{\oplus}$ around M-dwarfs compared to FGK-dwarfs (see Fig.\ref{fig:MR_MKFGdwarfs}).
    
\end{enumerate}

Regarding the first point, in our PlanetS calatog, planets falling on the 50\% condensed water line have a minimum mass of approximately 1.8 M$_{\oplus}$ for M-planets (M$_{\star} \sim 0.33~\text{M}_\odot$), of 3.4 M$_{\oplus}$ for K-planets (M$_{\star} \sim 0.8~\text{M}_\odot$), and of 4.3 M$_{\oplus}$ for FG-planets (M$_{\star} \sim 0.96~\text{M}_\odot$) as noticed in Sect. \ref{sect:sample_selection_FGKdwarfs}. In planetesimal-based models, \citet{Burn2021} form water-rich planets with a minimum mass of 1 and 2 M$_{\oplus}$ for M$_{\star}= 0.5  \, \text{and} \, 1.0 \, \text{M}_{\odot}$, respectively. In a similar way, but for pebble-based models, \citet{Venturini2024} find a minimum mass of "water-worlds" of 1.4, 2.8, and 4.5 M$_{\oplus}$ for M$_{\star}= 0.4, 0.7, \, \text{and} \, 1.0 \, \text{M}_{\odot}$, respectively. We note that particularly the minimum masses of the pebble-based models are in very good agreement with our PlanetS catalog. As we already mentioned in Sect.\ref{sect:M_formation}, this trend of an increasing planet mass of "water-worlds" with stellar mass in the theoretical models may be a consequence of type-I migration, which acts more efficiently for low-mass stars. This also explains point 3, the filling of the radius valley when moving toward M-dwarfs \citep{Venturini2024}. However, these minimum values may be influenced by observational biases that render this parameter space incomplete. Despite finding planets smaller and less massive than these typical limits across spectral types, we are still approaching the current detection limits, which are around $\sim$1 m/s in radial velocity precision. Nonetheless, the increasing trend appears reliable as it persists in our sample test with increased precision in mass and radius (as mentioned in Sect.~\ref{sect:sample_selection_FGKdwarfs}).


Regarding the maximum mass of super-Earths as a function of host star mass, our findings remain consistent with a limit of 10~M$_\oplus$ for all the spectral types. Instead, theoretical studies of planet formation and evolution show a dependence on this maximum mass of super-Earths with stellar mass. Indeed, \citet{Burn2021} find values of approximately 10 and 20 M$_{\oplus}$ for M$_{\star}= 0.5 \,\text{and} \, 1.0 \, \text{M}_{\odot}$, respectively for planetesimal-based models. \citet{Venturini2024} find lower values for pebble-based models, and a flatter dependence, of 5, 6, and 7 M$_{\oplus}$ for M$_{\star}= 0.4, 0.7, \, \text{and} \, 1.0 \, \text{M}_{\odot}$, respectively. It is worth to mention that the differences between the two models might not be due to the assumption of planetesimals vs. pebbles alone. The simulations of \citet{Burn2021} consider 50 planetary seeds per disk and N-body interactions which result in planet merging, while the simulations of \citet{Venturini2024} consider 7 seeds per disk, and N-body interactions are not directly modeled. They are included "a posteriori" but assuming that each planet suffered from only one major giant impact. In any case, a constant maximum mass of rocky planets for stellar masses ranging from 0.3 to 1.0  M$_{\odot}$ as we find in the current study, seems to be difficult to reproduce with the current formation models. An interesting possibility to explore in the future is that such maximum mass of rocky planets is linked to evolutionary processes. The super-Earths with masses close to 10 Earth masses in our sample have equilibrium temperatures above 1000~K (see Fig.\ref{fig:MR_MKFGdwarfs}). At such high levels of irradiation, even a steam atmosphere could escape (see \citet{Venturini2024}, their App.B.2), rendering the planet as a remnant rocky core.

Concerning the final point (4), the results from \citet{Venturini2024} show the same qualitative trend. In that work, the deficit of sub-Neptunes with masses larger than 10~M$_\oplus$ around M-dwarfs is a result of the icy cores reaching smaller masses than around FGK, due to the formation in less massive disks.




\section{Conclusions}\label{sect:conclusions}

We have updated the PlanetS catalog, incorporating its extension and update. This has allowed us to re-assess the mass-radius (M-R) relationships, which are pivotal tools for transit follow-up radial velocity missions, to estimate the masses of objects detected through transit methods by space telescopes like \textit{Kepler}/K2, TESS, or the upcoming PLATO mission \citep{Rauer2014}.

Moving beyond the histogram, we delved into the study of density and radius distributions of small planets around M-dwarfs from our PlanetS catalog. Our approach involved utilizing Kernel density estimation coupled with a resampling technique allowing us to address the uncertainties associated with planetary parameters and to acknowledge the continuous nature of the distribution. The examination of the M-R diagram and KDEs revealed several trends:
\begin{enumerate}
    \item There is a continuous transition of composition of the planets in the transition between super-Earths and sub-Neptunes around M-dwarfs, ranging from terrestrial to more volatile compositions.
    \item There is a dearth of sub-Neptunes around M-dwarfs relative to super-Earths.
    \item The radius valley of M-planets is fading.
    \item A Super-Mercuries population is absent (at least for planets bigger than Earth due to observational bias).
\end{enumerate}

Subsequently, these observations were correlated with theories of internal structure, formation, and evolution, aiming to gain a deeper understanding. Our investigation extended to questioning various composition line models in the M-R diagram and their implications for the potential water composition of sub-Neptunes orbiting M-dwarfs, introducing equilibrium temperature as a third crucial parameter. Regarding the first point (1), the sub-Neptunes in question exhibit a diverse range of compositions contingent on their level of insolation, and so do not constitute a distinct population of "water-worlds" but rather represent a continuum within the broader spectrum of small planets around M-dwarfs. Advanced models of interior and atmosphere suggest that these planets can possess a water-rich interior, but some of their observed radii can only be accounted for by the presence of an H/He-rich atmosphere. Ongoing theoretical and experimental studies are continually enhancing our comprehension of the complexity of planet interiors with possible compositional coupling between atmospheres, mantle, and iron cores. The composition lines resulting from the P-T profiles, linked to an advanced internal structure model, will be a subject of keen interest in the upcoming months. 
Recent investigations into planetary formation through pebble accretion around M-dwarfs reveal a more diverse range of water compositions compared to Solar-type stars. However, the absence of a compositional gap in our observations suggests that these small planets around M-dwarfs may be primarily formed through planetesimal accretion or a hybrid combination of pebble and planetesimal accretion processes.

Expanding our investigation, we included FGK-dwarfs to compare the M-R diagrams and KDEs across different spectral types and highlighted tendencies:

\begin{enumerate}
    \setcounter{enumi}{4}
    \item There is a scarcity of sub-Neptunes around M-dwarfs in contrast to FGK-dwarfs.
    \item Small sub-Neptunes (1.8 R$_\oplus$ < R$_\text{p}$ < 2.8 R$_\oplus$) observed around M-dwarfs exhibit significantly lower density compared to their counterparts around FGK-dwarfs.
    \item The minimum mass of volatile-rich sub-Neptunes increases as stellar masses increase.
    \item The maximum mass of a planet with a terrestrial composition appears consistent across spectral types, remaining close to 10~M$_\oplus$.
\end{enumerate}

These demographic trends were then also interpreted in the context of formation and evolution models.
The discrepancy of the point (5) might be attributed to the intense XUV flux emanating from M-dwarfs in their youth, potentially leading to heightened atmospheric evaporation from planets in these systems. 
The point (6) may be explained by the fact that M-planets are richer in ice than FGK-planets, and migrated more efficiently after accreting most of their solids beyond the ice line.
The point (7) may be an effect of type-I migration, but the exact values of minimum mass could still be affected by observational biases. This migration effect is also responsible for the fading of the radius valley around M-dwarfs compared with FGKs. 
However, pinpointing this value precisely of 10~M$_\oplus$ for the point (8) through formation and evolution models poses challenges.

Ultimately, the constrained sample sizes pose limitations, hindering a comprehensive interpretation and differentiation of various theoretical phenomena underlying these distributions and diagrams. We saw that the significant 30\% increase in the number of M-planets within the sample has notable implications for the conclusions drawn regarding planetary populations. This underscores the critical need for more data in a parameter space that inherently lacks completeness and exhibits biases, such as the M-R diagram. We need to enlarge this parameter space to ensure the robustness of further statistical studies. On the one hand, instruments such as NIRPS (\citealt{Bouchy2017}), the new infrared spectrograph installed on ESO's 3.6m telescope at La Silla and dedicated to the study of M-dwarfs and their planetary systems, can help us achieve this. We also need to explore planets with longer orbital periods, as the median for this sample is 3.4 days. The NIRPS Guaranteed Time of Observation (GTO), and in particular Work Package 2 (WP2) devoted to radial velocity follow-up of transiting planets, is focused to determining the mass of planets around M-dwarfs in this parameter space. On the other hand, the \textit{James Webb} Space Telescope (JWST, \citealt{JWST}) emerges as a pivotal player in helping to lift the degeneracy of the planetary compositions. It is poised to significantly contribute to this endeavor through its capabilities in unveiling the atmospheric complexities of these sub-Neptunes. We calculate the transmission spectroscopy metric (TSM) proposed by \citet{Kempton2018} for the sub-Neptunes of our samples (R$_\text{p} \ge $ 1.5 R$_\oplus$), and find out that with a TSM ranging from 20 to 457, with a median of 71, the sub-Neptunes around M-dwarfs therefore stands out to be excellent candidates for atmospheric characterization via transmission spectroscopy. For the sub-Neptunes of the planets orbiting FGK-stars (R$_\text{p} \ge $ 2 R$_\oplus$), they are less good candidates for atmospheric characterization via transmission spectroscopy with a TSM ranging from 1 to 240, with a median of 34. High-resolution atmospheric spectroscopic observations by instruments such as ESPRESSO and NIRPS complement space telescopes, including the potential detection of crucial components such as water, helium, carbon dioxide, carbon monoxide or methane. Finally, PLATO  will play a crucial role in advancing our understanding of host stars. For instance, it will contribute to determining the age of the stars: this information will be crucial for unraveling the temporal evolution of the stars and their associated planetary systems.

\begin{acknowledgements}
We thank the anonymous referee for valuable comments that helped improve the manuscript. This work has been carried out within the framework of the NCCR PlanetS supported by the Swiss National Science Foundation under grants 51NF40\_182901 and 51NF40\_205606. J.V. acknowledges support from the Swiss National Science Foundation (SNSF) under grant PZ00P2$\_$208945. C.D. acknowledges support from the Swiss National Science Foundation under grant TMSGI2\_211313. This publication makes use of The Data \& Analysis Center for Exoplanets (DACE), which is a facility based at the University of Geneva (CH) dedicated to extrasolar planets data visualization, exchange and analysis. DACE is a platform of the Swiss National Center of Competence in Research (NCCR) PlanetS, federating the Swiss expertise in Exoplanet research. The DACE platform is available at \url{https://dace.unige.ch}. This research has made use of the NASA Exoplanet Archive, which is operated by the California Institute of Technology, under contract with the National Aeronautics and Space Administration under the Exoplanet Exploration Program. This work presents results from the European Space Agency (ESA) space mission Gaia. Gaia data are being processed by the Gaia Data Processing and Analysis Consortium (DPAC). Funding for the DPAC is provided by national institutions, in particular the institutions participating in the Gaia MultiLateral Agreement (MLA). The Gaia mission website is \url{https://www.cosmos.esa.int/gaia}. The Gaia archive website is \url{https://archives.esac.esa.int/gaia}.
\end{acknowledgements}

%
%

\bibliographystyle{aa} 
\bibliography{bib}

\begin{thebibliography}{117}
\expandafter\ifx\csname natexlab\endcsname\relax\def\natexlab#1{#1}\fi

\bibitem[{{Aguichine} {et~al.}(2021){Aguichine}, {Mousis}, {Deleuil}, \&
  {Marcq}}]{Aguichine2021}
{Aguichine}, A., {Mousis}, O., {Deleuil}, M., \& {Marcq}, E. 2021, \apj, 914,
  84

\bibitem[{{Alibert}(2017)}]{Alibert2017b}
{Alibert}, Y. 2017, \aap, 606, A69

\bibitem[{{Alibert} \& {Benz}(2017)}]{Alibert2017}
{Alibert}, Y. \& {Benz}, W. 2017, \aap, 598, L5

\bibitem[{{Alibert} {et~al.}(2018){Alibert}, {Venturini}, {Helled}, {Ataiee},
  {Burn}, {Senecal}, {Benz}, {Mayer}, {Mordasini}, {Quanz}, \&
  {Sch{\"o}nb{\"a}chler}}]{Alibert18}
{Alibert}, Y., {Venturini}, J., {Helled}, R., {et~al.} 2018, Nature Astronomy,
  2, 873

\bibitem[{{Artigau} {et~al.}(2014){Artigau}, {Kouach}, {Donati}, {Doyon},
  {Delfosse}, {Baratchart}, {Lacombe}, {Moutou}, {Rabou}, {Par{\`e}s},
  {Micheau}, {Thibault}, {Reshetov}, {Dubois}, {Hernandez}, {Vall{\'e}e},
  {Wang}, {Dolon}, {Pepe}, {Bouchy}, {Striebig}, {H{\'e}nault}, {Loop},
  {Saddlemyer}, {Barrick}, {Vermeulen}, {Dupieux}, {H{\'e}brard}, {Boisse},
  {Martioli}, {Alencar}, {do Nascimento}, \& {Figueira}}]{SPIRou}
{Artigau}, {\'E}., {Kouach}, D., {Donati}, J.-F., {et~al.} 2014, in Society of
  Photo-Optical Instrumentation Engineers (SPIE) Conference Series, Vol. 9147,
  Ground-based and Airborne Instrumentation for Astronomy V, ed. S.~K.
  {Ramsay}, I.~S. {McLean}, \& H.~{Takami}, 914715

\bibitem[{{Baglin} {et~al.}(2006){Baglin}, {Auvergne}, {Boisnard}, {Lam-Trong},
  {Barge}, {Catala}, {Deleuil}, {Michel}, \& {Weiss}}]{Baglin2006}
{Baglin}, A., {Auvergne}, M., {Boisnard}, L., {et~al.} 2006, in 36th COSPAR
  Scientific Assembly, Vol.~36, 3749

\bibitem[{{Bashi} {et~al.}(2017){Bashi}, {Helled}, {Zucker}, \&
  {Mordasini}}]{Bashi2017}
{Bashi}, D., {Helled}, R., {Zucker}, S., \& {Mordasini}, C. 2017, \aap, 604,
  A83

\bibitem[{{Batalha} {et~al.}(2011){Batalha}, {Borucki}, {Bryson}, {Buchhave},
  {Caldwell}, {Christensen-Dalsgaard}, {Ciardi}, {Dunham}, {Fressin},
  {Gautier}, {Gilliland}, {Haas}, {Howell}, {Jenkins}, {Kjeldsen}, {Koch},
  {Latham}, {Lissauer}, {Marcy}, {Rowe}, {Sasselov}, {Seager}, {Steffen},
  {Torres}, {Basri}, {Brown}, {Charbonneau}, {Christiansen}, {Clarke},
  {Cochran}, {Dupree}, {Fabrycky}, {Fischer}, {Ford}, {Fortney}, {Girouard},
  {Holman}, {Johnson}, {Isaacson}, {Klaus}, {Machalek}, {Moorehead},
  {Morehead}, {Ragozzine}, {Tenenbaum}, {Twicken}, {Quinn}, {VanCleve},
  {Walkowicz}, {Welsh}, {Devore}, \& {Gould}}]{Batalha2011}
{Batalha}, N.~M., {Borucki}, W.~J., {Bryson}, S.~T., {et~al.} 2011, \apj, 729,
  27

\bibitem[{{Batygin} {et~al.}(2011){Batygin}, {Stevenson}, \&
  {Bodenheimer}}]{Batygin2011}
{Batygin}, K., {Stevenson}, D.~J., \& {Bodenheimer}, P.~H. 2011, \apj, 738, 1

\bibitem[{{Berger} {et~al.}(2020){Berger}, {Huber}, {Gaidos}, {van Saders}, \&
  {Weiss}}]{Berger2020}
{Berger}, T.~A., {Huber}, D., {Gaidos}, E., {van Saders}, J.~L., \& {Weiss},
  L.~M. 2020, \aj, 160, 108

\bibitem[{{Bluhm} {et~al.}(2020){Bluhm}, {Luque}, {Espinoza}, {Pall{\'e}},
  {Caballero}, {Dreizler}, {Livingston}, {Mathur}, {Quirrenbach}, {Stock}, {Van
  Eylen}, {Nowak}, {L{\'o}pez}, {Csizmadia}, {Zapatero Osorio}, {Sch{\"o}fer},
  {Lillo-Box}, {Oshagh}, {Gonz{\'a}lez-{\'A}lvarez}, {Amado}, {Barrado},
  {B{\'e}jar}, {Cale}, {Chaturvedi}, {Cifuentes}, {Cochran}, {Collins},
  {Collins}, {Cort{\'e}s-Contreras}, {D{\'\i}ez Alonso}, {El Mufti},
  {Ercolino}, {Fridlund}, {Gaidos}, {Garc{\'\i}a}, {Georgieva},
  {Gonz{\'a}lez-Cuesta}, {Guerra}, {Hatzes}, {Henning}, {Herrero}, {Hidalgo},
  {Isopi}, {Jeffers}, {Jenkins}, {Jensen}, {K{\'a}bath}, {Kaminski}, {Kemmer},
  {Korth}, {Kossakowski}, {K{\"u}rster}, {Lafarga}, {Mallia}, {Montes},
  {Morales}, {Morales-Calder{\'o}n}, {Murgas}, {Narita}, {Passegger}, {Pedraz},
  {Persson}, {Plavchan}, {Rauer}, {Redfield}, {Reffert}, {Reiners}, {Ribas},
  {Ricker}, {Rodr{\'\i}guez-L{\'o}pez}, {Santos}, {Seager}, {Schlecker},
  {Schweitzer}, {Shan}, {Soto}, {Subjak}, {Tal-Or}, {Trifonov}, {Vanaverbeke},
  {Vanderspek}, {Wittrock}, {Zechmeister}, \& {Zohrabi}}]{Bluhm2020}
{Bluhm}, P., {Luque}, R., {Espinoza}, N., {et~al.} 2020, \aap, 639, A132

\bibitem[{Boels {et~al.}(2019)Boels, Bakker, Van~Dooren, \&
  Drijvers}]{Boels2019}
Boels, L., Bakker, A., Van~Dooren, W., \& Drijvers, P. 2019, Educational
  Research Review, 28, 100291

\bibitem[{{Bolmont} {et~al.}(2016){Bolmont}, {Libert}, {Leconte}, \&
  {Selsis}}]{Bolmont2016}
{Bolmont}, E., {Libert}, A.-S., {Leconte}, J., \& {Selsis}, F. 2016, \aap, 591,
  A106

\bibitem[{{Bonfils} {et~al.}(2015){Bonfils}, {Almenara}, {Jocou}, {Wunsche},
  {Kern}, {Delboulb{\'e}}, {Delfosse}, {Feautrier}, {Forveille}, {Gluck},
  {Lafrasse}, {Magnard}, {Maurel}, {Moulin}, {Murgas}, {Rabou}, {Rochat},
  {Roux}, \& {Stadler}}]{ExTrA}
{Bonfils}, X., {Almenara}, J.~M., {Jocou}, L., {et~al.} 2015, in Society of
  Photo-Optical Instrumentation Engineers (SPIE) Conference Series, Vol. 9605,
  Techniques and Instrumentation for Detection of Exoplanets VII, ed.
  S.~{Shaklan}, 96051L

\bibitem[{{Bonfils} {et~al.}(2013){Bonfils}, {Delfosse}, {Udry}, {Forveille},
  {Mayor}, {Perrier}, {Bouchy}, {Gillon}, {Lovis}, {Pepe}, {Queloz}, {Santos},
  {S{\'e}gransan}, \& {Bertaux}}]{Bonfils2013}
{Bonfils}, X., {Delfosse}, X., {Udry}, S., {et~al.} 2013, \aap, 549, A109

\bibitem[{{Borucki} {et~al.}(2010){Borucki}, {Koch}, {Basri}, {Batalha},
  {Brown}, {Caldwell}, {Caldwell}, {Christensen-Dalsgaard}, {Cochran},
  {DeVore}, {Dunham}, {Dupree}, {Gautier}, {Geary}, {Gilliland}, {Gould},
  {Howell}, {Jenkins}, {Kondo}, {Latham}, {Marcy}, {Meibom}, {Kjeldsen},
  {Lissauer}, {Monet}, {Morrison}, {Sasselov}, {Tarter}, {Boss}, {Brownlee},
  {Owen}, {Buzasi}, {Charbonneau}, {Doyle}, {Fortney}, {Ford}, {Holman},
  {Seager}, {Steffen}, {Welsh}, {Rowe}, {Anderson}, {Buchhave}, {Ciardi},
  {Walkowicz}, {Sherry}, {Horch}, {Isaacson}, {Everett}, {Fischer}, {Torres},
  {Johnson}, {Endl}, {MacQueen}, {Bryson}, {Dotson}, {Haas}, {Kolodziejczak},
  {Van Cleve}, {Chandrasekaran}, {Twicken}, {Quintana}, {Clarke}, {Allen},
  {Li}, {Wu}, {Tenenbaum}, {Verner}, {Bruhweiler}, {Barnes}, \&
  {Prsa}}]{Kepler}
{Borucki}, W.~J., {Koch}, D., {Basri}, G., {et~al.} 2010, Science, 327, 977

\bibitem[{{Bouchy} {et~al.}(2017){Bouchy}, {Doyon}, {Artigau}, {Melo},
  {Hernandez}, {Wildi}, {Delfosse}, {Lovis}, {Figueira}, {Canto Martins},
  {Gonz{\'a}lez Hern{\'a}ndez}, {Thibault}, {Reshetov}, {Pepe}, {Santos}, {de
  Medeiros}, {Rebolo}, {Abreu}, {Adibekyan}, {Bandy}, {Benz}, {Blind},
  {Bohlender}, {Boisse}, {Bovay}, {Broeg}, {Brousseau}, {Cabral}, {Chazelas},
  {Cloutier}, {Coelho}, {Conod}, {Cumming}, {Delabre}, {Genolet}, {Hagelberg},
  {Jayawardhana}, {K{\"a}ufl}, {Lafreni{\`e}re}, {de Castro Le{\~a}o}, {Malo},
  {de Medeiros Martins}, {Matthews}, {Metchev}, {Oshagh}, {Ouellet}, {Parro},
  {Rasilla Pi{\~n}eiro}, {Santos}, {Sarajlic}, {Segovia}, {Sordet}, {Udry},
  {Valencia}, {Vall{\'e}e}, {Venn}, {Wade}, \& {Saddlemyer}}]{Bouchy2017}
{Bouchy}, F., {Doyon}, R., {Artigau}, {\'E}., {et~al.} 2017, The Messenger,
  169, 21

\bibitem[{{Brady} \& {Bean}(2022)}]{Brady2022}
{Brady}, M.~T. \& {Bean}, J.~L. 2022, \aj, 163, 255

\bibitem[{{Br{\"u}gger} {et~al.}(2020){Br{\"u}gger}, {Burn}, {Coleman},
  {Alibert}, \& {Benz}}]{Brugger2020}
{Br{\"u}gger}, N., {Burn}, R., {Coleman}, G.~A.~L., {Alibert}, Y., \& {Benz},
  W. 2020, \aap, 640, A21

\bibitem[{{Burn} {et~al.}(2024){Burn}, {Mordasini}, {Mishra}, {Haldemann},
  {Venturini}, {Emsenhuber}, \& {Henning}}]{Burn24}
{Burn}, R., {Mordasini}, C., {Mishra}, L., {et~al.} 2024, Nature Astronomy, 8,
  463

\bibitem[{{Burn} {et~al.}(2021){Burn}, {Schlecker}, {Mordasini}, {Emsenhuber},
  {Alibert}, {Henning}, {Klahr}, \& {Benz}}]{Burn2021}
{Burn}, R., {Schlecker}, M., {Mordasini}, C., {et~al.} 2021, \aap, 656, A72

\bibitem[{{Burt} {et~al.}(2021){Burt}, {Dragomir}, {Molli{\`e}re},
  {Youngblood}, {Garc{\'\i}a Mu{\~n}oz}, {McCann}, {Kreidberg}, {Huang},
  {Collins}, {Eastman}, {Abe}, {Almenara}, {Crossfield}, {Ziegler},
  {Rodriguez}, {Mamajek}, {Stassun}, {Halverson}, {Villanueva}, {Butler},
  {Wang}, {Schwarz}, {Ricker}, {Vanderspek}, {Latham}, {Seager}, {Winn},
  {Jenkins}, {Agabi}, {Bonfils}, {Ciardi}, {Cointepas}, {Crane}, {Crouzet},
  {Dransfield}, {Feng}, {Furlan}, {Guillot}, {Gupta}, {Howell}, {Jensen},
  {Law}, {Mann}, {Marie-Sainte}, {Matson}, {Matthews}, {M{\'e}karnia},
  {Pepper}, {Scott}, {Shectman}, {Schlieder}, {Schmider}, {Stevens}, {Teske},
  {Triaud}, {Charbonneau}, {Berta-Thompson}, {Burke}, {Daylan}, {Barclay},
  {Wohler}, \& {Brasseur}}]{Burt2021}
{Burt}, J.~A., {Dragomir}, D., {Molli{\`e}re}, P., {et~al.} 2021, \aj, 162, 87

\bibitem[{{Chen} \& {Kipping}(2017)}]{Chen2017}
{Chen}, J. \& {Kipping}, D. 2017, \apj, 834, 17

\bibitem[{{Cloutier} {et~al.}(2021){Cloutier}, {Charbonneau}, {Deming},
  {Bonfils}, \& {Astudillo-Defru}}]{Cloutier2021b}
{Cloutier}, R., {Charbonneau}, D., {Deming}, D., {Bonfils}, X., \&
  {Astudillo-Defru}, N. 2021, \aj, 162, 174

\bibitem[{{Cloutier} \& {Menou}(2020)}]{Cloutier2020}
{Cloutier}, R. \& {Menou}, K. 2020, \aj, 159, 211

\bibitem[{{Delrez} {et~al.}(2018){Delrez}, {Gillon}, {Queloz}, {Demory},
  {Almleaky}, {de Wit}, {Jehin}, {Triaud}, {Barkaoui}, {Burdanov}, {Burgasser},
  {Ducrot}, {McCormac}, {Murray}, {Silva Fernandes}, {Sohy}, {Thompson}, {Van
  Grootel}, {Alonso}, {Benkhaldoun}, \& {Rebolo}}]{Delrez2018}
{Delrez}, L., {Gillon}, M., {Queloz}, D., {et~al.} 2018, in Society of
  Photo-Optical Instrumentation Engineers (SPIE) Conference Series, Vol. 10700,
  Ground-based and Airborne Telescopes VII, ed. H.~K. {Marshall} \&
  J.~{Spyromilio}, 107001I

\bibitem[{{Demangeon} {et~al.}(2021){Demangeon}, {Zapatero Osorio}, {Alibert},
  {Barros}, {Adibekyan}, {Tabernero}, {Antoniadis-Karnavas}, {Camacho},
  {Su{\'a}rez Mascare{\~n}o}, {Oshagh}, {Micela}, {Sousa}, {Lovis}, {Pepe},
  {Rebolo}, {Cristiani}, {Santos}, {Allart}, {Allende Prieto}, {Bossini},
  {Bouchy}, {Cabral}, {Damasso}, {Di Marcantonio}, {D'Odorico}, {Ehrenreich},
  {Faria}, {Figueira}, {G{\'e}nova Santos}, {Haldemann}, {Hara}, {Gonz{\'a}lez
  Hern{\'a}ndez}, {Lavie}, {Lillo-Box}, {Lo Curto}, {Martins}, {M{\'e}gevand},
  {Mehner}, {Molaro}, {Nunes}, {Pall{\'e}}, {Pasquini}, {Poretti}, {Sozzetti},
  \& {Udry}}]{Demangeon2021}
{Demangeon}, O.~D.~S., {Zapatero Osorio}, M.~R., {Alibert}, Y., {et~al.} 2021,
  \aap, 653, A41

\bibitem[{{Dietrich} {et~al.}(2023){Dietrich}, {Apai}, {Schlecker},
  {Hardegree-Ullman}, {Rackham}, {Kurtovic}, {Molaverdikhani}, {Gabor},
  {Henning}, {Chen}, {Mancini}, {Bixel}, {Gibbs}, {Boyle}, {Brown-Sevilla},
  {Burn}, {Delage}, {Flores-Rivera}, {Franceschi}, {Pichierri}, {Savvidou},
  {Syed}, {Bruni}, {Ip}, {Ngeow}, {Tsai}, {Lin}, {Hou}, {Hsiao}, {Lin}, {Lin},
  {Basant}, \& {EDEN Project}}]{Dietrich2023}
{Dietrich}, J., {Apai}, D., {Schlecker}, M., {et~al.} 2023, \aj, 165, 149

\bibitem[{{Dorn} \& {Lichtenberg}(2021)}]{Dorn2021}
{Dorn}, C. \& {Lichtenberg}, T. 2021, \apjl, 922, L4

\bibitem[{{Dressing} \& {Charbonneau}(2013)}]{Dressing2013}
{Dressing}, C.~D. \& {Charbonneau}, D. 2013, \apj, 767, 95

\bibitem[{{Dressing} \& {Charbonneau}(2015)}]{Dressing2015}
{Dressing}, C.~D. \& {Charbonneau}, D. 2015, \apj, 807, 45

\bibitem[{{Edmondson} {et~al.}(2023){Edmondson}, {Norris}, \&
  {Kerins}}]{Edmonson2023}
{Edmondson}, K., {Norris}, J., \& {Kerins}, E. 2023, submitted to OJAp,
  arXiv:2310.16733

\bibitem[{{Emsenhuber} {et~al.}(2021){Emsenhuber}, {Mordasini}, {Burn},
  {Alibert}, {Benz}, \& {Asphaug}}]{Emsenhuber2021}
{Emsenhuber}, A., {Mordasini}, C., {Burn}, R., {et~al.} 2021, \aap, 656, A70

\bibitem[{{Fulton} \& {Petigura}(2018)}]{Fulton2018}
{Fulton}, B.~J. \& {Petigura}, E.~A. 2018, \aj, 156, 264

\bibitem[{{Fulton} {et~al.}(2017){Fulton}, {Petigura}, {Howard}, {Isaacson},
  {Marcy}, {Cargile}, {Hebb}, {Weiss}, {Johnson}, {Morton}, {Sinukoff},
  {Crossfield}, \& {Hirsch}}]{Fulton2017}
{Fulton}, B.~J., {Petigura}, E.~A., {Howard}, A.~W., {et~al.} 2017, \aj, 154,
  109

\bibitem[{{Gaia Collaboration} {et~al.}(2016){Gaia Collaboration}, {Prusti},
  {de Bruijne}, {Brown}, {Vallenari}, {Babusiaux}, {Bailer-Jones}, {Bastian},
  {Biermann}, {Evans}, {Eyer}, {Jansen}, {Jordi}, {Klioner}, {Lammers},
  {Lindegren}, {Luri}, {Mignard}, {Milligan}, {Panem}, {Poinsignon},
  {Pourbaix}, {Randich}, {Sarri}, {Sartoretti}, {Siddiqui}, {Soubiran},
  {Valette}, {van Leeuwen}, {Walton}, {Aerts}, {Arenou}, {Cropper}, {Drimmel},
  {H{\o}g}, {Katz}, {Lattanzi}, {O'Mullane}, {Grebel}, {Holland}, {Huc},
  {Passot}, {Bramante}, {Cacciari}, {Casta{\~n}eda}, {Chaoul}, {Cheek}, {De
  Angeli}, {Fabricius}, {Guerra}, {Hern{\'a}ndez}, {Jean-Antoine-Piccolo},
  {Masana}, {Messineo}, {Mowlavi}, {Nienartowicz}, {Ord{\'o}{\~n}ez-Blanco},
  {Panuzzo}, {Portell}, {Richards}, {Riello}, {Seabroke}, {Tanga},
  {Th{\'e}venin}, {Torra}, {Els}, {Gracia-Abril}, {Comoretto},
  {Garcia-Reinaldos}, {Lock}, {Mercier}, {Altmann}, {Andrae}, {Astraatmadja},
  {Bellas-Velidis}, {Benson}, {Berthier}, {Blomme}, {Busso}, {Carry},
  {Cellino}, {Clementini}, {Cowell}, {Creevey}, {Cuypers}, {Davidson}, {De
  Ridder}, {de Torres}, {Delchambre}, {Dell'Oro}, {Ducourant}, {Fr{\'e}mat},
  {Garc{\'\i}a-Torres}, {Gosset}, {Halbwachs}, {Hambly}, {Harrison}, {Hauser},
  {Hestroffer}, {Hodgkin}, {Huckle}, {Hutton}, {Jasniewicz}, {Jordan},
  {Kontizas}, {Korn}, {Lanzafame}, {Manteiga}, {Moitinho}, {Muinonen},
  {Osinde}, {Pancino}, {Pauwels}, {Petit}, {Recio-Blanco}, {Robin}, {Sarro},
  {Siopis}, {Smith}, {Smith}, {Sozzetti}, {Thuillot}, {van Reeven}, {Viala},
  {Abbas}, {Abreu Aramburu}, {Accart}, {Aguado}, {Allan}, {Allasia},
  {Altavilla}, {{\'A}lvarez}, {Alves}, {Anderson}, {Andrei}, {Anglada Varela},
  {Antiche}, {Antoja}, {Ant{\'o}n}, {Arcay}, {Atzei}, {Ayache}, {Bach},
  {Baker}, {Balaguer-N{\'u}{\~n}ez}, {Barache}, {Barata}, {Barbier}, {Barblan},
  {Baroni}, {Barrado y Navascu{\'e}s}, {Barros}, {Barstow}, {Becciani},
  {Bellazzini}, {Bellei}, {Bello Garc{\'\i}a}, {Belokurov}, {Bendjoya},
  {Berihuete}, {Bianchi}, {Bienaym{\'e}}, {Billebaud}, {Blagorodnova},
  {Blanco-Cuaresma}, {Boch}, {Bombrun}, {Borrachero}, {Bouquillon}, {Bourda},
  {Bouy}, {Bragaglia}, {Breddels}, {Brouillet}, {Br{\"u}semeister},
  {Bucciarelli}, {Budnik}, {Burgess}, {Burgon}, {Burlacu}, {Busonero}, {Buzzi},
  {Caffau}, {Cambras}, {Campbell}, {Cancelliere}, {Cantat-Gaudin}, {Carlucci},
  {Carrasco}, {Castellani}, {Charlot}, {Charnas}, {Charvet}, {Chassat},
  {Chiavassa}, {Clotet}, {Cocozza}, {Collins}, {Collins}, {Costigan}, {Crifo},
  {Cross}, {Crosta}, {Crowley}, {Dafonte}, {Damerdji}, {Dapergolas}, {David},
  {David}, {De Cat}, {de Felice}, {de Laverny}, {De Luise}, {De March}, {de
  Martino}, {de Souza}, {Debosscher}, {del Pozo}, {Delbo}, {Delgado},
  {Delgado}, {di Marco}, {Di Matteo}, {Diakite}, {Distefano}, {Dolding}, {Dos
  Anjos}, {Drazinos}, {Dur{\'a}n}, {Dzigan}, {Ecale}, {Edvardsson}, {Enke},
  {Erdmann}, {Escolar}, {Espina}, {Evans}, {Eynard Bontemps}, {Fabre},
  {Fabrizio}, {Faigler}, {Falc{\~a}o}, {Farr{\`a}s Casas}, {Faye}, {Federici},
  {Fedorets}, {Fern{\'a}ndez-Hern{\'a}ndez}, {Fernique}, {Fienga}, {Figueras},
  {Filippi}, {Findeisen}, {Fonti}, {Fouesneau}, {Fraile}, {Fraser}, {Fuchs},
  {Furnell}, {Gai}, {Galleti}, {Galluccio}, {Garabato}, {Garc{\'\i}a-Sedano},
  {Gar{\'e}}, {Garofalo}, {Garralda}, {Gavras}, {Gerssen}, {Geyer}, {Gilmore},
  {Girona}, {Giuffrida}, {Gomes}, {Gonz{\'a}lez-Marcos},
  {Gonz{\'a}lez-N{\'u}{\~n}ez}, {Gonz{\'a}lez-Vidal}, {Granvik}, {Guerrier},
  {Guillout}, {Guiraud}, {G{\'u}rpide}, {Guti{\'e}rrez-S{\'a}nchez}, {Guy},
  {Haigron}, {Hatzidimitriou}, {Haywood}, {Heiter}, {Helmi}, {Hobbs},
  {Hofmann}, {Holl}, {Holland}, {Hunt}, {Hypki}, {Icardi}, {Irwin}, {Jevardat
  de Fombelle}, {Jofr{\'e}}, {Jonker}, {Jorissen}, {Julbe}, {Karampelas},
  {Kochoska}, {Kohley}, {Kolenberg}, {Kontizas}, {Koposov}, {Kordopatis},
  {Koubsky}, {Kowalczyk}, {Krone-Martins}, {Kudryashova}, {Kull}, {Bachchan},
  {Lacoste-Seris}, {Lanza}, {Lavigne}, {Le Poncin-Lafitte}, {Lebreton},
  {Lebzelter}, {Leccia}, {Leclerc}, {Lecoeur-Taibi}, {Lemaitre}, {Lenhardt},
  {Leroux}, {Liao}, {Licata}, {Lindstr{\o}m}, {Lister}, {Livanou}, {Lobel},
  {L{\"o}ffler}, {L{\'o}pez}, {Lopez-Lozano}, {Lorenz}, {Loureiro},
  {MacDonald}, {Magalh{\~a}es Fernandes}, {Managau}, {Mann}, {Mantelet},
  {Marchal}, {Marchant}, {Marconi}, {Marie}, {Marinoni}, {Marrese},
  {Marschalk{\'o}}, {Marshall}, {Mart{\'\i}n-Fleitas}, {Martino}, {Mary},
  {Matijevi{\v{c}}}, {Mazeh}, {McMillan}, {Messina}, {Mestre}, {Michalik},
  {Millar}, {Miranda}, {Molina}, {Molinaro}, {Molinaro}, {Moln{\'a}r},
  {Moniez}, {Montegriffo}, {Monteiro}, {Mor}, {Mora}, {Morbidelli}, {Morel},
  {Morgenthaler}, {Morley}, {Morris}, {Mulone}, {Muraveva}, {Musella},
  {Narbonne}, {Nelemans}, {Nicastro}, {Noval}, {Ord{\'e}novic},
  {Ordieres-Mer{\'e}}, {Osborne}, {Pagani}, {Pagano}, {Pailler}, {Palacin},
  {Palaversa}, {Parsons}, {Paulsen}, {Pecoraro}, {Pedrosa}, {Pentik{\"a}inen},
  {Pereira}, {Pichon}, {Piersimoni}, {Pineau}, {Plachy}, {Plum}, {Poujoulet},
  {Pr{\v{s}}a}, {Pulone}, {Ragaini}, {Rago}, {Rambaux}, {Ramos-Lerate},
  {Ranalli}, {Rauw}, {Read}, {Regibo}, {Renk}, {Reyl{\'e}}, {Ribeiro},
  {Rimoldini}, {Ripepi}, {Riva}, {Rixon}, {Roelens}, {Romero-G{\'o}mez},
  {Rowell}, {Royer}, {Rudolph}, {Ruiz-Dern}, {Sadowski}, {Sagrist{\`a}
  Sell{\'e}s}, {Sahlmann}, {Salgado}, {Salguero}, {Sarasso}, {Savietto},
  {Schnorhk}, {Schultheis}, {Sciacca}, {Segol}, {Segovia}, {Segransan},
  {Serpell}, {Shih}, {Smareglia}, {Smart}, {Smith}, {Solano}, {Solitro},
  {Sordo}, {Soria Nieto}, {Souchay}, {Spagna}, {Spoto}, {Stampa}, {Steele},
  {Steidelm{\"u}ller}, {Stephenson}, {Stoev}, {Suess}, {S{\"u}veges}, {Surdej},
  {Szabados}, {Szegedi-Elek}, {Tapiador}, {Taris}, {Tauran}, {Taylor},
  {Teixeira}, {Terrett}, {Tingley}, {Trager}, {Turon}, {Ulla}, {Utrilla},
  {Valentini}, {van Elteren}, {Van Hemelryck}, {van Leeuwen}, {Varadi},
  {Vecchiato}, {Veljanoski}, {Via}, {Vicente}, {Vogt}, {Voss}, {Votruba},
  {Voutsinas}, {Walmsley}, {Weiler}, {Weingrill}, {Werner}, {Wevers},
  {Whitehead}, {Wyrzykowski}, {Yoldas}, {{\v{Z}}erjal}, {Zucker}, {Zurbach},
  {Zwitter}, {Alecu}, {Allen}, {Allende Prieto}, {Amorim},
  {Anglada-Escud{\'e}}, {Arsenijevic}, {Azaz}, {Balm}, {Beck}, {Bernstein},
  {Bigot}, {Bijaoui}, {Blasco}, {Bonfigli}, {Bono}, {Boudreault}, {Bressan},
  {Brown}, {Brunet}, {Bunclark}, {Buonanno}, {Butkevich}, {Carret}, {Carrion},
  {Chemin}, {Ch{\'e}reau}, {Corcione}, {Darmigny}, {de Boer}, {de Teodoro}, {de
  Zeeuw}, {Delle Luche}, {Domingues}, {Dubath}, {Fodor}, {Fr{\'e}zouls},
  {Fries}, {Fustes}, {Fyfe}, {Gallardo}, {Gallegos}, {Gardiol}, {Gebran},
  {Gomboc}, {G{\'o}mez}, {Grux}, {Gueguen}, {Heyrovsky}, {Hoar}, {Iannicola},
  {Isasi Parache}, {Janotto}, {Joliet}, {Jonckheere}, {Keil}, {Kim},
  {Klagyivik}, {Klar}, {Knude}, {Kochukhov}, {Kolka}, {Kos}, {Kutka}, {Lainey},
  {LeBouquin}, {Liu}, {Loreggia}, {Makarov}, {Marseille}, {Martayan},
  {Martinez-Rubi}, {Massart}, {Meynadier}, {Mignot}, {Munari}, {Nguyen},
  {Nordlander}, {Ocvirk}, {O'Flaherty}, {Olias Sanz}, {Ortiz}, {Osorio},
  {Oszkiewicz}, {Ouzounis}, {Palmer}, {Park}, {Pasquato}, {Peltzer}, {Peralta},
  {P{\'e}turaud}, {Pieniluoma}, {Pigozzi}, {Poels}, {Prat}, {Prod'homme},
  {Raison}, {Rebordao}, {Risquez}, {Rocca-Volmerange}, {Rosen}, {Ruiz-Fuertes},
  {Russo}, {Sembay}, {Serraller Vizcaino}, {Short}, {Siebert}, {Silva},
  {Sinachopoulos}, {Slezak}, {Soffel}, {Sosnowska}, {Strai{\v{z}}ys}, {ter
  Linden}, {Terrell}, {Theil}, {Tiede}, {Troisi}, {Tsalmantza}, {Tur},
  {Vaccari}, {Vachier}, {Valles}, {Van Hamme}, {Veltz}, {Virtanen}, {Wallut},
  {Wichmann}, {Wilkinson}, {Ziaeepour}, \& {Zschocke}}]{gaia2016}
{Gaia Collaboration}, {Prusti}, T., {de Bruijne}, J.~H.~J., {et~al.} 2016,
  \aap, 595, A1

\bibitem[{{Gaia Collaboration} {et~al.}(2023){Gaia Collaboration}, {Vallenari},
  {Brown}, {Prusti}, {de Bruijne}, {Arenou}, {Babusiaux}, {Biermann},
  {Creevey}, {Ducourant}, {Evans}, {Eyer}, {Guerra}, {Hutton}, {Jordi},
  {Klioner}, {Lammers}, {Lindegren}, {Luri}, {Mignard}, {Panem}, {Pourbaix},
  {Randich}, {Sartoretti}, {Soubiran}, {Tanga}, {Walton}, {Bailer-Jones},
  {Bastian}, {Drimmel}, {Jansen}, {Katz}, {Lattanzi}, {van Leeuwen}, {Bakker},
  {Cacciari}, {Casta{\~n}eda}, {De Angeli}, {Fabricius}, {Fouesneau},
  {Fr{\'e}mat}, {Galluccio}, {Guerrier}, {Heiter}, {Masana}, {Messineo},
  {Mowlavi}, {Nicolas}, {Nienartowicz}, {Pailler}, {Panuzzo}, {Riclet}, {Roux},
  {Seabroke}, {Sordo}, {Th{\'e}venin}, {Gracia-Abril}, {Portell}, {Teyssier},
  {Altmann}, {Andrae}, {Audard}, {Bellas-Velidis}, {Benson}, {Berthier},
  {Blomme}, {Burgess}, {Busonero}, {Busso}, {C{\'a}novas}, {Carry}, {Cellino},
  {Cheek}, {Clementini}, {Damerdji}, {Davidson}, {de Teodoro}, {Nu{\~n}ez
  Campos}, {Delchambre}, {Dell'Oro}, {Esquej}, {Fern{\'a}ndez-Hern{\'a}ndez},
  {Fraile}, {Garabato}, {Garc{\'\i}a-Lario}, {Gosset}, {Haigron}, {Halbwachs},
  {Hambly}, {Harrison}, {Hern{\'a}ndez}, {Hestroffer}, {Hodgkin}, {Holl},
  {Jan{\ss}en}, {Jevardat de Fombelle}, {Jordan}, {Krone-Martins}, {Lanzafame},
  {L{\"o}ffler}, {Marchal}, {Marrese}, {Moitinho}, {Muinonen}, {Osborne},
  {Pancino}, {Pauwels}, {Recio-Blanco}, {Reyl{\'e}}, {Riello}, {Rimoldini},
  {Roegiers}, {Rybizki}, {Sarro}, {Siopis}, {Smith}, {Sozzetti}, {Utrilla},
  {van Leeuwen}, {Abbas}, {{\'A}brah{\'a}m}, {Abreu Aramburu}, {Aerts},
  {Aguado}, {Ajaj}, {Aldea-Montero}, {Altavilla}, {{\'A}lvarez}, {Alves},
  {Anders}, {Anderson}, {Anglada Varela}, {Antoja}, {Baines}, {Baker},
  {Balaguer-N{\'u}{\~n}ez}, {Balbinot}, {Balog}, {Barache}, {Barbato},
  {Barros}, {Barstow}, {Bartolom{\'e}}, {Bassilana}, {Bauchet}, {Becciani},
  {Bellazzini}, {Berihuete}, {Bernet}, {Bertone}, {Bianchi}, {Binnenfeld},
  {Blanco-Cuaresma}, {Blazere}, {Boch}, {Bombrun}, {Bossini}, {Bouquillon},
  {Bragaglia}, {Bramante}, {Breedt}, {Bressan}, {Brouillet}, {Brugaletta},
  {Bucciarelli}, {Burlacu}, {Butkevich}, {Buzzi}, {Caffau}, {Cancelliere},
  {Cantat-Gaudin}, {Carballo}, {Carlucci}, {Carnerero}, {Carrasco},
  {Casamiquela}, {Castellani}, {Castro-Ginard}, {Chaoul}, {Charlot}, {Chemin},
  {Chiaramida}, {Chiavassa}, {Chornay}, {Comoretto}, {Contursi}, {Cooper},
  {Cornez}, {Cowell}, {Crifo}, {Cropper}, {Crosta}, {Crowley}, {Dafonte},
  {Dapergolas}, {David}, {David}, {de Laverny}, {De Luise}, {De March}, {De
  Ridder}, {de Souza}, {de Torres}, {del Peloso}, {del Pozo}, {Delbo},
  {Delgado}, {Delisle}, {Demouchy}, {Dharmawardena}, {Di Matteo}, {Diakite},
  {Diener}, {Distefano}, {Dolding}, {Edvardsson}, {Enke}, {Fabre}, {Fabrizio},
  {Faigler}, {Fedorets}, {Fernique}, {Fienga}, {Figueras}, {Fournier},
  {Fouron}, {Fragkoudi}, {Gai}, {Garcia-Gutierrez}, {Garcia-Reinaldos},
  {Garc{\'\i}a-Torres}, {Garofalo}, {Gavel}, {Gavras}, {Gerlach}, {Geyer},
  {Giacobbe}, {Gilmore}, {Girona}, {Giuffrida}, {Gomel}, {Gomez},
  {Gonz{\'a}lez-N{\'u}{\~n}ez}, {Gonz{\'a}lez-Santamar{\'\i}a},
  {Gonz{\'a}lez-Vidal}, {Granvik}, {Guillout}, {Guiraud},
  {Guti{\'e}rrez-S{\'a}nchez}, {Guy}, {Hatzidimitriou}, {Hauser}, {Haywood},
  {Helmer}, {Helmi}, {Sarmiento}, {Hidalgo}, {Hilger}, {H{\l}adczuk}, {Hobbs},
  {Holland}, {Huckle}, {Jardine}, {Jasniewicz}, {Jean-Antoine Piccolo},
  {Jim{\'e}nez-Arranz}, {Jorissen}, {Juaristi Campillo}, {Julbe}, {Karbevska},
  {Kervella}, {Khanna}, {Kontizas}, {Kordopatis}, {Korn}, {K{\'o}sp{\'a}l},
  {Kostrzewa-Rutkowska}, {Kruszy{\'n}ska}, {Kun}, {Laizeau}, {Lambert},
  {Lanza}, {Lasne}, {Le Campion}, {Lebreton}, {Lebzelter}, {Leccia}, {Leclerc},
  {Lecoeur-Taibi}, {Liao}, {Licata}, {Lindstr{\o}m}, {Lister}, {Livanou},
  {Lobel}, {Lorca}, {Loup}, {Madrero Pardo}, {Magdaleno Romeo}, {Managau},
  {Mann}, {Manteiga}, {Marchant}, {Marconi}, {Marcos}, {Marcos Santos},
  {Mar{\'\i}n Pina}, {Marinoni}, {Marocco}, {Marshall}, {Martin Polo},
  {Mart{\'\i}n-Fleitas}, {Marton}, {Mary}, {Masip}, {Massari},
  {Mastrobuono-Battisti}, {Mazeh}, {McMillan}, {Messina}, {Michalik}, {Millar},
  {Mints}, {Molina}, {Molinaro}, {Moln{\'a}r}, {Monari}, {Mongui{\'o}},
  {Montegriffo}, {Montero}, {Mor}, {Mora}, {Morbidelli}, {Morel}, {Morris},
  {Muraveva}, {Murphy}, {Musella}, {Nagy}, {Noval}, {Oca{\~n}a}, {Ogden},
  {Ordenovic}, {Osinde}, {Pagani}, {Pagano}, {Palaversa}, {Palicio},
  {Pallas-Quintela}, {Panahi}, {Payne-Wardenaar}, {Pe{\~n}alosa Esteller},
  {Penttil{\"a}}, {Pichon}, {Piersimoni}, {Pineau}, {Plachy}, {Plum}, {Poggio},
  {Pr{\v{s}}a}, {Pulone}, {Racero}, {Ragaini}, {Rainer}, {Raiteri}, {Rambaux},
  {Ramos}, {Ramos-Lerate}, {Re Fiorentin}, {Regibo}, {Richards}, {Rios Diaz},
  {Ripepi}, {Riva}, {Rix}, {Rixon}, {Robichon}, {Robin}, {Robin}, {Roelens},
  {Rogues}, {Rohrbasser}, {Romero-G{\'o}mez}, {Rowell}, {Royer}, {Ruz Mieres},
  {Rybicki}, {Sadowski}, {S{\'a}ez N{\'u}{\~n}ez}, {Sagrist{\`a} Sell{\'e}s},
  {Sahlmann}, {Salguero}, {Samaras}, {Sanchez Gimenez}, {Sanna},
  {Santove{\~n}a}, {Sarasso}, {Schultheis}, {Sciacca}, {Segol}, {Segovia},
  {S{\'e}gransan}, {Semeux}, {Shahaf}, {Siddiqui}, {Siebert}, {Siltala},
  {Silvelo}, {Slezak}, {Slezak}, {Smart}, {Snaith}, {Solano}, {Solitro},
  {Souami}, {Souchay}, {Spagna}, {Spina}, {Spoto}, {Steele},
  {Steidelm{\"u}ller}, {Stephenson}, {S{\"u}veges}, {Surdej}, {Szabados},
  {Szegedi-Elek}, {Taris}, {Taylor}, {Teixeira}, {Tolomei}, {Tonello}, {Torra},
  {Torra}, {Torralba Elipe}, {Trabucchi}, {Tsounis}, {Turon}, {Ulla}, {Unger},
  {Vaillant}, {van Dillen}, {van Reeven}, {Vanel}, {Vecchiato}, {Viala},
  {Vicente}, {Voutsinas}, {Weiler}, {Wevers}, {Wyrzykowski}, {Yoldas}, {Yvard},
  {Zhao}, {Zorec}, {Zucker}, \& {Zwitter}}]{gaia2023}
{Gaia Collaboration}, {Vallenari}, A., {Brown}, A.~G.~A., {et~al.} 2023, \aap,
  674, A1

\bibitem[{{Gaidos} {et~al.}(2016){Gaidos}, {Mann}, {Kraus}, \&
  {Ireland}}]{Gaidos2016}
{Gaidos}, E., {Mann}, A.~W., {Kraus}, A.~L., \& {Ireland}, M. 2016, \mnras,
  457, 2877

\bibitem[{{Gardner} {et~al.}(2023){Gardner}, {Mather}, {Abbott}, {Abell},
  {Abernathy}, {Abney}, {Abraham}, {Abraham}, {Abul-Huda}, {Acton}, {Adams},
  {Adams}, {Adler}, {Adriaensen}, {Aguilar}, {Ahmed}, {Ahmed}, {Ahmed},
  {Albat}, {Albert}, {Alberts}, {Aldridge}, {Allen}, {Allen}, {Altenburg},
  {Altunc}, {Alvarez}, {{\'A}lvarez-M{\'a}rquez}, {Alves de Oliveira},
  {Ambrose}, {Anandakrishnan}, {Andersen}, {Anderson}, {Anderson}, {Anderson},
  {Anderson}, {Aprea}, {Archer}, {Arenberg}, {Argyriou}, {Arribas}, {Artigau},
  {Arvai}, {Atcheson}, {Atkinson}, {Averbukh}, {Aymergen}, {Bacinski},
  {Baggett}, {Bagnasco}, {Baker}, {Balzano}, {Banks}, {Baran}, {Barker},
  {Barrett}, {Barringer}, {Barto}, {Bast}, {Baudoz}, {Baum}, {Beatty},
  {Beaulieu}, {Bechtold}, {Beck}, {Beddard}, {Beichman}, {Bellagama}, {Bely},
  {Berger}, {Bergeron}, {Bernier}, {Bertch}, {Beskow}, {Betz}, {Biagetti},
  {Birkmann}, {Bjorklund}, {Blackwood}, {Blazek}, {Blossfeld}, {Bluth},
  {Boccaletti}, {Boegner}, {Bohlin}, {Boia}, {B{\"o}ker}, {Bonaventura},
  {Bond}, {Bosley}, {Boucarut}, {Bouchet}, {Bouwman}, {Bower}, {Bowers},
  {Bowers}, {Boyce}, {Boyer}, {Boyer}, {Boyer}, {Boyer}, {Bradley}, {Brady},
  {Brandl}, {Brannen}, {Breda}, {Bremmer}, {Brennan}, {Bresnahan}, {Bright},
  {Broiles}, {Bromenschenkel}, {Brooks}, {Brooks}, {Brown}, {Brown}, {Brown},
  {Bruce}, {Bryson}, {Bujanda}, {Bullock}, {Bunker}, {Bureo}, {Burt}, {Bush},
  {Bushouse}, {Bussman}, {Cabaud}, {Cale}, {Calhoon}, {Calvani}, {Canipe},
  {Caputo}, {Cara}, {Carey}, {Case}, {Cesari}, {Cetorelli}, {Chance},
  {Chandler}, {Chaney}, {Chapman}, {Charlot}, {Chayer}, {Cheezum}, {Chen},
  {Chen}, {Cherinka}, {Chichester}, {Chilton}, {Chittiraibalan}, {Clampin},
  {Clark}, {Clark}, {Clark}, {Claybrooks}, {Cleveland}, {Cohen}, {Cohen},
  {Col{\'o}n}, {Coleman}, {Colina}, {Comber}, {Comeau}, {Comer}, {Conde Reis},
  {Connolly}, {Conroy}, {Contos}, {Contreras}, {Cook}, {Cooper}, {Cooper},
  {Correia}, {Correnti}, {Cossou}, {Costanza}, {Coulais}, {Cox}, {Coyle},
  {Cracraft}, {Crew}, {Curtis}, {Cusveller}, {Da Costa Maciel}, {Dailey},
  {Daugeron}, {Davidson}, {Davies}, {Davis}, {Davis}, {Day}, {de Chambure}, {de
  Jong}, {De Marchi}, {Dean}, {Decker}, {Delisa}, {Dell}, {Dellagatta},
  {Dembinska}, {Demosthenes}, {Dencheva}, {Deneu}, {DePriest}, {Deschenes},
  {Dethienne}, {Detre}, {Diaz}, {Dicken}, {DiFelice}, {Dillman}, {Disharoon},
  {Dixon}, {Doggett}, {Dominguez}, {Donaldson}, {Doria-Warner}, {Santos},
  {Doty}, {Douglas}, {Doyon}, {Dressler}, {Driggers}, {Driggers}, {Dunn},
  {DuPrie}, {Dupuis}, {Durning}, {Dutta}, {Earl}, {Eccleston}, {Ecobichon},
  {Egami}, {Ehrenwinkler}, {Eisenhamer}, {Eisenhower}, {Eisenstein}, {El
  Hamel}, {Elie}, {Elliott}, {Elliott}, {Engesser}, {Espinoza}, {Etienne},
  {Etxaluze}, {Evans}, {Fabreguettes}, {Falcolini}, {Falini}, {Fatig},
  {Feeney}, {Feinberg}, {Fels}, {Ferdous}, {Ferguson}, {Ferrarese}, {Ferreira},
  {Ferruit}, {Ferry}, {Filippazzo}, {Firre}, {Fix}, {Flagey}, {Flanagan},
  {Fleming}, {Florian}, {Flynn}, {Foiadelli}, {Fontaine}, {Fontanella},
  {Forshay}, {Fortner}, {Fox}, {Framarini}, {Francisco}, {Franck}, {Franx},
  {Franz}, {Friedman}, {Friend}, {Frost}, {Fu}, {Fullerton}, {Gaillard},
  {Galkin}, {Gallagher}, {Galyer}, {Garc{\'\i}a Mar{\'\i}n}, {Gardner},
  {Garland}, {Garrett}, {Gasman}, {G{\'a}sp{\'a}r}, {Gastaud}, {Gaudreau},
  {Gauthier}, {Geers}, {Geithner}, {Gennaro}, {Gerber}, {Gereau}, {Giampaoli},
  {Giardino}, {Gibbons}, {Gilbert}, {Gilman}, {Girard}, {Giuliano}, {Gkountis},
  {Glasse}, {Glassmire}, {Glauser}, {Glazer}, {Goldberg}, {Golimowski},
  {Gonzaga}, {Gordon}, {Gordon}, {Goudfrooij}, {Gough}, {Graham}, {Grau},
  {Green}, {Greene}, {Greene}, {Greenfield}, {Greenhouse}, {Greve}, {Greville},
  {Grimaldi}, {Groe}, {Groebner}, {Grumm}, {Grundy}, {G{\"u}del}, {Guillard},
  {Guldalian}, {Gunn}, {Gurule}, {Gutman}, {Guy}, {Guyot}, {Hack}, {Haderlein},
  {Hagan}, {Hagedorn}, {Hainline}, {Haley}, {Hami}, {Hamilton}, {Hammann},
  {Hammel}, {Hanley}, {Hansen}, {Hardy}, {Harnisch}, {Harr}, {Harris}, {Hart},
  {Hartig}, {Hasan}, {Hashim}, {Hashimoto}, {Haskins}, {Hawkins}, {Hayden},
  {Hayden}, {Healy}, {Hecht}, {Heeg}, {Hejal}, {Helm}, {Hengemihle}, {Henning},
  {Henry}, {Henry}, {Henshaw}, {Hernandez}, {Herrington}, {Heske}, {Hesman},
  {Hickey}, {Hilbert}, {Hines}, {Hinz}, {Hirsch}, {Hitcho}, {Hodapp}, {Hodge},
  {Hoffman}, {Holfeltz}, {Holler}, {Hoppa}, {Horner}, {Howard}, {Howard},
  {Huber}, {Hunkeler}, {Hunter}, {Hunter}, {Hurd}, {Hurst}, {Hutchings},
  {Hylan}, {Ignat}, {Illingworth}, {Irish}, {Isaacs}, {Jackson}, {Jaffe},
  {Jahic}, {Jahromi}, {Jakobsen}, {James}, {James}, {James}, {Jamieson},
  {Jandra}, {Jayawardhana}, {Jedrzejewski}, {Jeffers}, {Jensen}, {Joanne},
  {Johns}, {Johnson}, {Johnson}, {Johnson}, {Johnson}, {Johnson}, {Johnson},
  {Johnstone}, {Jollet}, {Jones}, {Jones}, {Jones}, {Jones}, {Jones}, {Jordan},
  {Jordan}, {Jue}, {Jurkowski}, {Justis}, {Justtanont}, {Kaleida}, {Kalirai},
  {Kalmanson}, {Kaltenegger}, {Kammerer}, {Kan}, {Kanarek}, {Kao}, {Karakla},
  {Karl}, {Kassin}, {Kauffman}, {Kavanagh}, {Kelley}, {Kelly}, {Kendrew},
  {Kennedy}, {Kenny}, {Keski-Kuha}, {Keyes}, {Khan}, {Kidwell}, {Kimble},
  {King}, {King}, {Kinzel}, {Kirk}, {Kirkpatrick}, {Klaassen}, {Klingemann},
  {Klintworth}, {Knapp}, {Knight}, {Knollenberg}, {Knutsen}, {Koehler},
  {Koekemoer}, {Kofler}, {Kontson}, {Kovacs}, {Kozhurina-Platais}, {Krause},
  {Kriss}, {Krist}, {Kristoffersen}, {Krogel}, {Krueger}, {Kulp}, {Kumari},
  {Kwan}, {Kyprianou}, {Labador}, {Labiano}, {Lafreni{\`e}re}, {Lagage},
  {Laidler}, {Laine}, {Laird}, {Lajoie}, {Lallo}, {Lam}, {LaMassa}, {Lambros},
  {Lampenfield}, {Lander}, {Langston}, {Larson}, {Larson}, {LaVerghetta},
  {Law}, {Lawrence}, {Lee}, {Lee}, {Lee}, {Leisenring}, {Leveille}, {Levenson},
  {Levi}, {Levine}, {Lewis}, {Lewis}, {Lewis}, {Libralato}, {Lidon},
  {Liebrecht}, {Lightsey}, {Lilly}, {Lim}, {Lim}, {Ling}, {Link}, {Link},
  {Lipinski}, {Liu}, {Lo}, {Lobmeyer}, {Logue}, {Long}, {Long}, {Long}, {Long},
  {L{\'o}pez-Caniego}, {Lotz}, {Love-Pruitt}, {Lubskiy}, {Luers}, {Luetgens},
  {Luevano}, {Lui}, {Lund}, {Lundquist}, {Lunine}, {L{\"u}tzgendorf}, {Lynch},
  {MacDonald}, {MacDonald}, {Macias}, {Macklis}, {Maghami}, {Maharaja},
  {Maiolino}, {Makrygiannis}, {Malla}, {Malumuth}, {Manjavacas}, {Marini},
  {Marrione}, {Marston}, {Martel}, {Martin}, {Martin}, {Martinez}, {Maschmann},
  {Masci}, {Masetti}, {Maszkiewicz}, {Matthews}, {Matuskey}, {McBrayer},
  {McCarthy}, {McCaughrean}, {McClare}, {McClare}, {McCloskey}, {McClurg},
  {McCoy}, {McElwain}, {McGregor}, {McGuffey}, {McKay}, {McKenzie}, {McLean},
  {McMaster}, {McNeil}, {De Meester}, {Mehalick}, {Meixner}, {Mel{\'e}ndez},
  {Menzel}, {Menzel}, {Merz}, {Mesterharm}, {Meyer}, {Meyett}, {Meza},
  {Midwinter}, {Milam}, {Miller}, {Miller}, {Miskey}, {Misselt}, {Mitchell},
  {Mohan}, {Montoya}, {Moran}, {Morishita}, {Moro-Mart{\'\i}n}, {Morrison},
  {Morrison}, {Morse}, {Moschos}, {Moseley}, {Mosier}, {Mosner}, {Mountain},
  {Muckenthaler}, {Mueller}, {Mueller}, {Muhiem}, {M{\"u}hlmann}, {Mullally},
  {Mullen}, {Munger}, {Murphy}, {Murray}, {Muzerolle}, {Mycroft}, {Myers},
  {Myers}, {Myers}, {Myers}, {Myrick}, {Nagle}, {Nayak}, {Naylor}, {Neff},
  {Nelan}, {Nella}, {Nguyen}, {Nguyen}, {Nickson}, {Nidhiry}, {Niedner},
  {Nieto-Santisteban}, {Nikolov}, {Nishisaka}, {Noriega-Crespo}, {Nota},
  {O'Mara}, {Oboryshko}, {O'Brien}, {Ochs}, {Offenberg}, {Ogle}, {Ohl},
  {Olmsted}, {Osborne}, {O'Shaughnessy}, {{\"O}stlin}, {O'Sullivan}, {Otor},
  {Ottens}, {Ouellette}, {Outlaw}, {Owens}, {Pacifici}, {Page}, {Paranilam},
  {Park}, {Parrish}, {Paschal}, {Patapis}, {Patel}, {Patrick}, {Pattishall},
  {Paul}, {Paul}, {Pauly}, {Pavlovsky}, {Pe{\~n}a-Guerrero}, {Pedder}, {Peek},
  {Pelham}, {Penanen}, {Perriello}, {Perrin}, {Perrine}, {Perrygo}, {Peslier},
  {Petach}, {Peterson}, {Pfarr}, {Pierson}, {Pietraszkiewicz}, {Pilchen},
  {Pipher}, {Pirzkal}, {Pitman}, {Player}, {Plesha}, {Plitzke}, {Pohner},
  {Poletis}, {Pollizzi}, {Polster}, {Pontius}, {Pontoppidan}, {Porges},
  {Potter}, {Prescott}, {Proffitt}, {Pueyo}, {Quispe Neira}, {Radich}, {Rager},
  {Rameau}, {Ramey}, {Ramos Alarcon}, {Rampini}, {Rapp}, {Rashford},
  {Rauscher}, {Ravindranath}, {Rawle}, {Rawlings}, {Ray}, {Regan}, {Rehm},
  {Rehm}, {Reid}, {Reis}, {Renk}, {Reoch}, {Ressler}, {Rest}, {Reynolds},
  {Richon}, {Richon}, {Ridgaway}, {Riedel}, {Rieke}, {Rieke}, {Rifelli},
  {Rigby}, {Riggs}, {Ringel}, {Ritchie}, {Rix}, {Robberto}, {Robinson},
  {Robinson}, {Robinson}, {Rock}, {Rodriguez}, {Rodr{\'\i}guez del Pino},
  {Roellig}, {Rohrbach}, {Roman}, {Romelfanger}, {Romo}, {Rosales}, {Rose},
  {Roteliuk}, {Roth}, {Rothwell}, {Rouzaud}, {Rowe}, {Rowlands}, {Roy},
  {Royer}, {Rui}, {Rumler}, {Rumpl}, {Russ}, {Ryan}, {Ryan}, {Saad}, {Sabata},
  {Sabatino}, {Sabbi}, {Sabelhaus}, {Sabia}, {Sahu}, {Saif}, {Salvignol},
  {Samara-Ratna}, {Samuelson}, {Sanders}, {Sappington}, {Sargent}, {Sauer},
  {Savadkin}, {Sawicki}, {Schappell}, {Scheffer}, {Scheithauer}, {Scherer},
  {Schiff}, {Schlawin}, {Schmeitzky}, {Schmitz}, {Schmude}, {Schneider},
  {Schreiber}, {Schroeven-Deceuninck}, {Schultz}, {Schwab}, {Schwartz},
  {Scoccimarro}, {Scott}, {Scott}, {Seaton}, {Seely}, {Seery}, {Seidleck},
  {Sembach}, {Shanahan}, {Shaughnessy}, {Shaw}, {Shay}, {Sheehan}, {Sheth},
  {Shih}, {Shivaei}, {Siegel}, {Sienkiewicz}, {Simmons}, {Simon}, {Sirianni},
  {Sivaramakrishnan}, {Slade}, {Sloan}, {Slocum}, {Slowinski}, {Smith},
  {Smith}, {Smith}, {Smith}, {Smith}, {Smith}, {Smolik}, {Soderblom}, {Sohn},
  {Sokol}, {Sonneborn}, {Sontag}, {Sooy}, {Soummer}, {Southwood}, {Spain},
  {Sparmo}, {Speer}, {Spencer}, {Sprofera}, {Stallcup}, {Stanley},
  {Stansberry}, {Stark}, {Starr}, {Stassi}, {Steck}, {Steeley}, {Stephens},
  {Stephenson}, {Stewart}, {Stiavelli}, {}, {Strada}, {Straughn}, {Streetman},
  {Strickland}, {Strobele}, {Stuhlinger}, {Stys}, {Such}, {Sukhatme},
  {Sullivan}, {Sullivan}, {Sumner}, {Sun}, {Sunnquist}, {Swade}, {Swam},
  {Swenton}, {Swoish}, {Tam Litten}, {Tamas}, {Tao}, {Taylor}, {Taylor}, {te
  Plate}, {Van Tea}, {Teague}, {Telfer}, {Temim}, {Texter}, {Thatte},
  {Thompson}, {Thompson}, {Thomson}, {Thronson}, {Tierney}, {Tikkanen},
  {Tinnin}, {Tippet}, {Todd}, {Tran}, {Trauger}, {Trejo}, {Vinh Truong},
  {Tsukamoto}, {Tufail}, {Tumlinson}, {Tustain}, {Tyra}, {Ubeda}, {Underwood},
  {Uzzo}, {Vaclavik}, {Valenduc}, {Valenti}, {Van Campen}, {van de Wetering},
  {Van Der Marel}, {van Haarlem}, {Vandenbussche}, {van Dishoeck},
  {Vanterpool}, {Vernoy}, {Vila Costas}, {Volk}, {Voorzaat}, {Voyton}, {Vydra},
  {Waddy}, {Waelkens}, {Wahlgren}, {Walker}, {Wander}, {Warfield}, {Warner},
  {Wasiak}, {Wasiak}, {Wehner}, {Weiler}, {Weilert}, {Weiss}, {Wells}, {Welty},
  {Wheate}, {Wheeler}, {White}, {Whitehouse}, {Whiteleather}, {Whitman},
  {Williams}, {Willmer}, {Willott}, {Willoughby}, {Wilson}, {Wilson}, {Wilson},
  {Windhorst}, {Wislowski}, {Wolfe}, {Wolfe}, {Wolff}, {Wondel}, {Woo},
  {Woods}, {Worden}, {Workman}, {Wright}, {Wu}, {Wu}, {Wun}, {Wymer},
  {Yadetie}, {Yan}, {Yang}, {Yates}, {Yeager}, {Yerger}, {Young}, {Young},
  {Yu}, {Yu}, {Zak}, {Zeidler}, {Zepp}, {Zhou}, {Zincke}, {Zonak}, \&
  {Zondag}}]{JWST}
{Gardner}, J.~P., {Mather}, J.~C., {Abbott}, R., {et~al.} 2023, \pasp, 135,
  068001

\bibitem[{{Ginzburg} {et~al.}(2018){Ginzburg}, {Schlichting}, \&
  {Sari}}]{Ginzburg2018}
{Ginzburg}, S., {Schlichting}, H.~E., \& {Sari}, R. 2018, \mnras, 476, 759

\bibitem[{{Goffo} {et~al.}(2023){Goffo}, {Gandolfi}, {Egger}, {Mustill},
  {Albrecht}, {Hirano}, {Kochukhov}, {Astudillo-Defru}, {Barragan}, {Serrano},
  {Hatzes}, {Alibert}, {Guenther}, {Dai}, {Lam}, {Csizmadia}, {Smith},
  {Fossati}, {Luque}, {Rodler}, {Winther}, {R{\o}rsted}, {Alarcon}, {Bonfils},
  {Cochran}, {Deeg}, {Jenkins}, {Korth}, {Livingston}, {Meech}, {Murgas},
  {Orell-Miquel}, {Osborne}, {Palle}, {Persson}, {Redfield}, {Ricker},
  {Seager}, {Vanderspek}, {Van Eylen}, \& {Winn}}]{Goffo2023}
{Goffo}, E., {Gandolfi}, D., {Egger}, J.~A., {et~al.} 2023, \apjl, 955, L3

\bibitem[{{Guilera} {et~al.}(2020){Guilera}, {S{\'a}ndor}, {Ronco},
  {Venturini}, \& {Miller Bertolami}}]{Guilera20}
{Guilera}, O.~M., {S{\'a}ndor}, Z., {Ronco}, M.~P., {Venturini}, J., \& {Miller
  Bertolami}, M.~M. 2020, \aap, 642, A140

\bibitem[{{Gupta} \& {Schlichting}(2019)}]{Gupta2019}
{Gupta}, A. \& {Schlichting}, H.~E. 2019, \mnras, 487, 24

\bibitem[{{Hadden} \& {Lithwick}(2017)}]{HaddenLithwick2017}
{Hadden}, S. \& {Lithwick}, Y. 2017, \aj, 154, 5

\bibitem[{{Hatzes} \& {Rauer}(2015)}]{Hatzes2015}
{Hatzes}, A.~P. \& {Rauer}, H. 2015, \apjl, 810, L25

\bibitem[{{Heidenreich} {et~al.}(2013){Heidenreich}, {Schindler}, \&
  {Sperlich}}]{heidenreich_bandwidth_2013}
{Heidenreich}, N.-B., {Schindler}, A., \& {Sperlich}, S. 2013, AStA, 97, 403

\bibitem[{{Helled}(2023)}]{Helled2023}
{Helled}, R. 2023, \aap, 675, L8

\bibitem[{{Ho} {et~al.}(2024){Ho}, {Rogers}, {Van Eylen}, {Owen}, \&
  {Schlichting}}]{Ho2024}
{Ho}, C. S.~K., {Rogers}, J.~G., {Van Eylen}, V., {Owen}, J.~E., \&
  {Schlichting}, H.~E. 2024, \mnras, 531, 3698

\bibitem[{{Ho} \& {Van Eylen}(2023)}]{Ho2023}
{Ho}, C. S.~K. \& {Van Eylen}, V. 2023, \mnras, 519, 4056

\bibitem[{{Howell} {et~al.}(2014){Howell}, {Sobeck}, {Haas}, {Still},
  {Barclay}, {Mullally}, {Troeltzsch}, {Aigrain}, {Bryson}, {Caldwell},
  {Chaplin}, {Cochran}, {Huber}, {Marcy}, {Miglio}, {Najita}, {Smith},
  {Twicken}, \& {Fortney}}]{K2}
{Howell}, S.~B., {Sobeck}, C., {Haas}, M., {et~al.} 2014, \pasp, 126, 398

\bibitem[{{Jehin} {et~al.}(2011){Jehin}, {Gillon}, {Queloz}, {Magain},
  {Manfroid}, {Chantry}, {Lendl}, {Hutsem{\'e}kers}, \& {Udry}}]{TRAPPIST}
{Jehin}, E., {Gillon}, M., {Queloz}, D., {et~al.} 2011, The Messenger, 145, 2

\bibitem[{{Jin} \& {Mordasini}(2018)}]{Jin2018}
{Jin}, S. \& {Mordasini}, C. 2018, \apj, 853, 163

\bibitem[{{Kanodia} {et~al.}(2019){Kanodia}, {Wolfgang}, {Stefansson}, {Ning},
  \& {Mahadevan}}]{Kanodia2019}
{Kanodia}, S., {Wolfgang}, A., {Stefansson}, G.~K., {Ning}, B., \& {Mahadevan},
  S. 2019, \apj, 882, 38

\bibitem[{{Kaye} {et~al.}(2022){Kaye}, {Vissapragada}, {G{\"u}nther},
  {Aigrain}, {Mikal-Evans}, {Jensen}, {Parviainen}, {Pozuelos}, {Abe}, {Acton},
  {Agabi}, {Alves}, {Anderson}, {Armstrong}, {Barkaoui}, {Barrag{\'a}n},
  {Benneke}, {Boyd}, {Brahm}, {Bruni}, {Bryant}, {Burleigh}, {Casewell},
  {Ciardi}, {Cloutier}, {Collins}, {Collins}, {Conti}, {Crossfield}, {Crouzet},
  {Daylan}, {Dragomir}, {Dransfield}, {Fabrycky}, {Fausnaugh},
  {Fu{\H{u}}r{\'e}sz}, {Gan}, {Gill}, {Gillon}, {Goad}, {Gorjian},
  {Greklek-McKeon}, {Guerrero}, {Guillot}, {Jehin}, {Jenkins}, {Lendl},
  {Kamler}, {Kane}, {Kielkopf}, {Kunimoto}, {Marie-Sainte}, {McCormac},
  {M{\'e}karnia}, {Morales}, {Moyano}, {Palle}, {Parmentier}, {Relles},
  {Schmider}, {Schwarz}, {Seager}, {Smith}, {Tan}, {Taylor}, {Triaud},
  {Twicken}, {Udry}, {Vines}, {Wang}, {Wheatley}, \& {Winn}}]{Kaye2022}
{Kaye}, L., {Vissapragada}, S., {G{\"u}nther}, M.~N., {et~al.} 2022, \mnras,
  510, 5464

\bibitem[{{Kempton} {et~al.}(2018){Kempton}, {Bean}, {Louie}, {Deming}, {Koll},
  {Mansfield}, {Christiansen}, {L{\'o}pez-Morales}, {Swain}, {Zellem},
  {Ballard}, {Barclay}, {Barstow}, {Batalha}, {Beatty}, {Berta-Thompson},
  {Birkby}, {Buchhave}, {Charbonneau}, {Cowan}, {Crossfield}, {de Val-Borro},
  {Doyon}, {Dragomir}, {Gaidos}, {Heng}, {Hu}, {Kane}, {Kreidberg}, {Mallonn},
  {Morley}, {Narita}, {Nascimbeni}, {Pall{\'e}}, {Quintana}, {Rauscher},
  {Seager}, {Shkolnik}, {Sing}, {Sozzetti}, {Stassun}, {Valenti}, \& {von
  Essen}}]{Kempton2018}
{Kempton}, E. M.~R., {Bean}, J.~L., {Louie}, D.~R., {et~al.} 2018, \pasp, 130,
  114401

\bibitem[{{Kite} {et~al.}(2019){Kite}, {Fegley}, {Schaefer}, \&
  {Ford}}]{Kite2019}
{Kite}, E.~S., {Fegley}, Bruce, J., {Schaefer}, L., \& {Ford}, E.~B. 2019,
  \apjl, 887, L33

\bibitem[{{Kubyshkina} \& {Vidotto}(2021)}]{Kubyshkina2021}
{Kubyshkina}, D. \& {Vidotto}, A.~A. 2021, \mnras, 504, 2034

\bibitem[{{Lammer} {et~al.}(2003){Lammer}, {Selsis}, {Ribas}, {Guinan},
  {Bauer}, \& {Weiss}}]{Lammer2003}
{Lammer}, H., {Selsis}, F., {Ribas}, I., {et~al.} 2003, \apjl, 598, L121

\bibitem[{{Lee} {et~al.}(2022){Lee}, {Karalis}, \& {Thorngren}}]{Lee2022}
{Lee}, E.~J., {Karalis}, A., \& {Thorngren}, D.~P. 2022, \apj, 941, 186

\bibitem[{{L{\'e}ger} {et~al.}(2004){L{\'e}ger}, {Selsis}, {Sotin}, {Guillot},
  {Despois}, {Mawet}, {Ollivier}, {Lab{\`e}que}, {Valette}, {Brachet},
  {Chazelas}, \& {Lammer}}]{Leger2004}
{L{\'e}ger}, A., {Selsis}, F., {Sotin}, C., {et~al.} 2004, \icarus, 169, 499

\bibitem[{{Leleu} {et~al.}(2023){Leleu}, {Delisle}, {Udry}, {Mardling},
  {Turbet}, {Egger}, {Alibert}, {Chatel}, {Eggenberger}, \&
  {Stalport}}]{Leleu2023}
{Leleu}, A., {Delisle}, J.~B., {Udry}, S., {et~al.} 2023, \aap, 669, A117

\bibitem[{{Li} {et~al.}(2020){Li}, {Vo{\v{c}}adlo}, {Sun}, \&
  {Brodholt}}]{Li2020}
{Li}, Y., {Vo{\v{c}}adlo}, L., {Sun}, T., \& {Brodholt}, J.~P. 2020, Nature
  Geoscience, 13, 453

\bibitem[{{Lithwick} {et~al.}(2012){Lithwick}, {Xie}, \& {Wu}}]{Lithwick2012}
{Lithwick}, Y., {Xie}, J., \& {Wu}, Y. 2012, \apj, 761, 122

\bibitem[{{Lopez} \& {Fortney}(2014)}]{Lopez2014}
{Lopez}, E.~D. \& {Fortney}, J.~J. 2014, \apj, 792, 1

\bibitem[{{Lovis} \& {Fischer}(2010)}]{Lovis2010}
{Lovis}, C. \& {Fischer}, D. 2010, in Exoplanets, ed. S.~{Seager}, 27--53

\bibitem[{{Luger} \& {Barnes}(2015)}]{Luger2015}
{Luger}, R. \& {Barnes}, R. 2015, Astrobiology, 15, 119

\bibitem[{{Luo} {et~al.}(2024){Luo}, {Dorn}, \& {Deng}}]{LuoDorn2024}
{Luo}, H., {Dorn}, C., \& {Deng}, J. 2024, submitted to Nature Astronomy,
  arXiv:2401.16394

\bibitem[{{Luque} \& {Pall{\'e}}(2022)}]{Luque2022}
{Luque}, R. \& {Pall{\'e}}, E. 2022, Science, 377, 1211

\bibitem[{Mann \& Whitney(1947)}]{Mann1947}
Mann, H.~B. \& Whitney, D.~R. 1947, The Annals of Mathematical Statistics, 18,
  50

\bibitem[{{M{\'e}ndez} \& {Rivera-Valent{\'\i}n}(2017)}]{Mendez2017}
{M{\'e}ndez}, A. \& {Rivera-Valent{\'\i}n}, E.~G. 2017, \apjl, 837, L1

\bibitem[{{Ment} \& {Charbonneau}(2023)}]{MentCharbonneau2023}
{Ment}, K. \& {Charbonneau}, D. 2023, \aj, 165, 265

\bibitem[{{Miguel} {et~al.}(2020){Miguel}, {Cridland}, {Ormel}, {Fortney}, \&
  {Ida}}]{Miguel2020}
{Miguel}, Y., {Cridland}, A., {Ormel}, C.~W., {Fortney}, J.~J., \& {Ida}, S.
  2020, \mnras, 491, 1998

\bibitem[{{Mousis} {et~al.}(2020){Mousis}, {Deleuil}, {Aguichine}, {Marcq},
  {Naar}, {Aguirre}, {Brugger}, \& {Gon{\c{c}}alves}}]{Mousis2020}
{Mousis}, O., {Deleuil}, M., {Aguichine}, A., {et~al.} 2020, \apjl, 896, L22

\bibitem[{{Mulders} {et~al.}(2015){Mulders}, {Pascucci}, \&
  {Apai}}]{Mulders2015}
{Mulders}, G.~D., {Pascucci}, I., \& {Apai}, D. 2015, \apj, 798, 112

\bibitem[{{M{\"u}ller} {et~al.}(2024){M{\"u}ller}, {Baron}, {Helled}, {Bouchy},
  \& {Parc}}]{Muller2023}
{M{\"u}ller}, S., {Baron}, J., {Helled}, R., {Bouchy}, F., \& {Parc}, L. 2024,
  \aap, 686, A296

\bibitem[{{Nowak} {et~al.}(2020){Nowak}, {Luque}, {Parviainen}, {Pall{\'e}},
  {Molaverdikhani}, {B{\'e}jar}, {Lillo-Box}, {Rodr{\'\i}guez-L{\'o}pez},
  {Caballero}, {Zechmeister}, {Passegger}, {Cifuentes}, {Schweitzer}, {Narita},
  {Cale}, {Espinoza}, {Murgas}, {Hidalgo}, {Zapatero Osorio}, {Pozuelos},
  {Aceituno}, {Amado}, {Barkaoui}, {Barrado}, {Bauer}, {Benkhaldoun},
  {Caldwell}, {Casasayas Barris}, {Chaturvedi}, {Chen}, {Collins}, {Collins},
  {Cort{\'e}s-Contreras}, {Crossfield}, {de Le{\'o}n}, {D{\'\i}ez Alonso},
  {Dreizler}, {El Mufti}, {Esparza-Borges}, {Essack}, {Fukui}, {Gaidos},
  {Gillon}, {Gonzales}, {Guerra}, {Hatzes}, {Henning}, {Herrero}, {Hesse},
  {Hirano}, {Howell}, {Jeffers}, {Jehin}, {Jenkins}, {Kaminski}, {Kemmer},
  {Kielkopf}, {Kossakowski}, {Kotani}, {K{\"u}rster}, {Lafarga}, {Latham},
  {Law}, {Lissauer}, {Lodieu}, {Madrigal-Aguado}, {Mann}, {Massey}, {Matson},
  {Matthews}, {Monta{\~n}{\'e}s-Rodr{\'\i}guez}, {Montes}, {Morales}, {Mori},
  {Nagel}, {Oshagh}, {Pedraz}, {Plavchan}, {Pollacco}, {Quirrenbach},
  {Reffert}, {Reiners}, {Ribas}, {Ricker}, {Rose}, {Schlecker}, {Schlieder},
  {Seager}, {Stangret}, {Stock}, {Tamura}, {Tanner}, {Teske}, {Trifonov},
  {Twicken}, {Vanderspek}, {Watanabe}, {Wittrock}, {Ziegler}, \&
  {Zohrabi}}]{Nowak2020}
{Nowak}, G., {Luque}, R., {Parviainen}, H., {et~al.} 2020, \aap, 642, A173

\bibitem[{{Otegi} {et~al.}(2020){Otegi}, {Bouchy}, \& {Helled}}]{Otegi2020}
{Otegi}, J.~F., {Bouchy}, F., \& {Helled}, R. 2020, \aap, 634, A43

\bibitem[{{Owen} \& {Wu}(2017)}]{Owen2017}
{Owen}, J.~E. \& {Wu}, Y. 2017, \apj, 847, 29

\bibitem[{{Paardekooper} {et~al.}(2010){Paardekooper}, {Baruteau}, {Crida}, \&
  {Kley}}]{Paardekooper2010}
{Paardekooper}, S.~J., {Baruteau}, C., {Crida}, A., \& {Kley}, W. 2010, \mnras,
  401, 1950

\bibitem[{{Parviainen} {et~al.}(2024){Parviainen}, {Luque}, \&
  {Palle}}]{Parviainen2023}
{Parviainen}, H., {Luque}, R., \& {Palle}, E. 2024, \mnras, 527, 5693

\bibitem[{{Peacock} {et~al.}(2020){Peacock}, {Barman}, {Shkolnik}, {Loyd},
  {Schneider}, {Pagano}, \& {Meadows}}]{Peacock2020}
{Peacock}, S., {Barman}, T., {Shkolnik}, E.~L., {et~al.} 2020, \apj, 895, 5

\bibitem[{{Pecaut} \& {Mamajek}(2013)}]{Pecaut2013}
{Pecaut}, M.~J. \& {Mamajek}, E.~E. 2013, \apjs, 208, 9

\bibitem[{{Pierrehumbert}(2023)}]{Pierrehumbert2023}
{Pierrehumbert}, R.~T. 2023, \apj, 944, 20

\bibitem[{{Quirrenbach} {et~al.}(2020){Quirrenbach}, {CARMENES Consortium},
  {Amado}, {Ribas}, {Reiners}, {Caballero}, {Aceituno}, {Alacid},
  {Alonso-Floriano}, {Anglada-Escud{\'e}}, {Azzaro}, {Baroch}, {Bauer},
  {Becerril}, {B{\'e}jar}, {Bluhm}, {Calvo Ortega}, {Cardona Guill{\'e}n},
  {Casasayas-Barris}, {Chaturvedi}, {Cifuentes}, {Colom{\'e}}, {Conte},
  {Cort{\'e}s-Contreras}, {Czesla}, {D{\'\i}ez-Alonso}, {Dom{\'\i}nguez
  Fern{\'a}ndez}, {Dreizler}, {Duque-Arribas}, {Espinoza}, {Fuhrmeister},
  {Galad{\'\i}-Enr{\'\i}quez}, {Gar{\textasciiacute}a Quintana},
  {Gonz{\'a}lez-Alvare}, {Gonz{\'a}lez Cuesta}, {Gonz{\'a}lez Hern{\'a}ndez},
  {Guenther}, {de Guindos}, {Hatzes}, {Henning}, {Herbort}, {Herrero}, {Hintz},
  {Iglesias-P{\'a}ra}, {Jeffers}, {Johnson}, {de Juan}, {Kaminski}, {Kemmer},
  {Khaimova}, {Khalafinejad}, {Klahr}, {Kossakowski}, {Kreidberg},
  {K{\"u}rster}, {Labarga}, {Lafarga}, {Lamp{\'o}n}, {Lara}, {Lillo-Box},
  {Lodieu}, {L{\'o}pez Gallifa}, {L{\'o}pez Gonz{\'a}lez}, {L{\'o}pez-Puertas},
  {Luque}, {Marfil}, {Mart{\'\i}n-Ruiz}, {Matth{\'e}}, {Molaverdikhani},
  {Montes}, {Morales}, {Morales-Calder{\'o}on}, {Nagel}, {Nortmann}, {Nowak},
  {Ofir}, {Oshaghi}, {Pall{\'e}}, {Passegger}, {Pavlov}, {Pedraz},
  {Perdelwitz}, {Perger}, {Reffert}, {Revilla}, {Rodr{\'\i}guez},
  {Rodr{\'\i}guez L{\'o}pez}, {Sabotta}, {Sadegi}, {Sairam}, {Salz},
  {S{\'a}nchez-L{\'o}pez}, {Sanz-Forcada}, {Sarkis}, {Sch{\"a}fer}, {Schiller},
  {Schlecker}, {Schmitt}, {Sch{\"o}fer}, {Schweitzer}, {Seiferta}, {Shan},
  {Shulyak}, {Skrzypinski}, {Solano}, {Soto}, {Stahl}, {Stangret}, {Stock},
  {Strachan}, {Stuber}, {St{\"u}rmer}, {Tabernero}, {Tal-Or}, {Tala-Pinto},
  {Trifonov}, {Vanaverbeke}, {Yan}, {Zapatero Osorio}, \&
  {Zechmeister}}]{CARMENES}
{Quirrenbach}, A., {CARMENES Consortium}, {Amado}, P.~J., {et~al.} 2020, in
  Society of Photo-Optical Instrumentation Engineers (SPIE) Conference Series,
  Vol. 11447, Society of Photo-Optical Instrumentation Engineers (SPIE)
  Conference Series, 114473C

\bibitem[{{Rauer} {et~al.}(2014){Rauer}, {Catala}, {Aerts}, {Appourchaux},
  {Benz}, {Brandeker}, {Christensen-Dalsgaard}, {Deleuil}, {Gizon}, {Goupil},
  {G{\"u}del}, {Janot-Pacheco}, {Mas-Hesse}, {Pagano}, {Piotto}, {Pollacco},
  {Santos}, {Smith}, {Su{\'a}rez}, {Szab{\'o}}, {Udry}, {Adibekyan}, {Alibert},
  {Almenara}, {Amaro-Seoane}, {Eiff}, {Asplund}, {Antonello}, {Barnes},
  {Baudin}, {Belkacem}, {Bergemann}, {Bihain}, {Birch}, {Bonfils}, {Boisse},
  {Bonomo}, {Borsa}, {Brand{\~a}o}, {Brocato}, {Brun}, {Burleigh}, {Burston},
  {Cabrera}, {Cassisi}, {Chaplin}, {Charpinet}, {Chiappini}, {Church},
  {Csizmadia}, {Cunha}, {Damasso}, {Davies}, {Deeg}, {D{\'\i}az}, {Dreizler},
  {Dreyer}, {Eggenberger}, {Ehrenreich}, {Eigm{\"u}ller}, {Erikson}, {Farmer},
  {Feltzing}, {de Oliveira Fialho}, {Figueira}, {Forveille}, {Fridlund},
  {Garc{\'\i}a}, {Giommi}, {Giuffrida}, {Godolt}, {Gomes da Silva}, {Granzer},
  {Grenfell}, {Grotsch-Noels}, {G{\"u}nther}, {Haswell}, {Hatzes},
  {H{\'e}brard}, {Hekker}, {Helled}, {Heng}, {Jenkins}, {Johansen},
  {Khodachenko}, {Kislyakova}, {Kley}, {Kolb}, {Krivova}, {Kupka}, {Lammer},
  {Lanza}, {Lebreton}, {Magrin}, {Marcos-Arenal}, {Marrese}, {Marques},
  {Martins}, {Mathis}, {Mathur}, {Messina}, {Miglio}, {Montalban}, {Montalto},
  {Monteiro}, {Moradi}, {Moravveji}, {Mordasini}, {Morel}, {Mortier},
  {Nascimbeni}, {Nelson}, {Nielsen}, {Noack}, {Norton}, {Ofir}, {Oshagh},
  {Ouazzani}, {P{\'a}pics}, {Parro}, {Petit}, {Plez}, {Poretti}, {Quirrenbach},
  {Ragazzoni}, {Raimondo}, {Rainer}, {Reese}, {Redmer}, {Reffert},
  {Rojas-Ayala}, {Roxburgh}, {Salmon}, {Santerne}, {Schneider}, {Schou},
  {Schuh}, {Schunker}, {Silva-Valio}, {Silvotti}, {Skillen}, {Snellen}, {Sohl},
  {Sousa}, {Sozzetti}, {Stello}, {Strassmeier}, {{\v{S}}vanda}, {Szab{\'o}},
  {Tkachenko}, {Valencia}, {Van Grootel}, {Vauclair}, {Ventura}, {Wagner},
  {Walton}, {Weingrill}, {Werner}, {Wheatley}, \& {Zwintz}}]{Rauer2014}
{Rauer}, H., {Catala}, C., {Aerts}, C., {et~al.} 2014, Experimental Astronomy,
  38, 249

\bibitem[{{Ribas} {et~al.}(2016){Ribas}, {Bolmont}, {Selsis}, {Reiners},
  {Leconte}, {Raymond}, {Engle}, {Guinan}, {Morin}, {Turbet}, {Forget}, \&
  {Anglada-Escud{\'e}}}]{Ribas2016}
{Ribas}, I., {Bolmont}, E., {Selsis}, F., {et~al.} 2016, \aap, 596, A111

\bibitem[{{Ricker} {et~al.}(2015){Ricker}, {Winn}, {Vanderspek}, {Latham},
  {Bakos}, {Bean}, {Berta-Thompson}, {Brown}, {Buchhave}, {Butler}, {Butler},
  {Chaplin}, {Charbonneau}, {Christensen-Dalsgaard}, {Clampin}, {Deming},
  {Doty}, {De Lee}, {Dressing}, {Dunham}, {Endl}, {Fressin}, {Ge}, {Henning},
  {Holman}, {Howard}, {Ida}, {Jenkins}, {Jernigan}, {Johnson}, {Kaltenegger},
  {Kawai}, {Kjeldsen}, {Laughlin}, {Levine}, {Lin}, {Lissauer}, {MacQueen},
  {Marcy}, {McCullough}, {Morton}, {Narita}, {Paegert}, {Palle}, {Pepe},
  {Pepper}, {Quirrenbach}, {Rinehart}, {Sasselov}, {Sato}, {Seager},
  {Sozzetti}, {Stassun}, {Sullivan}, {Szentgyorgyi}, {Torres}, {Udry}, \&
  {Villasenor}}]{TESS}
{Ricker}, G.~R., {Winn}, J.~N., {Vanderspek}, R., {et~al.} 2015, Journal of
  Astronomical Telescopes, Instruments, and Systems, 1, 014003

\bibitem[{{Rogers} {et~al.}(2023){Rogers}, {Schlichting}, \&
  {Owen}}]{Rogers2023}
{Rogers}, J.~G., {Schlichting}, H.~E., \& {Owen}, J.~E. 2023, \apjl, 947, L19

\bibitem[{{Rogers}(2015)}]{Rogers2015}
{Rogers}, L.~A. 2015, \apj, 801, 41

\bibitem[{{Seabold} \& {Perktold}(2010)}]{seabold2010statsmodels}
{Seabold}, S. \& {Perktold}, J. 2010, in 9th Python in Science Conference

\bibitem[{{Seifahrt} {et~al.}(2016){Seifahrt}, {Bean}, {St{\"u}rmer}, {Gers},
  {Grobler}, {Reed}, \& {Jones}}]{MAROONX}
{Seifahrt}, A., {Bean}, J.~L., {St{\"u}rmer}, J., {et~al.} 2016, in Society of
  Photo-Optical Instrumentation Engineers (SPIE) Conference Series, Vol. 9908,
  Ground-based and Airborne Instrumentation for Astronomy VI, ed. C.~J.
  {Evans}, L.~{Simard}, \& H.~{Takami}, 990818

\bibitem[{{Selsis} {et~al.}(2023){Selsis}, {Leconte}, {Turbet}, {Chaverot}, \&
  {Bolmont}}]{Selsis2023}
{Selsis}, F., {Leconte}, J., {Turbet}, M., {Chaverot}, G., \& {Bolmont}, {\'E}.
  2023, \nat, 620, 287

\bibitem[{{Sestovic} {et~al.}(2018){Sestovic}, {Demory}, \&
  {Queloz}}]{Sestovic2018}
{Sestovic}, M., {Demory}, B.-O., \& {Queloz}, D. 2018, \aap, 616, A76

\bibitem[{{Shkolnik} \& {Barman}(2014)}]{Shkolnik2014}
{Shkolnik}, E.~L. \& {Barman}, T.~S. 2014, \aj, 148, 64

\bibitem[{Silverman(1981)}]{Silverman1981}
Silverman, B.~W. 1981, Journal of the Royal Statistical Society: Series B
  (Methodological), 43, 97

\bibitem[{Silverman(2018)}]{Silverman2018}
Silverman, B.~W. 2018, Density estimation for statistics and data analysis
  (Routledge)

\bibitem[{{Southworth}(2011)}]{Southworth2011}
{Southworth}, J. 2011, \mnras, 417, 2166

\bibitem[{{Stefansson} {et~al.}(2020){Stefansson}, {Mahadevan}, {Maney},
  {Ninan}, {Robertson}, {Rajagopal}, {Haase}, {Allen}, {Ford}, {Winn},
  {Wolfgang}, {Dawson}, {Wisniewski}, {Bender}, {Ca{\~n}as}, {Cochran},
  {Diddams}, {Fredrick}, {Halverson}, {Hearty}, {Hebb}, {Kanodia}, {Levi},
  {Metcalf}, {Monson}, {Ramsey}, {Roy}, {Schwab}, {Terrien}, \&
  {Wright}}]{Stefansson2020}
{Stefansson}, G., {Mahadevan}, S., {Maney}, M., {et~al.} 2020, \aj, 160, 192

\bibitem[{Sturges(1926)}]{Sturges1926}
Sturges, H.~A. 1926, Journal of the American Statistical Association

\bibitem[{{Tagawa} {et~al.}(2021){Tagawa}, {Sakamoto}, {Hirose}, {Yokoo},
  {Hernlund}, {Ohishi}, \& {Yurimoto}}]{Tagawa2021}
{Tagawa}, S., {Sakamoto}, N., {Hirose}, K., {et~al.} 2021, Nature
  Communications, 12, 2588

\bibitem[{{Thorngren} \& {Fortney}(2018)}]{Thorngren2018}
{Thorngren}, D.~P. \& {Fortney}, J.~J. 2018, \aj, 155, 214

\bibitem[{{Turbet} {et~al.}(2020){Turbet}, {Bolmont}, {Ehrenreich}, {Gratier},
  {Leconte}, {Selsis}, {Hara}, \& {Lovis}}]{Turbet2020}
{Turbet}, M., {Bolmont}, E., {Ehrenreich}, D., {et~al.} 2020, \aap, 638, A41

\bibitem[{{Van Eylen} {et~al.}(2018){Van Eylen}, {Agentoft}, {Lundkvist},
  {Kjeldsen}, {Owen}, {Fulton}, {Petigura}, \& {Snellen}}]{VanEylen2018}
{Van Eylen}, V., {Agentoft}, C., {Lundkvist}, M.~S., {et~al.} 2018, \mnras,
  479, 4786

\bibitem[{{Van Eylen} {et~al.}(2021){Van Eylen}, {Astudillo-Defru}, {Bonfils},
  {Livingston}, {Hirano}, {Luque}, {Lam}, {Justesen}, {Winn}, {Gandolfi},
  {Nowak}, {Palle}, {Albrecht}, {Dai}, {Campos Estrada}, {Owen},
  {Foreman-Mackey}, {Fridlund}, {Korth}, {Mathur}, {Forveille}, {Mikal-Evans},
  {Osborne}, {Ho}, {Almenara}, {Artigau}, {Barrag{\'a}n}, {Barros}, {Bouchy},
  {Cabrera}, {Caldwell}, {Charbonneau}, {Chaturvedi}, {Cochran}, {Csizmadia},
  {Damasso}, {Delfosse}, {De Medeiros}, {D{\'\i}az}, {Doyon}, {Esposito},
  {F{\H{u}}r{\'e}sz}, {Figueira}, {Georgieva}, {Goffo}, {Grziwa}, {Guenther},
  {Hatzes}, {Jenkins}, {Kabath}, {Knudstrup}, {Latham}, {Lavie}, {Lovis},
  {Mennickent}, {Mullally}, {Murgas}, {Narita}, {Pepe}, {Persson}, {Redfield},
  {Ricker}, {Santos}, {Seager}, {Serrano}, {Smith}, {Su{\'a}rez Mascare{\~n}o},
  {Subjak}, {Twicken}, {Udry}, {Vanderspek}, \& {Zapatero
  Osorio}}]{VanEylen2021}
{Van Eylen}, V., {Astudillo-Defru}, N., {Bonfils}, X., {et~al.} 2021, \mnras,
  507, 2154

\bibitem[{{Venturini} {et~al.}(2020){Venturini}, {Guilera}, {Haldemann},
  {Ronco}, \& {Mordasini}}]{Venturini2020}
{Venturini}, J., {Guilera}, O.~M., {Haldemann}, J., {Ronco}, M.~P., \&
  {Mordasini}, C. 2020, \aap, 643, L1

\bibitem[{{Venturini} {et~al.}(2024){Venturini}, {Ronco}, {Guilera},
  {Haldemann}, {Mordasini}, \& {Miller Bertolami}}]{Venturini2024}
{Venturini}, J., {Ronco}, M.~P., {Guilera}, O.~M., {et~al.} 2024, \aap, 686, L9

\bibitem[{{Vidal-Madjar} {et~al.}(2004){Vidal-Madjar}, {D{\'e}sert},
  {Lecavelier des Etangs}, {H{\'e}brard}, {Ballester}, {Ehrenreich}, {Ferlet},
  {McConnell}, {Mayor}, \& {Parkinson}}]{Vidal2004}
{Vidal-Madjar}, A., {D{\'e}sert}, J.~M., {Lecavelier des Etangs}, A., {et~al.}
  2004, \apjl, 604, L69

\bibitem[{{Vivien} {et~al.}(2022){Vivien}, {Aguichine}, {Mousis}, {Deleuil}, \&
  {Marcq}}]{Vivien2022}
{Vivien}, H.~G., {Aguichine}, A., {Mousis}, O., {Deleuil}, M., \& {Marcq}, E.
  2022, \apj, 931, 143

\bibitem[{Wand \& Jones(1995)}]{Wand1995}
Wand, M.~P. \& Jones, M.~C. 1995, Kernel Smoothing

\bibitem[{{Weiss} \& {Marcy}(2014)}]{Weiss2014}
{Weiss}, L.~M. \& {Marcy}, G.~W. 2014, \apjl, 783, L6

\bibitem[{{Weiss} {et~al.}(2013){Weiss}, {Marcy}, {Rowe}, {Howard}, {Isaacson},
  {Fortney}, {Miller}, {Demory}, {Fischer}, {Adams}, {Dupree}, {Howell},
  {Kolbl}, {Johnson}, {Horch}, {Everett}, {Fabrycky}, \& {Seager}}]{Weiss2013}
{Weiss}, L.~M., {Marcy}, G.~W., {Rowe}, J.~F., {et~al.} 2013, \apj, 768, 14

\bibitem[{Wilcoxon(1945)}]{Wilcoxon1945}
Wilcoxon, F. 1945, Biometrics Bulletin, 1, 80

\bibitem[{{Winters} {et~al.}(2015){Winters}, {Henry}, {Lurie}, {Hambly}, {Jao},
  {Bartlett}, {Boyd}, {Dieterich}, {Finch}, {Hosey}, {Ianna}, {Riedel},
  {Slatten}, \& {Subasavage}}]{Winters2015}
{Winters}, J.~G., {Henry}, T.~J., {Lurie}, J.~C., {et~al.} 2015, \aj, 149, 5

\bibitem[{{Wolfgang} {et~al.}(2016){Wolfgang}, {Rogers}, \&
  {Ford}}]{Wolfgang2016}
{Wolfgang}, A., {Rogers}, L.~A., \& {Ford}, E.~B. 2016, IAU Focus Meeting, 29A,
  223

\bibitem[{{Zeng} {et~al.}(2019){Zeng}, {Jacobsen}, {Sasselov}, {Petaev},
  {Vanderburg}, {Lopez-Morales}, {Perez-Mercader}, {Mattsson}, {Li}, {Heising},
  {Bonomo}, {Damasso}, {Berger}, {Cao}, {Levi}, \& {Wordsworth}}]{Zeng2019}
{Zeng}, L., {Jacobsen}, S.~B., {Sasselov}, D.~D., {et~al.} 2019, Proceedings of
  the National Academy of Science, 116, 9723

\bibitem[{{Zeng} {et~al.}(2016){Zeng}, {Sasselov}, \& {Jacobsen}}]{Zeng2016}
{Zeng}, L., {Sasselov}, D.~D., \& {Jacobsen}, S.~B. 2016, \apj, 819, 127

\bibitem[{{Zhang} {et~al.}(2024){Zhang}, {Hu}, {Inglis}, {Dai}, {Bean},
  {Knutson}, {Lam}, {Goffo}, \& {Gandolfi}}]{Zhang2024}
{Zhang}, M., {Hu}, R., {Inglis}, J., {et~al.} 2024, \apjl, 961, L44

\end{thebibliography}

\begin{appendix} 
\section{Classical calculations in the PlanetS catalog}\label{appendix:calculations}

\subsection{Insolation Flux}\label{appendix:calculations_F}
To calculate the flux received by the planet in a homogeneous way, we use the following formula \citep{Bolmont2016}:

\begin{equation}
    F = \frac{L_\star}{4\pi a^2\sqrt{1-e^2}}
\end{equation}
\noindent
where L$_\star$ is the star's luminosity, $a$ is the semi-major axis, and $e$ is the eccentricity of the orbit. The luminosity used is that taken from Gaia DR3 if it exists, otherwise the value from the reference article is used. If neither exists, the value is not calculated.

\subsection{Equilibrium Temperature}\label{appendix:calculations_Teq}

We use the insolation flux to calculate the equilibrium temperature with the formula from \citet{Mendez2017}:

\begin{equation}
    T_{\text{eq}} = \left[ \frac{(1 - A)F}{4\sigma}\right]^{\frac{1}{4}}
\end{equation}
\noindent
where $F$ is the insolation flux received calculated in \ref{appendix:calculations_F}, $A$ the planet albedo or Bond Albedo (we use A = 0), and $\sigma$ the Stefan-Boltzmann constant. We consider a full heat redistribution. 
\subsection{Expected Radial Velocity Semi-Amplitude}

In the PlanetS catalog, we also compute the expected radial-velocity semi-amplitude considering the inclination, the eccentricity, the semi-major axis, and the masses of the system’s bodys with the formula from \citet{Lovis2010} :

\begin{equation}
    K=\frac{28.4329 \mathrm{~m} \mathrm{~s}^{-1}}{\sqrt{1-e^2}} \frac{M_p \sin i}{M_{\mathrm{Jup}}}\left(\frac{M_\star+M_p}{M_{\odot}}\right)^{-1 / 2}\left(\frac{a}{1 \mathrm{AU}}\right)^{-1 / 2}
\end{equation}
\noindent
where $M_p$ and $M_\star$ are the masses of the planet and star respectively, $a$ is the semi-major axis, $i$ is the inclination, and $e$ is the eccentricity of the orbit.

\newpage
\section{Comparison of the samples : LP22 vs this work}\label{appendix:comparison_samples_M}
\begin{figure}[h]
  \centering
    \resizebox{\hsize}{!}{\includegraphics{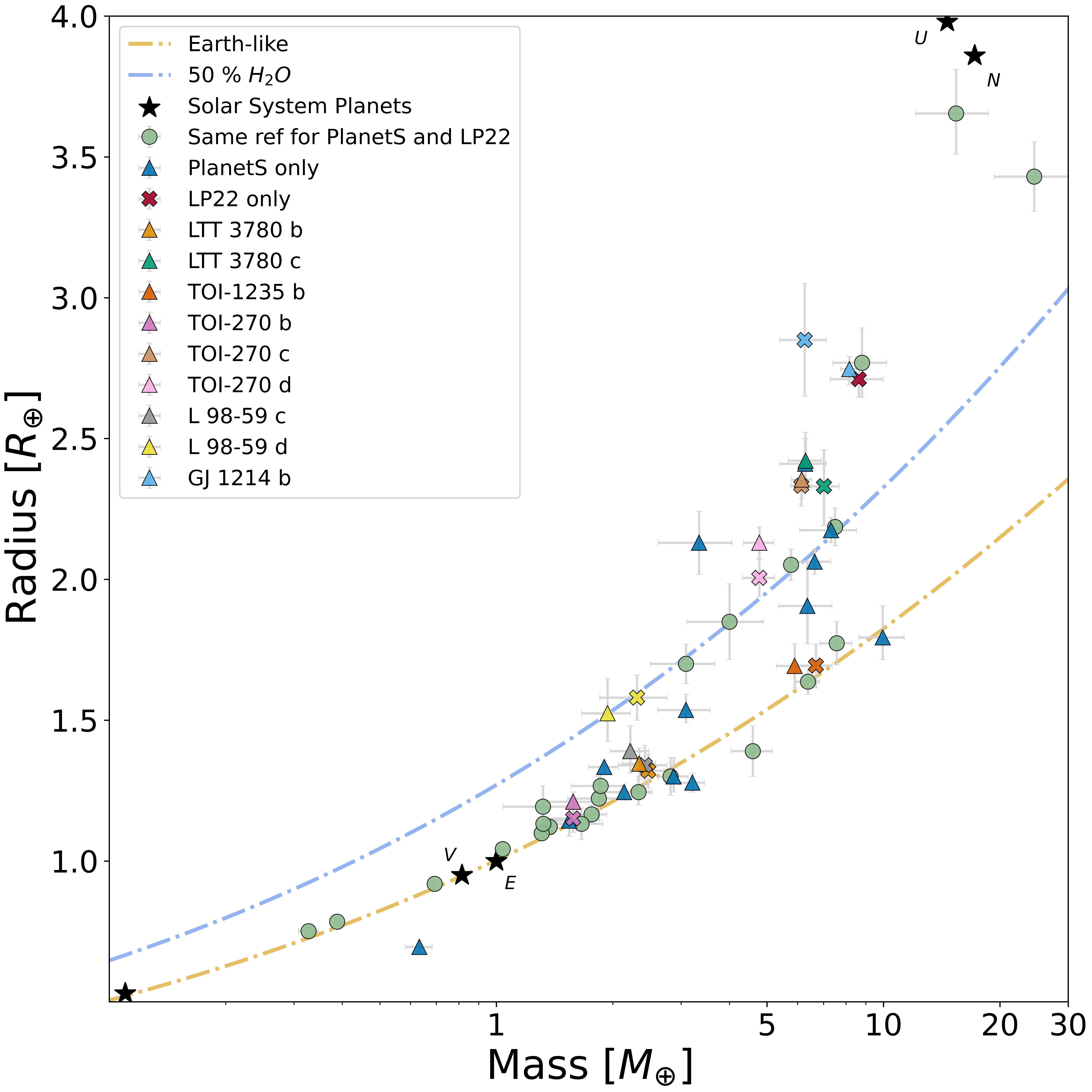}}
      \caption{Mass-radius diagram comparing the samples from \citetalias{Luque2022} and this work. The light green circles represent the planets having the same reference and parameters in both studies. The blue triangles correspond to planets only in the PlanetS catalog (new characterizations). Red crosses are planets only in the sample of \citetalias{Luque2022}. Then, the planets with different references and/or parameters in both samples are in different colors with a triangle for the parameters from the PlanetS catalog and a cross for the parameters from  \citetalias{Luque2022} sample. The composition lines of Earth-like planets (yellow) and 50\% water (blue) from \citet{Zeng2019} are displayed.}
         \label{fig:appendix_comparison_M}
\end{figure}

To complete the Section \ref{sect:sample_selection_Mdwarfs}, this appendix offers a comparison between the sample from \citetalias{Luque2022} and from this work, from the PlanetS catalog.  Fig.~\ref{fig:appendix_comparison_M} shows the differences in the planets and parameters chosen in a mass-radius diagram. The sample of \citetalias{Luque2022} study contains 34 planets and the sample from this work, from the PlanetS catalog, contains 46 planets. 24 planets have the same reference and parameters between the two samples (light green circles). Our sample has 12 new planets corresponding to characterizations dating from after the \citetalias{Luque2022} catalog creation (July 2021) (blue triangles). 1 planet is only in the \citetalias{Luque2022} catalog (red cross): it is K2-18 b because there is no reference with the required precision, or rather, those that do are flagged controversial by NASA Exoplanet Archive due to the controversial nature of planet c in the system, which in turn influences the analysis of planet b's mass in radial velocities. 9 planets are in both samples, but with different parameters as a result of a choice of a different reference. LTT 3780, TOI-1235, L 98-59 and GJ 1214 systems was re-analyzed by \citetalias{Luque2022} but the uncertainties of the derived parameters are larger than those the references used in the PlanetS catalog : \citet{Nowak2020,Bluhm2020,Demangeon2021,Cloutier2021b} respectively. A special case is the TOI-270 system and its 3 planets. The reference paper for both samples is \citet{VanEylen2021}, but the radii of the planets don't match\protect\footnote{Since this publication dates from October 2021 and the \citetalias{Luque2022} catalog from July 2021, we suspect that these are preliminary study values that have been taken.}. In our sample, the radii are $R_b = 1.21 \pm 0.03$, $R_c = 2.35 \pm 0.07$, $R_d = 2.13 \pm 0.06$, whereas \citetalias{Luque2022} considers $R_b = 1.15 \pm 0.05$, $R_c = 2.33 \pm 0.07$, $R_d = 2.00 \pm 0.07$. This difference between the reference paper and the values taken by \citetalias{Luque2022} leads to significant errors in their results, given that planet d in this system is part of their "water-world" population. \citet{Kaye2022} re-analyzed this system but the masses were determined using TTVs and the analysis do not respect our robustness criteria (described Sect.~\ref{sect:planets_catalog}). All the planets with different parameters between \citetalias{Luque2022} and this study are shown in different colors in the figure, with a triangle for PlanetS catalog parameters and a cross for \citetalias{Luque2022} catalog parameters. All the parameters used for this work are online via the PlanetS catalog\protect\footnote{\url{https://dace.unige.ch/exoplanets/}\label{catalogfootnote}} using the selection criteria set out in Sect.~\ref{sect:sample_selection_Mdwarfs}.

\newpage
\section{The limitations of histograms and KDEs with small samples}\label{appendix:hist_limits}

\begin{figure*}
\centering
   \includegraphics[width=17cm]{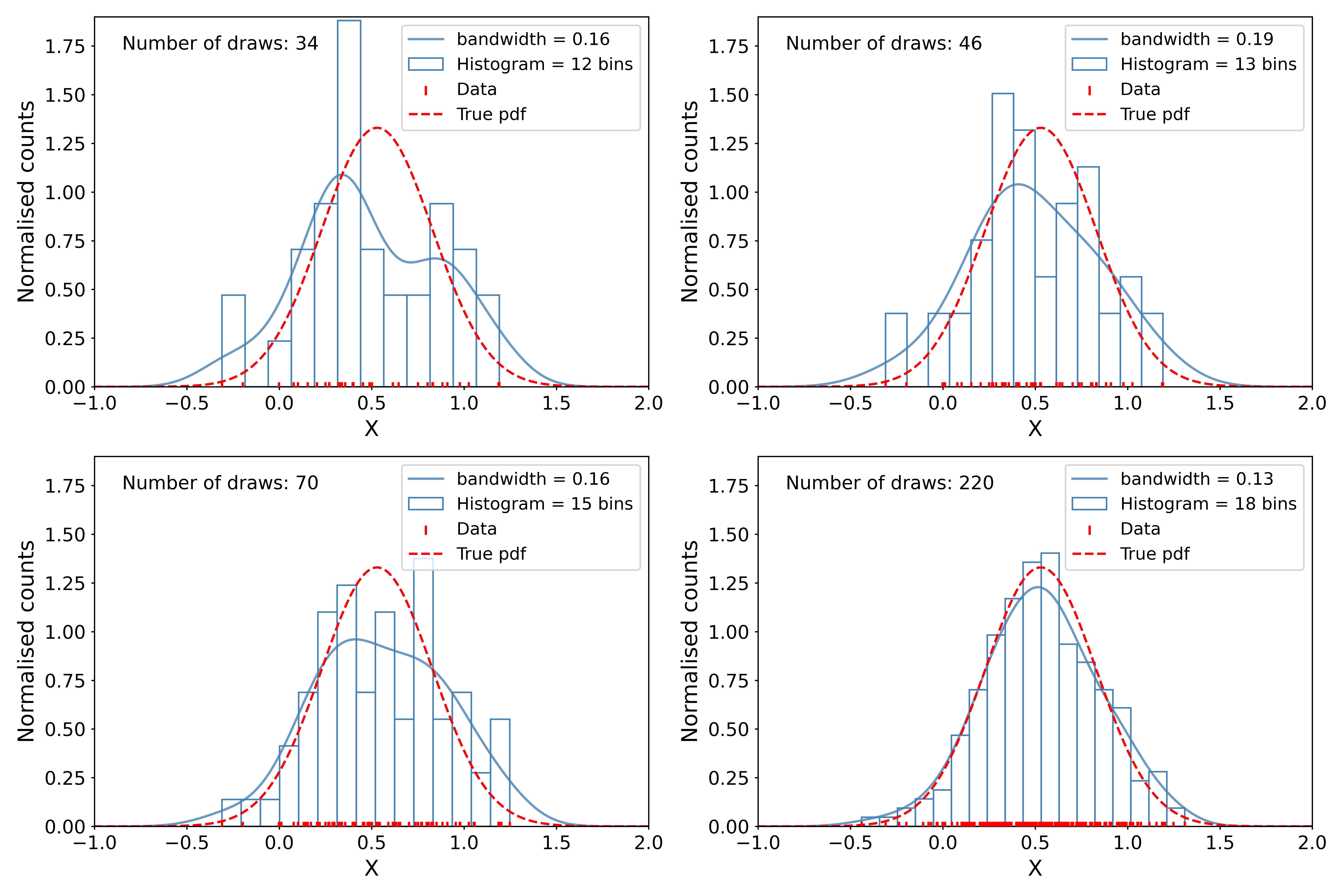}
     \caption{Comparison of histogram and KDE performance in representing a random draw in a Gaussian distribution. Each panel corresponds to a specific number of draws in the Gaussian distribution (n=34, 46, 70, 220). The actual distribution is depicted in red, with red dashed lines indicating randomly drawn points from this real distribution (using a fixed seed for repeatability). The histogram is presented with blue outlines, and the KDE is represented by a blue line. The legend provides information about the bins and bandwidth, determined by Sturges' rule and cross-validation, respectively, for each case.}
     \label{fig:appendix_limitations}
\end{figure*}
Despite an increase of 10 planets in our sample of small planets around M-dwarfs compared to LP22, our sample size is a limiting factor for robust statistical analysis. The purpose of this appendix is to show the limitations of histograms and KDEs in relation to this number.  

To do this, we drew 34, 46, 70 and 220 bins from a Gaussian distribution with mean 0.53 and standard deviation 0.3 (representing the density distribution found in Fig.\ref{fig:KDE_Mdwarfs}). We then plotted histograms using a Sturges rule \citep{Sturges1926} to define the number of bins. We also calculated the KDEs in the same way as explained in Section \ref{sect:kde_method}. The results are shown in Fig. \ref{fig:appendix_limitations}.

In the first two cases, with sample sizes of 34 and 46, a "false" bimodality emerges in the histogram. However, for the Kernel density estimation, bimodality is only apparent in the first case. The continuous nature of KDE aids in representing the actual distribution more accurately, and as the number of draws in the Gaussian distribution increases, the observed distribution converges to the real one.

To quantify this convergence, we use the Kullback-Leibler (K-L) divergence formula between two probability distributions P and Q, here P is the actual distribution and Q the calculated KDE : 

\begin{equation}
D_{KL}(P \, || \, Q) = \sum_{i} \left( p_i \cdot \log\left(\frac{p_i}{q_i}\right) \right)
\end{equation}
where p$_i$ and q$_i$ represent the respective probabilities of events in the P and Q distributions.

\begin{figure}[h]
  \centering
    \resizebox{\hsize}{!}{\includegraphics{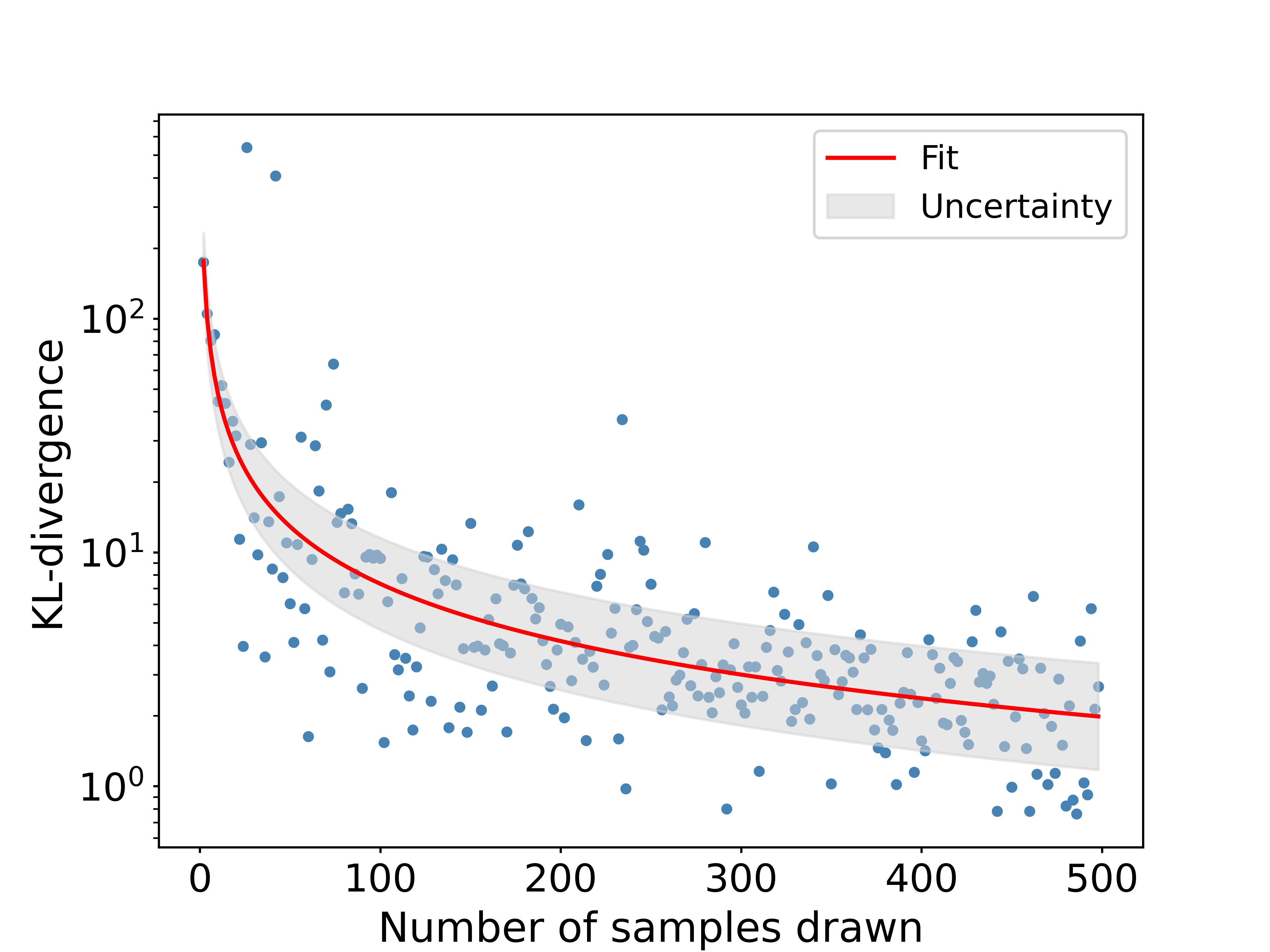}}
      \caption{K-L divergence as a function of the number of samples drawn from the Gaussian distribution. The K-L divergence is calculated between this latter and the KDE calculated from the drawn samples. The red line correspond to the fit by the power law and the gray filled area the uncertainty of the fit.}
         \label{fig:appendix_kldiv}
\end{figure}

We have then plotted in Fig. \ref{fig:appendix_kldiv} the calculation of this divergence between the actual pdf of the distribution and the KDE as a function of the number of samples drawn from the distribution. This can be modeled as a power-law evolution :
\begin{equation}
    KL = 310.67 \times n^{-0.81}
\end{equation}
with KL the resulting K-L divergence and n the number of samples drawn.

The two distributions can be considered close when the value of the divergence of K-L is less than 10. This is the case here for a number of samples drawn greater than 68$\pm$8.

In conclusion, the KDEs proves to be more effective in representing a continuous distribution with a low number of elements compared to the histogram. A larger sample size is essential for a more accurate quantification of the resulting distribution. Moreover, in the real case, uncertainties in the measurements of the parameter under consideration further emphasize the need for a larger sample size.
\end{appendix}

\end{document}